\documentclass[aps,prx,twocolumn,superscriptaddress,showpacs,longbibliography]{revtex4-2}
\usepackage{graphicx}
\usepackage{color} 
\usepackage{txfonts} 

\usepackage{ulem}
\usepackage{xspace}  
\usepackage{multirow}
\usepackage{dcolumn} %centers to the decimal point in tables using ``d''
\usepackage{booktabs}

\usepackage{xspace}  
\newcommand{\etal}{\textit{et al.}\xspace}

\makeatletter
\renewcommand{\p@subsection}{}
\makeatother

\begin{document}

%%%%   T I T L E   of  paper   %%%%
%\title{Prevailing Triaxial Shapes in Heavy Nuclei Driven by Nuclear Tensor Force}
\title{Prevailing Triaxial Shapes in Atomic Nuclei and a Quantum Theory of Rotation of Composite Objects}


\newcommand{\ariken}{      \affiliation{RIKEN Nishina Center, 2-1 Hirosawa, Wako, Saitama 351-0198, Japan}}
\newcommand{\aut}{         \affiliation{Department of Physics, The University of Tokyo, 7-3-1 Hongo, Bunkyo, Tokyo 113-0033, Japan}}
\newcommand{\atud}{         \affiliation{Institut f\"ur Kernphysik, Technische Universit\"at Darmstadt, D-64289 Darmstadt, Germany}}
\newcommand{\acns}{        \affiliation{Center for Nuclear Study, The University of Tokyo, 7-3-1 Hongo, Bunkyo, Tokyo 113-0033, Japan}}
\newcommand{\akul}{         \affiliation{KU Leuven, Instituut voor Kern- en Stralingsfysica, 3000 Leuven, Belgium}}
\newcommand{\atsuk}{         \affiliation{Center for Computational Sciences, University of Tsukuba, 1-1-1 Tennodai, Tsukuba, Ibaraki, 305-8577, Japan}}
\newcommand{\ajar}{         \affiliation{Advanced Science Research Center, Japan Atomic Energy Agency, Tokai, Ibaraki 319-1195, Japan} }
\newcommand{\ake}{         \affiliation{Quantum Computing Center, Keio University, 3-14-1 Hiyoshi, Kohoku-ku, Yokohama 223-8522, Japan} }

\newcommand{\aemp}{\email{otsuka@phys.s.u-tokyo.ac.jp}}  % author email
 
\author{T.~Otsuka}   \aemp  \aut \ariken \atud \akul%\aemt  $^*$
\author{Y.~Tsunoda}    \acns  \atsuk
\author{N.~Shimizu}    \atsuk  
\author{Y.~Utsuno}    \ajar \acns
\author{T.~Abe}   \ake \ariken
\author{H.~Ueno} \ariken

\date{\today}

\begin{abstract}     
Virtually any object can rotate: the rotation of a rod or a linear molecule appears evident, but a number of objects, including a simple example of H$_2$O molecule, are of complex shapes and their rotation is of great interest. For atomic nuclei, rotational bands have been observed in many nuclei, and their basic picture is considered to have been established in the 1950’s. We, however, show that \textcolor{black}{there may be substantial changes in the basic picture of nuclear rotation:}  
In the traditional view, as stressed by Aage Bohr in his Nobel lecture with an example of $^{166}$Er nucleus, \textcolor{black}{a large fraction of heavy (mass number $A$$>$150)} nuclei are like axially-symmetric prolate ellipsoids (i.e., with two shorter axes of equal length), rotating about one of the short axes, like a rod.   In \textcolor{black}{an alternative} picture, however, the lengths of these three axes are all different, called triaxial.  The triaxial shape yields more complex rotations.  \textcolor{black}{This alternative picture was also discussed in the past, but has not been recognized as a major picture.  We show that substantially triaxial shapes occur in a large number of heavy deformed nuclei.  Such prevailing triaxiality results in salient descriptions of experimental data over many nuclei,} as confirmed by state-of-the-art Configuration Interaction calculations.  Two origins are suggested for the triaxiality \textcolor{black}{in the heavy deformed nuclei}: (i) binding-energy gain by the symmetry restoration for triaxial shapes, and (ii) another gain by specific components of the nuclear force, like tensor force and high-multipole (e.g. hexadecupole) central force.  While the origin (i) produces \textcolor{black}{small} triaxiality for virtually all deformed nuclei, the origin (ii) produces \textcolor{black}{medium} triaxiality for a certain class of nuclei.  An example of the former is $^{154}$Sm, a typical showcase of axial symmetry but is now suggested to depict a \textcolor{black}{small}  yet finite triaxiality.   The \textcolor{black}{medium} triaxiality is discussed from various viewpoints for some exemplified nuclei including $^{166}$Er, and experimental findings, for instance, those by multiple Coulomb excitations decades ago, are re-evaluated to be supportive of the \textcolor{black}{medium} triaxiality.
Many-body structures of the $\gamma$ band and the double-$\gamma$ band are clarified, and the puzzles over them are solved.  
%There are other nuclei (e.g., $A$$<$ 150) with large triaxiality, but they are weakly deformed, excluding them from the present scope.  
%Regarding the general features of rotational states of deformed many-body systems including triaxial ones,   
\textcolor{black}{The well-known J(J+1)-K$^2$ formula of rotational excitation energies, which was derived by Taylor expansion in the past, is derived in an alternative way with polynomial property, including the previous work as an approximation.  The rotational states of strongly and triaxially deformed heavy nuclei are described within quantum many-body framework, with K quantum number shown to be practically conserved.   %The present picture of the rotation is robust and can be applied to various shapes or configurations, including clusters and molecules.    
Thus, two long-standing open problems for strongly deformed heavy nuclei, (i) occurrence and origins of triaxial shapes and (ii) quantum many-body description of their rotational bands classified by K quantum number are solved.  
\textcolor{black}{The picture of prevailing triaxial shapes thus emerges, where the empirically known rotational-band pattern appears with good K quantum number, but the internal structure is different from conventional picture {\it \`a la} A. Bohr. }
%and their rotation is described without resorting to the quantization of the free rotation of rigid body {\it \`a la} A. Bohr. 
As a feasible experimental approach to the triaxiality of the 0$^+$ ground state,} the Relativistic Heavy-Ion Collision is mentioned.  Davydov's claim of triaxial shapes over many nuclei and the validity of his rigid-triaxial-rotor model are separately assessed with high appreciation of the former.
%Substantial impacts on superheavy nuclei and fission due to the wide occurrence of triaxial shapes are mentioned.
%Their possible relations to the Nambu-Goldstone mode for the symmetry restoration are mentioned.    
\end{abstract}  

\maketitle

\tableofcontents

\section{Introduction Referring to Molecular Rotation \label{sec:1}} 

\begin{figure}[tb]
  \centering
  \includegraphics[width=8.5cm]{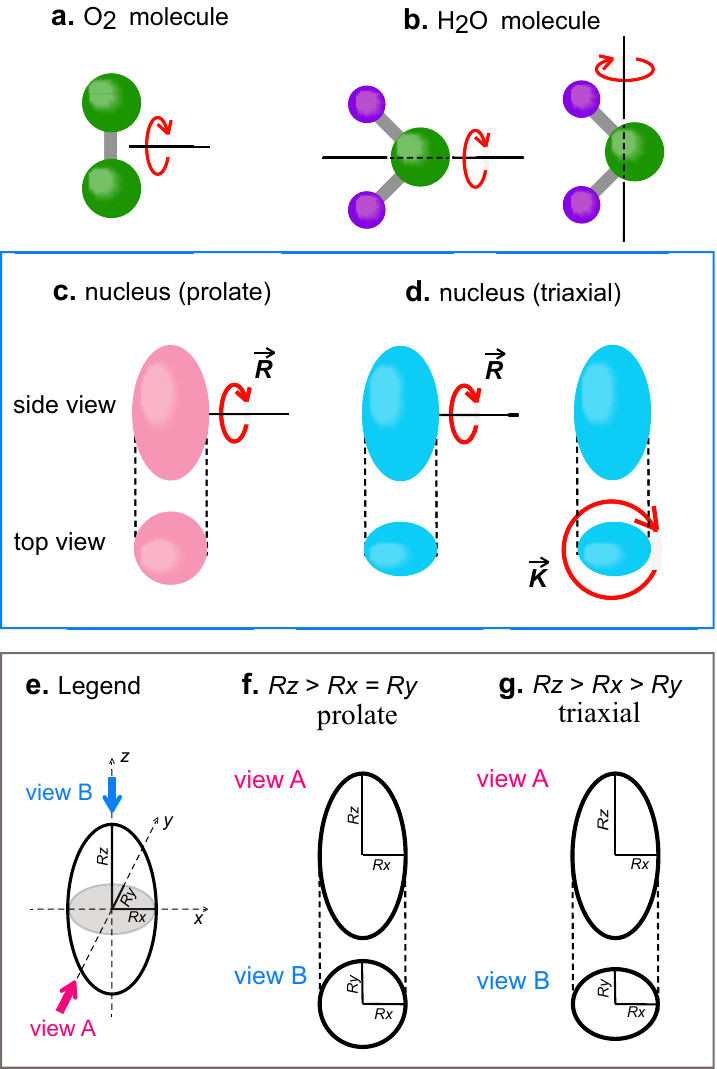}
    \caption{ {Schematic illustrations of the rotations of molecules and atomic nuclei.}
    {\bf a} O$_2$ molecule.    {\bf b} H$_2$O molecule.
    {\bf c} prolate and {\bf d} triaxial nuclear shapes, with associated rotations $\vec{R}$ and $\vec{K}$.
    {\bf e} Legend. {\bf f, g}   Principal axes of prolate and triaxial ellipsoids.
} 
  \label{fig:image}  
\end{figure}

The atomic nucleus comprises $Z$ protons and $N$ neutrons, which are collectively called nucleons.   With the mass number $A$=$Z$+$N$, a given nucleus is labeled as $^A$X where X denotes the element, e.g., $^{166}$Er for erbium-166 ($Z$=68).    The nucleus, as an assembly of many nucleons, exhibits a clear surface with a certain shape.  In many nuclei, the shape is an ellipsoid, which rotates \cite{rainwater_1950,bohr_1952,bohr_1953,bohr_mottelson_book2}.  There has been the conventional text-book picture of nuclear shape and rotation\cite{rowe_book,bohr_mottelson_book2,deshalit_book,ring_schuck_book}, as stressed by Aage Bohr in his Nobel-prize lecture in 1975 \cite{bohr_nobel}.  The present work, however, depicts a different picture supported by recent studies that were infeasible in earlier days.  Before an in-depth description of the major outcome, we overview the rotation of molecular systems, because it is easier to imagine.
Figures~\ref{fig:image}{\bf a-b} schematically display the rotation of diatomic (O$_2$) and triatomic (H$_2$O) molecules \cite{atkins}.   The O$_2$ molecule is like a rod and rotates about an axis perpendicular to the axis connecting the two O atoms (see Fig.~\ref{fig:image}{\bf a}).  It cannot rotate about the axis connecting two atoms, because the quantum state does not change by such rotation.  

%%%%%%%%%%%%  FIGURE 1  %%%%%%%%%%%%%
% Fig1: Classical cluster models

%%%%%%%%%%%%%%%%%%%%%%%%%%%%%%%%
 
The situation with the H$_2$O molecule is more complex.  As shown in Fig.~\ref{fig:image}{\bf b}, %in the view of classical mechanics, 
this molecule can rotate about more than one axis.  
%Although the rotational motions are different in quantum mechanics, the appearance of multiple rotational axes remains.  
Thus, if the molecule has a complex configuration of atoms, its rotational motion occurs about multiple axes.   

Nucleons form atomic nuclei with approximately constant nucleon density.  The surface of a nucleus can then be discussed because this density sharply drops down outside the surface in most of nuclei.  The surface forms a shape.  In a na\"ive picture with the surface tension {\it \`a la} the liquid drop, the shape should be a sphere.  However, the shape appears, in many nuclei, not to be a sphere but to be an ellipsoid or something close to it \cite{rainwater_1950,bohr_1952,bohr_1953}.  Thus, the deformation of the ellipsoid (from the sphere) is a general and basic subject of nuclear physics.  

A nuclear analogy to the O$_2$ molecule can be found in a nucleus with an ellipsoidal shape like Fig.~\ref{fig:image}{\bf c}.  This type of ellipsoid is stretched on the vertical axis in the paper plane (side view), with a nuclear quantum state invariant under the rotation about the vertical axis (top view).  
This invariance is called {\it axial symmetry}, and the vertical axis here is called the {\it symmetry axis}.  In quantum mechanics, no rotational motion arises about the symmetry axis because the state does not change, like a di-atomic molecule (see  Fig.~\ref{fig:image}{\bf a}).  On the other hand, the ellipsoid can rotate about an axis perpendicular to the symmetry axis (side view).  

The axial symmetry can be broken in reality.  The cross section of the ellipsoid then becomes an ellipse as shown in the top view of Fig.~\ref{fig:image}{\bf d}.  The rotation can now also occur about the axis perpendicular to this cross section (i.e., the z-axis in Fig.~\ref{fig:image}{\bf e}), as seen in the right drawing of Fig.~\ref{fig:image}{\bf d}.  Although details are different, this is basically similar to the H$_2$O molecule case: a complex object can rotate in multiple ways.  

This article is organized as shown in the Table of Contents.
Section \ref{sec:1} is an introduction referring to molecules.  Section \ref{sec:ellipsoidal deformation} presents a simple overview of the nuclear surface deformation from a sphere to an ellipsoid and the so-called rotational bands.   Section \ref{sec:CI} shows the state-of-the-art Configuration Interaction (CI) calculation (the most advanced version of the Monte Carlo Shell Model (MCSM)) on the structure of the nucleus $^{166}$Er, reproducing experimentally observed properties and exhibiting its triaxial shapes.  General and analytic discussions are presented in Secs. \ref{sec:K} and \ref{sec:J}, respectively, on the underlying mechanisms of the triaxial shapes and the rotational spectra, by referring to the restoration of two rotational symmetries.  
It is mentioned that, in Sec. \ref{sec:J}, the well-known $J(J+1)$ rule of the rotational spectrum is derived as a $J$-dependent effect of the nuclear Hamiltonian for a quantum many-body system.  This derivation definitely differs from the explanation of the apparently same $J(J+1)$ rule in terms of the quantization of the kinetic energy of a free rigid-body rotor, e.g. \cite{ring_schuck_book}.
Section \ref{sec:2nd source} shows the second major mechanism for the triaxial shapes based on characteristic properties of nuclear forces.   
An interim summary of Secs. \ref{sec:K}-\ref{sec:2nd source} is given in Sec. \ref{sec:7}, \textcolor{black}{with the two mechanisms for triaxiality, and the actual triaxiality of heavy deformed nuclei is sorted into small and medium triaxialities}.  
It will be deduced from these arguments that the intrinsic structure of a deformed nucleus is not completely decoupled from extrinsic features.  This may be in contrast to the typical  conventional view pointing to the opposite direction, a good separation between them, e.g. \cite{ring_schuck_book}.

Some related important topics are mentioned in Sec. \ref{sec:8}.  In Sec. \ref{sec:nuclei_around}, 13 heavy nuclei are described as nuclei with \textcolor{black}{medium} triaxiality and strong ellipsoidal  deformation in terms of the CI calculations, and more nuclei are empirically suggested to be so.  
Sec. \ref{sec:154Sm} presents the discussions on the structure of the nucleus $^{154}$Sm, as an example of nuclei that are not triaxial traditionally.  
%This will provide a comparison to the discussions in earlier sections.  
It is shown that even $^{154}$Sm shows a modest but 
non-vanishing triaxiality, that is the \textcolor{black}{small trixiality}.  \textcolor{black}{A very recent experimental result supporting this property is mentioned.} 
Sec. \ref{sec:summary} is a summary and prospects, which include concise sketches of the present picture of the nuclear rotation compared to the traditional one, %a possible relevance to the Nambu-Goldstone theorem,  
possible experimental observations of ground-state triaxiality, and remarks on deeply related work by A. S. Davydov and that by D. Cline.
There are four appendices.
%, of which Appendices \ref{Ap_Kproj}, \ref{Ap_reduced}, and \ref{Ap_frozen} present new but somewhat detailed results and arguments.

%%%  Nuclear ellipsoidal deformation  %%%%%%
\section{Nuclear Ellipsoidal Deformation \label{sec:ellipsoidal deformation}}

The ellipsoidal deformation of nuclear surface is the main subject of this work.  We start with a simple modeling with a uniform nucleon-density distribution inside the surface of the ellipsoid, which also means a discrete drop to zero density outside the surface.  The surface is then specified by axes $R_x, R_y$ and $R_z$ as shown in Fig.~\ref{fig:image}{\bf e}, where the standard convention for the axis lengths $R_z \ge R_x \ge R_y$ is taken.  The ellipsoid is viewed from two different angles, A and B. The prolate ellipsoid is shown in Fig.~\ref{fig:image}{\bf f} where $R_x$=$R_y$ holds.  The ellipsoid without the axial-symmetry is shown in Fig.~\ref{fig:image}{\bf g}, where the three axes take different lengths.  This feature is called {\it triaxial}.   The ellipsoid shows quadrupole moments: 
\begin{equation}
Q_0 \,=\, \langle 2 z^2-x^2-y^2\rangle,   %\propto
\label{eq:Q0}
\end{equation}
and 
\begin{equation}
Q_2 \,=\, \sqrt{3/2} \, \langle x^2-y^2\rangle,   %\propto
\label{eq:Q2}
\end{equation}
where $\langle  \rangle$ implies the integral of the designated quantity multiplied by the matter (or charge if appropriate) density.  
%In the case of uniform-density distribution, it counts the density inside the surface.
The prolate shape is characterized by positive $Q_0$ and vanishing $Q_2$, whereas the triaxiality is by finite $Q_2$.  We show, in this paper, the unexpected preponderance of the triaxiality, its consequences in nuclear rotation and its robustness rooted in symmetry restoration and nuclear forces.  
Although the oblate ellipsoid $R_z = R_x > R_y$ is another possible case, it emerges at rather high excitation energies in the nuclei to be discussed.  Consequently, the oblate shape will be discussed elsewhere.

%%%%%%%%%%%%  FIGURE 2  %%%%%%%%%%%%%

\begin{figure}[tb]
  \centering
  \includegraphics[width=8.5cm]{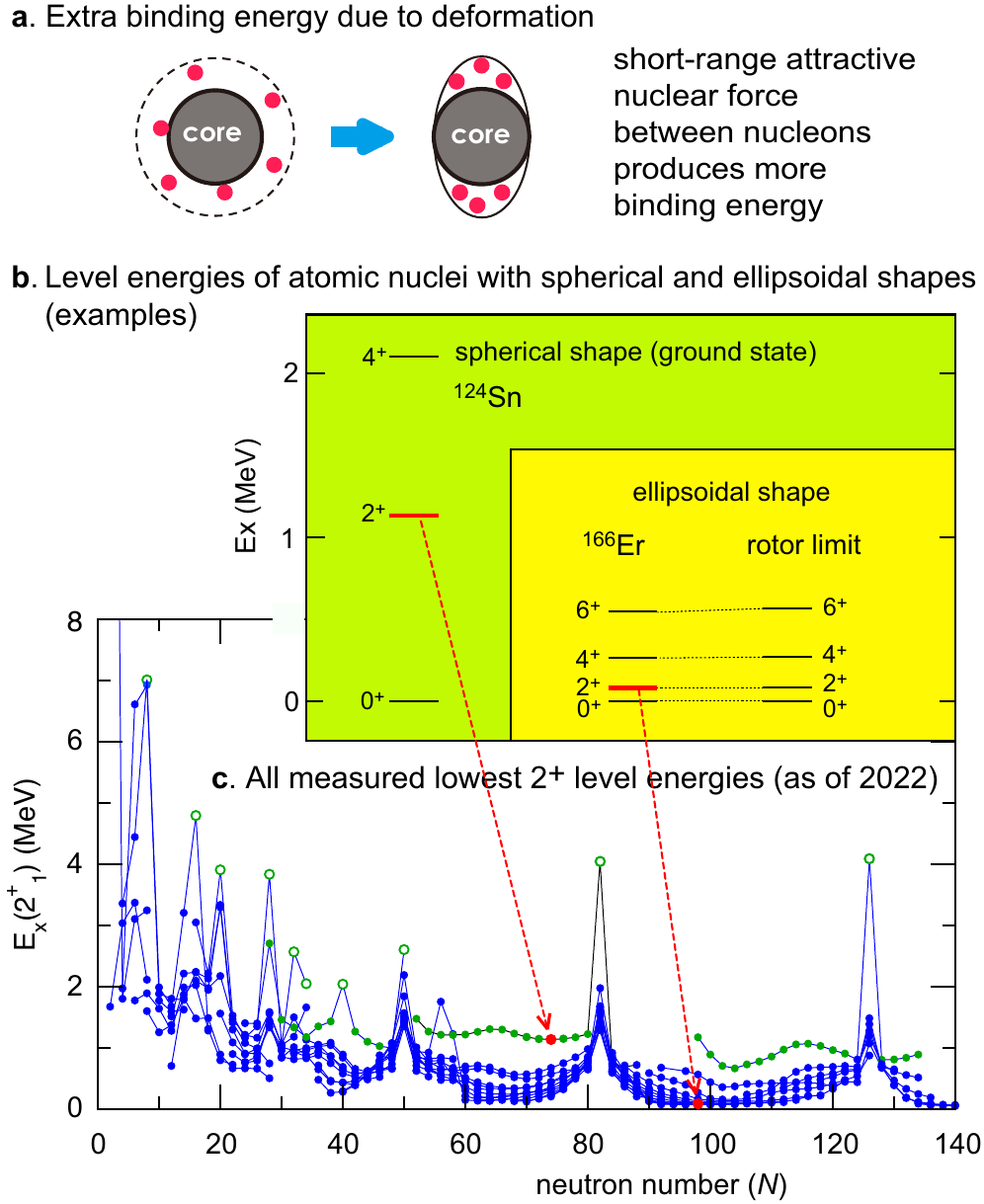}
    \caption{ {Overview of the 2$^+_1$ states.}
    {\bf  a} Intuitive picture of the major mechanism to drive ellipsoidal deformation. 
    {\bf  b} Examples of spherical and deformed cases with the rotational level energies.
    {\bf  c} All measured lowest 2$^+$ level energies of even-even nuclei as of the year 2022, based on the data taken from NuDat 3.0 \cite{nudat3}.  The open and closed circles indicate the individual 2$^+$ levels.  The circles within a given isotope chain are connected by lines to guide the eye. }    
  \label{fig:2+}  
\end{figure}  

%%%%%%%%%%%%%%%%%%%%%%%%%%%%%%%

The nuclear shape is studied, in the present work, as a quantum many-body problem.  The nuclear structure is described by the wave functions obtained by solving an appropriate many-body Schr\"odinger equation, where the Hamiltonian contains terms representing nuclear forces.  A possible interpretation of the major origin of the deformed nuclear shapes is %schematically 
depicted in Fig.~\ref{fig:2+}{\bf a}. Short-range attractive forces between nucleons \textcolor{black}{generally produce substantial binding energies of nuclei, and this effect can be generated by a 
density like the right} part of Fig.~\ref{fig:2+}{\bf a}.   \textcolor{black}{The mean-potential is often considered for basic understanding of  nuclear binding and structure, and such a mean potential, including deformed one in the body-fixed frame, is also a consequence of short-range attractive forces.}  If an isotropic density distribution is assumed, however, active nucleons are isotropically distributed in an outer layer of a certain thickness (see left part  of Fig.~\ref{fig:2+}{\bf a}) because of the shell structure: radial wave functions of single-particle orbitals are rather widely spread near the surface.  The nucleon density then becomes lower than that of the right part of Fig.~\ref{fig:2+}{\bf a}, resulting in a weaker binding due to fewer nucleons within the range of nuclear forces.  This interpretation appears to be consistent with results of quantum many-body calculations of various types including the present ones to be shown. 
%This interpretation is obtained based on a constant nucleon density, a shell structure and a short-range attractive force, where the short range refers to a distance shorter than the nuclear size. 
%An in-depth analysis of this interpretation seems to be an intriguing topic.  

In eigenstates of the Schr\"odinger equation of the current interest, nuclear shapes appear as consequences of many-body correlations (or collective motion), including Jahn-Teller effect \cite{jahn_1937}. \textcolor{black}{This Jahn-Teller effect produces superpositions of original single-particle states so as to constitute ellipsoidal shapes.   In this sense, nucleons (red circles) in the right part of Fig.~\ref{fig:2+}{\bf a} intuitively denote nucleons in such superposed states.}  

%To be more concrete, 
The nucleus is thus treated as a multi-nucleon quantum system, and the quadrupole moments and other related quantities are calculated from their wave functions.  Although the nucleus in such quantum mechanical descriptions shows a density approximately uniform and a surface with    imperfect sharpness, a nucleus can be connected to the afore-mentioned uniform-density classical ellipsoid with the same values of the quadrupole moments.  Such a mapping to the classical image serves as a useful intuitive way for grasping the essence of the structure of atomic nuclei, leaving other details untouched.  It is noted that all physical quantities are obtained in this article by quantum many-body calculations.

Attractive forces relevant to Fig.~\ref{fig:2+}{\bf a} are much stronger between a proton and a neutron than between two neutrons or between two protons.  This implies that the deformation becomes strong if there are sufficiently large numbers of active (or valence) protons and active neutrons, outside the inert core (see Fig.~\ref{fig:2+}{\bf a}).    On the other hand, if the protons (neutrons) form a closed shell, as the closed shell is spherical, the neutrons (protons) also form a spherical shape.  
%Note that the closed (inert) shell for deformed nuclei can be smaller than the closed shell, as is the case in this work.

Different nuclear shapes result in visible differences in observables.  
Figure~\ref{fig:2+}{\bf c} exhibits the \textcolor{black}{observed \cite{nudat3}} global systematics of the level energies of the lowest state of spin/parity $J^{P}$=2$^+$ for even-$Z$ and even-$N$ ({\it even-even}) nuclei, on top of the $J^{P}$=0$^+$ ground state.  
For doubly-magic nuclei, the excitation from the ground state can be made by moving a nucleon over a magic gap.  This requires a large amount of energy, meaning a high excitation energy.
There are high spikes in Fig.~\ref{fig:2+}{\bf c} with the excitation energies more than 2 MeV (see green open circles).  These spikes correspond to conventional ($N$=8, 20, 28, 50, 82, 126) or new ($N$=16, 32, 34, 40) magic numbers \cite{otsuka_2020, otsuka_2022_emerging}.
Beyond $N \sim$ 40, most of the points (blue closed circles) show low excitation energies down to several tens of keV.
An example of such low excitation energies is shown for the nucleus, $^{166}$Er, in the yellow-shaded part of Fig.~\ref{fig:2+}{\bf b}, which displays the lowest states of $J^{P}$=2$^+$, 4$^+$ and 6$^+$.  
The excitation energy for each $J^{P}$ is denoted $E_x(J^{P})$ hereafter.   Their experimental values are depicted in the left part in the yellow-shaded area of Fig.~\ref{fig:2+}{\bf b}.  In the right part, shown are $E_x(J)$'s in the rotor limit,
\begin{equation}
E_x(J^{P}) = \kappa J(J+1), 
\label{eq:rot_energy}
\end{equation}
where the value of $\kappa$ is adjusted to the measured 2$^+$ level energy, yielding a salient agreement for $J^{P}$=4$^+$ and 6$^+$.  
The $J(J+1)$ dependence of $E_x(J)$ in eq.~(\ref{eq:rot_energy}) can be explained by quantizing the free rotation of axially-symmetric classical rigid body, such as a symmetric top.   Note that this explanation is obtained by replacing a quantum many-body system with a classical rigid body.  Apart from this issue to be addressed later in this article, 
we come up with the image that a strongly deformed ellipsoid like the lower part of panel {\bf a}  ``rotates'' with $J^{P}$=2$^+$, 4$^+$, 6$^+$, ...  Similar rotational bands have been observed in many heavy nuclei.  
The systematics shown in Fig.~\ref{fig:2+}{\bf c} exhibits many of heavy nuclei with the first 2$^+$ state, denoted as 2$^+_1$, at a very low excitation energy.  This implies large moments of inertia indicative of strong ellipsoidal deformation.

The green shaded area of Fig.~\ref{fig:2+}{\bf b} displays observed level energies of $^{124}$Sn nucleus ($Z$=50, $N$=74), where the $Z$=50 proton closed shell results in a spherical shape for its ground state.  Its $J^{P}$=2$^+_1$ level is lying high, partly because of the absence of a mechanism like Fig.~\ref{fig:2+}{\bf a}.  This feature is depicted by closed green circles in Fig.~\ref{fig:2+}{\bf c}, forming a minor fraction of the points.  

%%%%%%%%%%%%  FIGURE 3  %%%%%%%%%%%%%

\begin{figure}[tb]
  \centering
  \includegraphics[width=8.5cm]{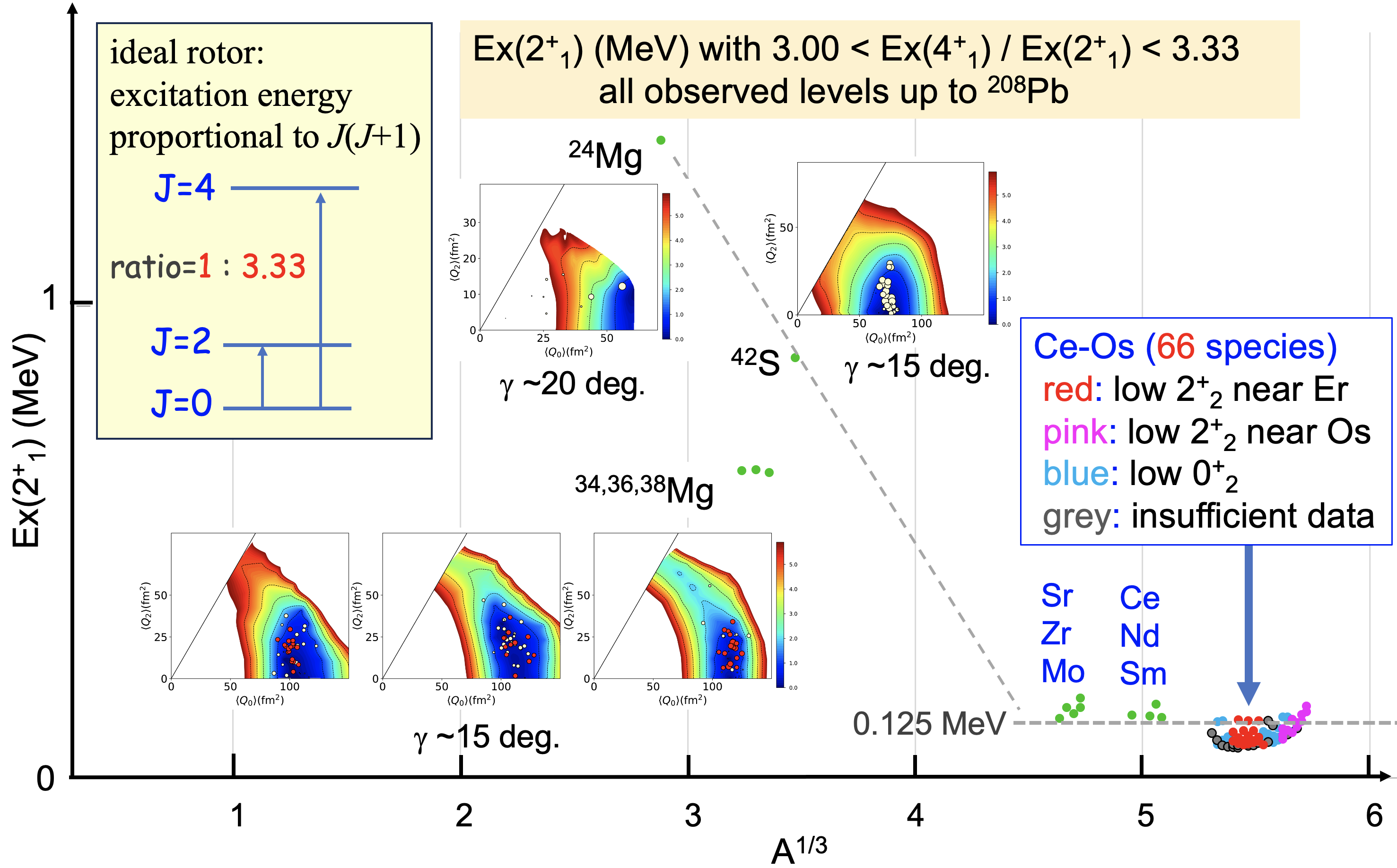}
    \caption{ Observed appearance of $E_x(2^+_1)$ for strongly deformed even-even nuclei, filtered by the criterion 3.00 $\le$ $E_x(4^+_1$)/$E_x(2^+_1$) $\le$ 3.33.  Each circle represents a nucleus plotted as a function of mass number, A, as scaled by A$^{1/3}$.  All cases up to $^{208}$Pb are included as of 2023.  Nuclei heavier than A$>$140 (A$^{1/3}>$5.2) are classified by their level structure as described in the figure.  The line of $E_x$=0.125 MeV is displayed to guide the eye, as well as the line showing the trend from $^{24}$Mg to $^{42}$S.  T-plot is displayed for the five lightest nuclei.  The inset shows the ideal case of rotational band. } 
  \label{fig:Ex2+}  
\end{figure}  

%%%%%%%%%%%%%%%%%%%%%%%%%%%%%%%%

Figure~\ref{fig:Ex2+} exhibits observed 2$^+_1$ excitation energies, $E_x(2^+_1$), for the nuclei filtered by the criterion 3.00 $\le$ $E_x(4^+_1$)/$E_x(2^+_1$) $\le$ 3.33, as of the year 2023 \cite{nudat3}.  Note that eq.~(\ref{eq:rot_energy}) produces this ratio=10/3.  Few nuclei fulfill this condition for the mass number, $A=Z+N \le$ 42, comprising stable $^{24}$Mg nucleus ($Z$=$N$=12) and exotic $^{42}$S nucleus ($Z$=16, $N$=26), as well as very neutron-rich exotic nuclei $^{34,36,38}$Mg ($Z$=12, $N$=22, 24, 26).  The next group consists of just nine nuclei, $^{100,102,104,106}$Zr ($Z$=40, $N$=60, 62, 64, 66), $^{106}$Mo ($Z$=42, $N$=64), $^{122}$Ce ($Z$=58, $N$=64), $^{128,130}$Nd ($Z$=60, $N$=68,70) and $^{132}$Sm ($Z$=62, $N$=70).  
Between $A$=150 and 208, however, a large number (66) of nuclei fulfill this severe condition with low $E_x(2^+_1$) values, mostly below 0.125 MeV.  This overall trend suggests that strong shape deformation primarily occurs in heavy nuclei with $A >$150.  
Naturally, the focus of this study is placed upon these nuclei, while the shapes of lighter nuclei are also interesting and will be discussed elsewhere from the present perspectives.

%%%  Multi-nucleon Structure clarified by large-scale CI calculations as exemplified with the nucleus $^{166}$Er
\section{Multi-nucleon Structure of $^{166}$Er nucleus clarified by CI calculations \label{sec:CI}}

\subsection{Rotational states of axially-symmetric and triaxial ellipsoids \label{subsec:R&K}}
%The strong ellipsoidal deformation is thus an important and dominating trend of heavy nuclei.  
Figures~\ref{fig:image}{\bf c, d} display two cases of the shape deformation.  Between the two, the axially symmetric shape (panel {\bf c}) has been considered to occur in the majority of deformed heavy ($A >$150) nuclei, such as those with low $E_x(2^+_1$) values in Fig.~\ref{fig:Ex2+}, since Bohr and Mottelson suggested \cite{bohr_mottelson_book2,bohr_nobel}.   
This picture has been a textbook item over 70 years \cite{rowe_book,deshalit_book,ring_schuck_book}, and many works were made
on top of it, for instance, the fission mechanism \cite{nix_fissionshape,ichikawa_fission}.  
As an example, M\"oller {\it et al.} presented, based on the finite-range liquid-drop model \cite{moller_1995,moller_2006}, effects of the axial-symmetry breaking on the nuclear mass, covering the Segr\`e (nuclear) chart, but no notable effects were found for the deformed heavy nuclei mentioned above (see Fig. 2 of \cite{moller_2006}). 

%%%%  R and K
The axially symmetric ellipsoid can rotate about one axis as shown in Fig.~\ref{fig:image}{\bf c}, with the corresponding angular momentum, $\vec{R}$.
This is nothing but the total angular momentum $\vec{J}$ (=$\vec{R}$).  Its magnitude, $J$,  and one of its components are conserved by the Hamiltonian.  If the triaxial shape sets in, another rotation of the ellipsoid emerges, because of an ellipse in the right part of Fig.~\ref{fig:image}{\bf d}, and is shown by $\vec{K}$,  
the angular momentum about the $z$ axis of the body-fixed frame.  The total angular momentum is intuitively expressed as $\vec{J}=\vec{R}+\vec{K}$.  The conservation law holds for $\vec{J}$, but not for $\vec{R}$ or $\vec{K}$ separately.
The $\vec{K}$ is pointing in the $z$ direction of the body-fixed frame, with its $z$ component denoted by $K$ as usual.
%, implying that different components are superposed in the eigenstate.  
Because the rotation by $\vec{K}$ actually occurs in the $xy$ plane of the body-fixed frame, it is essentially two-dimensional rotation and only the $z$-component, \textcolor{black}{denoted by $K$,} is relevant.    \textcolor{black}{The value of $K$ is not conserved in principle, and in fact, the present CI calculation is carried out without assuming  the conservation of $K$.  On the other side, $K$} \textcolor{black}{will be shown to be practically conserved in the cases of this article, as shown in Subsect.~\ref{subsec:Kmix}.
Although positive values of $K$ are usually mentioned, $K$ and $-K$ refer to essentially the same features.}   

%%%%%%%%%%%%  FIGURE 4  %%%%%%%%%%%%%

\begin{figure*}[tb]
  \centering
  \includegraphics[width=18cm]{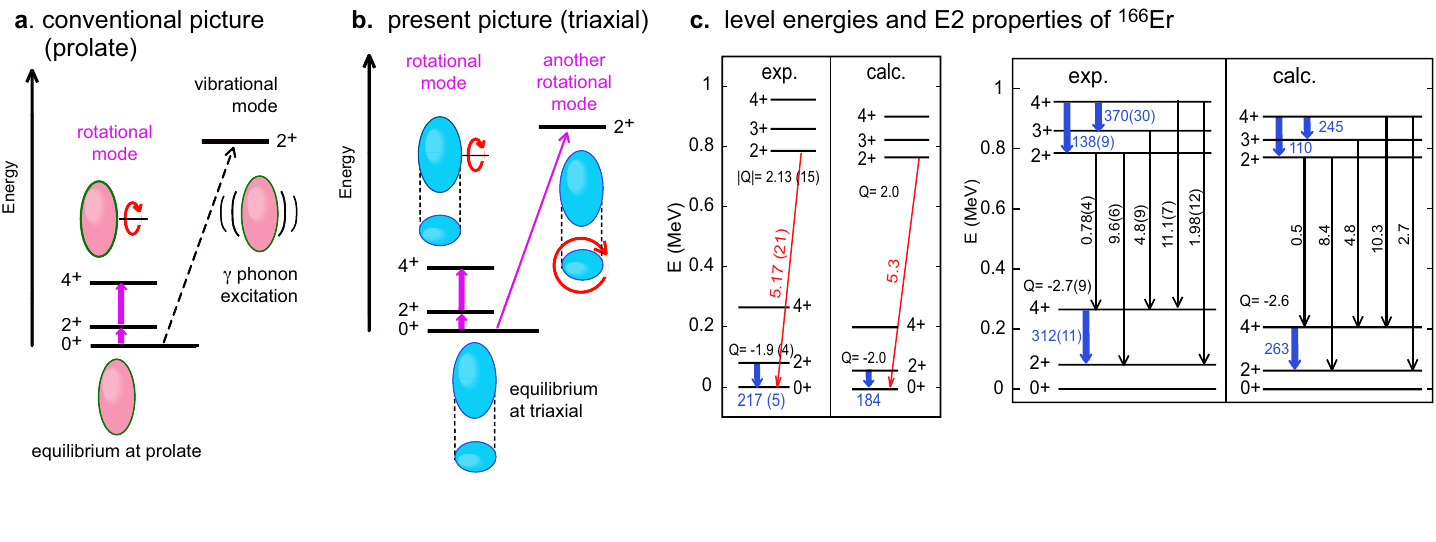}
 \vspace{-1.5cm}
    \caption{ {Schematic pictures of rotational motion and actual features.}
    {\bf  a} Conventional picture (prolate shape) and $\gamma$ vibration from it.
    {\bf  b} Present picture (triaxial shape).
    {\bf  c} Levels and E2 properties of $^{166}$Er ($Z$=68, $N$=98) compared to experimental data \cite{ensdf}.
    B(E2) values are in W.u., and spectroscopic electric quadrupole moments, Q, are in $e$ barn. 
      } 
  \label{fig:level}  
\end{figure*}  

%%%%%%%%%%%%%%%%%%%%%%%%%%%%%%%%

The surface tension of a liquid drop generally tends to minimize its surface area, ending up with a sphere in an extreme case.  The nuclear shape may be considered to be an equilibrium between this surface tension and the force causing the deformation {\it \`a la} Fig.~\ref{fig:2+}{\bf a}.   An axially-symmetric ellipsoid is then favored over a triaxial one, because of the constant curvature in the $xy$ plane. %(see Fig.~\ref{fig:image}{\bf f}).  
The rotation of such rigid bodies is described, for instance, in textbooks \cite{rowe_book,deshalit_book}.  
As a multi-nucleon problem, Kumar and Baranger \cite{kumar_baranger1968} performed the ``pairing-plus-quadrupole model'' calculation, as a then most advanced many-body calculation, and advocated the ``preponderance of axially symmetric shapes'' for many species of nuclei. 
A similar work was reported by Bes and Sorensen \cite{bes_sorensen1969}.   These works indicated that once many nucleons in many single-particle orbits of a mean potential coherently contribute, the axially-symmetric prolate deformation likely occurs.  The SU(3) model is a symmetry-based many-body approach, and favors the same feature in some cases \cite{elliott_1958a,elliott_1958b}, 
as well as its approximate extensions \cite{caurier_2005}.   
%\textcolor{black}{It is known by CI calculations that the lowest states of $^{20}$Ne provide a good example of axial symmetry, but the lowest states of other nuclei with similar $Z$ and $N$ may not show the axial symmetry, for instance those of $^{24}$Mg (see Appendix).} 
 
Figure~\ref{fig:level}{\bf a} shows a schematic picture of the prolate deformed ground state and its rotational excitations.   Figure~\ref{fig:level}{\bf a} includes another type of excitation, called $\gamma$ vibration \cite{bohr_mottelson_book2, bohr_nobel}.  
In this mode, the circle in the lower part of Fig.~\ref{fig:image}{\bf c} is distorted to an ellipse.  However, the distortion is not static, and is a vibrational motion.  This vibration occurs in the excited state shown in the right upper part of Fig.~\ref{fig:level}{\bf a}, but not in the members of the ground-state band (i.e., the rest of Fig.~\ref{fig:level}{\bf a}).    
This vibrational excitation is represented by $\gamma$ phonon in a quantum mechanical description, thus denoted as ``$\gamma$ phonon excitation'' in Fig.~\ref{fig:level}{\bf a}.  The 2$^+_2$ state is then identified as the one-$\gamma$-phonon state on top of the deformed equilibrium \cite{bohr_mottelson_book2, bohr_nobel}. %(see Fig.~\ref{fig:level}{\bf a}).

\subsection{Properties of $^{166}$Er nucleus by the present CI calculation}

Figure~\ref{fig:level}{\bf c} depicts measured excitation energies as well as electromagnetic decays and moments \cite{ensdf}, for $^{166}$Er ($Z$=68, $N$=98) as an example.  The decays are actually electric quadrupole (E2) transitions, and their strengths are expressed by squared E2 strengths called $B(E2;J^+\rightarrow J'^+)$ \cite{bohr_mottelson_book1,deshalit_book,ring_schuck_book}.  The 2$^+_2$ state decays to the ground state through relatively strong transition with $B(E2;2^+_2\rightarrow 0^+_1)$ = 5.17$\pm$0.21 (W.u.) (see panel {\bf c}), where W.u. implies the single-particle (Weisskopf) unit \cite{bohr_mottelson_book1}.  Such a moderately large $B(E2)$ value was interpreted as an indication that the 2$^+_2$ state is a phonon excitation which is collective to a certain extent \cite{bohr_nobel}.  

We present a recent work on this subject from a modern viewpoint combined with state-of-the-art configuration interaction (CI) calculations with a realistic effective nucleon-nucleon ($NN$) interaction.  The CI calculation in nuclear physics is called {\it shell model}, and some introductory details are found in Appendix~\ref{Ap_shell}. 
The CI calculation here means Monte Carlo Shell Model (MCSM) \cite{mcsm_1995,mcsm_1998,mcsm_2001,mcsm_2012,mcsm_2017}, which enabled us to carry out CI calculations far beyond the limit of the conventional CI approaches, on contemporary subjects \cite{togashi_2016, leoni_2017, togashi_2018, marsh_2018, ichikawa_2019, taniuchi_2019, otsuka_2019, marginean_2020, tsunoda_2020, abe_2021, otsuka_2022_nature}.  
Some details of the MCSM are presented in Appendix~\ref{Ap_MCSM}. 
A large number of single-particle orbits are taken as activated orbitals so that the ellipsoidal deformation can be described, 8 for protons and 10 for neutrons.  It is pointed out that the energetically lower part of the activated orbitals belong to the closed shell in  usual shell-model calculations, meaning that a smaller closed inert core is taken in this work.  The nucleus $^{166}$Er is first discussed as a typical example, with 28 active protons and 28 active neutrons.  The present work goes far beyond the earlier paper \cite{otsuka_2019}.   
%as new developments will become clear after all novel features are depicted.
The present CI calculations are carried out by the most advanced methodology in the MCSM, 
called Quasiparticle Vacua Shell Model (QVSM) \cite{shimizu_2021}, where the number-projected quasiparticle vacua are used as basis vectors, instead of Slater determinants adopted in usual MCSM calculations.  The number-projected quasiparticle vacua can be described, in simple terms, as the Hartree-Fock-Bogoliubov-type states with controlled $u$ and $v$ amplitudes \cite{shimizu_2021}.  Namely, the particle-number projection is incorporated in the sense of the variation-after-projection.  The basis vectors are selected from a large group of candidate states generated stochastically, so that the possible maximal lowering of the energy eigenvalues of interest can be obtained.  The selected basis vectors are further optimized by variational procedures \cite{mcsm_2001,shimizu_2021}.   \textcolor{black}{The angular-momentum and parity projections are also incorporated in the sense of the variation-after-projection for each basis vector.}  Some details of the computational methodology are presented in Appendix~\ref{Ap_MCSM}.    

The Hamiltonian for the present CI calculations is a slight modification of its prototype used in the earlier work \cite{otsuka_2019}.  
The proton-neutron (interaction) part was obtained from the monopole-based-universal  interaction, $V_{\rm MU}$ \cite{otsuka_2010}, which consists of the central force and the tensor force.   
The central force is a Gaussian interaction, where the parameters were fixed so as to simulate the shell-model and microscopic (G-matrix) interactions \cite{otsuka_2010}, and has been used in many studies, up to Hg isotopes \cite{marsh_2018}.  The tensor force was taken from \cite{osterfeld_1992}, where the strengths were obtained from $\pi$-nucleon scattering data in the free space.
The proton-proton and neutron-neutron interactions were fixed in \cite{otsuka_2019} mainly based on the microscopic interactions used in \cite{brown_2000_Pb}.  In the earlier work, the pairing correlations in wave functions may have been underestimated because of the Slater-determinant basis vectors in the original MCSM, and the strengths of pairing interactions were likely taken to be stronger, in order to compensate the possible underestimation of their effects.  This problem vanishes in the present calculation to a large extent, because the pairing energy can be better carried by individual basis vectors being of HFB (Hartree-Fock-Bogoliubov) (or BCS) type.   The pairing strengths are rescaled to be somewhat weaker.  The SPE values are also changed slightly.  Although the basic outcomes of the earlier work \cite{otsuka_2019} remain in the present work, many features emerge in the underlying theories and in the range of nuclei, and, in some cases, traditional arguments are corrected under \textcolor{black}{the views arising in this work}.

%%%%%%%%%%%%  FIGURE 5  %%%%%%%%%%%%%

\begin{figure}[tb]
  \centering
  \includegraphics[width=8.5cm]{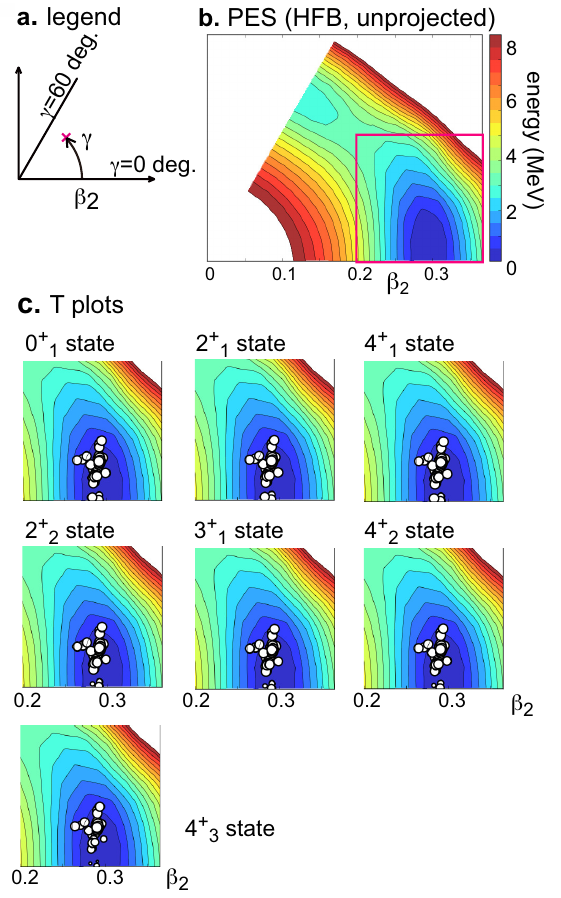}
    \caption{ {PES and T-plots for $^{166}$Er}.   
    {\bf a} Legend.   
    {\bf b} PES. 
    {\bf c} T-plots of ground and low-lying states for the red-square region of {\bf b}.
    %All T-plots are independently obtained, despite resulting resemblances.
   % {\bf d} T-plots for reduced crucial monopole interactions.
    %{\bf e} T-plots for monopole-frozen Hamiltonian.
    } 
  \label{fig:tplot}  
\end{figure}  

%%%%%%%%%%%%%%%%%%%%%%%%%%%%%%%
 
Figure~\ref{fig:level}{\bf c} depicts calculated level energies and E2 properties of $^{166}$Er, in a good agreement with experiment, including both large and small values of $B(E2)$.  The present value of $B(E2;2^+_2\rightarrow 0^+_1)$ is 5.3 W.u. in a salient agreement with experiment, and urges us to look into underlying multi-nucleon structure of the eigenstates.  

%%%%%%%%%%%%%%%%%%%%%%%%%%%%%%%%%%%%%%%%%%%%%%%

\subsection{Triaxiality of $^{166}$Er nucleus and its clarification by T-plot}
\label{subsec:166Er_Tplot}

For unveiling the underlying structure, deformation parameters $\beta_2$ and $\gamma$ are introduced \cite{bohr_mottelson_book2} to indicate how the sphere with the radius $R_0$ is deformed to ellipsoids (see textbook for example \cite{ring_schuck_book}).
The three axes of the ellipsoid in the classical uniform-density model are expressed as  
\begin{equation}
R_z =\{1+0.63 \beta_2 \cos \gamma\} \, R_0,
\label{eq:R1}
\end{equation}
\begin{equation}
R_x =  \{1+0.63 \beta_2 \sin(\gamma-30^\circ)\} \, R_0, 
\label{eq:R2}
\end{equation}
\begin{equation}
R_y = \{1-0.63 \beta_2 \cos(60^\circ-\gamma)\} \, R_0.
\label{eq:R3}
\end{equation}
Prolate shapes ($R_x$=$R_y$) correspond to $\gamma$=0$^{\circ}$, while triaxial shapes emerge for 0$^{\circ} < \gamma < 60^{\circ}$.  
Usually, $\beta_2 \ge 0$ and $0^{\circ} \le \gamma \le 60^{\circ}$ are considered, implying
$R_y \le R_x \le R_z$.   
The quadrupole moments, $Q_0$ and $Q_2$, are calculated for a given ellipsoid specified by $\beta_2$ and $\gamma$, with $R_0$ properly taken.  

In quantum multi-nucleon approaches, quadrupole moments $Q_0$ and $Q_2$ are obtained from relevant matrix elements of the state of interest.  
The corresponding $\beta_2$ and $\gamma$ values are obtained so that the classical uniform-density model yields the same $Q_0$ and $Q_2$ values with appropriate $\beta_2$ and $\gamma$ values.  We will use such $\beta_2$ and $\gamma$ values in the quantum many-body studies.

The $\beta_2$ parameter obviously stands for the magnitude of the quadrupole deformation, and  
\begin{equation}
\beta_2^2 \propto (Q_0^2 \,+\, 2 \, Q_2^2), 
\label{eq:beta2_QQ}
\end{equation}
holds in the lowest order.  Eq.~(\ref{eq:beta2_QQ}) is adopted as a good approximation.    Typically, $\beta_2$ is about 0.3 for heavy deformed nuclei being discussed.  
The deformation parameter (also called triaxial parameter) $\gamma$ is usually given by the simple relation (e.g., eq. (5-82) in \cite{bohr_mottelson_book2}), 
\begin{equation}
{\rm tan} \, \gamma = \sqrt{2} \,\, \frac{Q_2}{Q_0} .
\label{eq:gamma_def}
\end{equation}
This relation is also taken in this article.
We note that in the present work, in-medium polarization corrections to shell-model $Q_0$ and $Q_2$ values are incorporated \cite{utsuno_2015,otsuka_2022}, in order to obtain $\beta_2$ and $\gamma$ values.      

%computing $R_{x,y,z}$ values of the uniform ellipsoid having the same $Q_0$ and $Q_2$ values, as discussed in Sec.~\ref{sec:ellipsoidal deformation}.  

A constrained mean-field (more concretely HFB) calculation is carried out by imposing various values of $(\beta_2, \gamma)$ as constraints.  The obtained energy expectation value is displayed by contour plot in Fig.~\ref{fig:tplot}{\bf b}, with a legend in Fig.~\ref{fig:tplot}{\bf a}.  This plot is usually called the potential energy surface (PES).  The same Hamiltonian as in the present CI (i.e. MCSM (more precisely QVSM)) calculations is consistently used.  The PES in Fig.~\ref{fig:tplot}{\bf b} suggests that the bottom part of the PES spreads over finite $\gamma$ values, and the precise inspection actually indicates that the minimum is not at $\gamma$=0$^{\circ}$ but around $\gamma \sim$ 9$^{\circ}$.  This contradicts the prolate preponderance hypothesis believed over seven decades.  For $\beta_2=0.30$ and $\gamma$=9$^{\circ}$, the values of $R_{x,y,z}/R_0$ are 0.93, 0.88 and 1.19, respectively. 

The wave functions of the MCSM are expanded by so-called MCSM basis vectors, which are presently number-projected quasiparticle vacua \cite{shimizu_2021}.
% (see Appendix~\ref{Ap_MCSM} if interested in details).
Each of the basis vectors has %intrinsic 
quadrupole moments, from which the corresponding $(\beta_2, \gamma)$ value is obtained, after properly aligning the three axes for each basis vector.  The individual basis vector can be represented by a circle pinned down on the PES according to this $(\beta_2, \gamma)$ value, which now serves as a ``partial labeling'' of the basis vector.   
The importance of each basis is depicted by the circle's area proportional to the overlap probability with the eigenstate.  This visualization is called T-plot \cite{tsunoda_2014}, and turned out to be very useful (see Appendix~\ref{Ap_MCSM} for details).

If $K$ quantum number is conserved, E2 matrix elements follow certain regularities \cite{bohr_mottelson_book2}.  The actual values indeed follow them remarkably well.  For instance, the spectroscopic quadrupole moment vanishes for $K$=2 and $J$=3, and this is the case for the 3$^+_1$ state.  Combining such properties with strong E2 transitions expected within individual rotational bands, the band structure is assigned as $K$$\approx$0 for the 0$^+_1$, 2$^+_1$, 4$^+_1$ states forming the so-called {\it ground band}, $K$$\approx$2 for the 2$^+_2$, 3$^+_1$, 4$^+_2$ states forming the so-called {\it $\gamma$ band}, and $K$$\approx$4 for the 4$^+_3$ state.
The approximate conservation of $K$ quantum number is a well-known feature, for instance, \cite{polishK_2021}, and is discussed for $^{166}$Er also in \cite{tsunoda_2021}.

Figure~\ref{fig:tplot}{\bf c} displays the T-plots for the 0$^+_1$, 2$^+_1$, 4$^+_1$ states (ground-band members), and the 2$^+_2$, 3$^+_1$, 4$^+_2$ states ($\gamma$-band members), as well as the 4$^+_3$ state.  All of them exhibit remarkably similar T-plot patterns, {\it i.e.}, the mean position and spreading pattern around it. %fluctuation.  
Because the T-plot displays the corresponding shapes of individual basis vectors in the body-fixed (or intrinsic) frame, this similarity suggests a rather common intrinsic (i.e., body-fixed-frame) structure, which includes triaxial shape.  This observation is consistent with the picture shown in Fig.~\ref{fig:level}{\bf b}: all these states can be generated, to a very large extent, by rotating a common triaxial ellipsoidal state in multiple ways.   

The mean value of $\gamma$ is calculated, by the method described in \cite{otsuka_2022}, to be 8.2$^{\circ}$ for each member of the ground band, 9.1$^{\circ}$ for each member of the $\gamma$ band, and 9.5$^{\circ}$ for the 4$^+_3$ state, where small variations within a band (beyond two digits) are ignored.  
These values are almost the same, and in particular, are almost exactly so within a band.  This fact suggests that a fixed triaxial shape governs $^{166}$Er.  Fine details are consistent with the present band assignment of individual states.     
This property holds up to the 4$^+_3$ state, where the CI calculation stopped due to the computer resource.  

We stress that the values of $\beta_2$ and $\gamma$ can be evaluated, within the Kumar invariant approach, from quadrupole matrix elements between low-lying eigenstates \cite{kumar_1972}.  An application with the outcome of the present CI calculation shows $\beta_2 \sim$ 0.3 and $\gamma \sim$9.2$^{\circ}$, which are similar to the T-plot values mentioned above, implying that  the present way for estimating $\beta_2$ and $\gamma$ values is basically consistent with the Kumar invariant method.  

Very importantly, Davydov and his collaborators made a pioneering theoretical work of the systematic appearance of triaxial shapes in a number of nuclei, including strongly deformed ones such as $^{166}$Er, already around 1958 \cite{davydov1,davydov2}.  
\textcolor{black}{Yamazaki soon argued nuclear shapes along this line \cite{yamazaki_1963}.}   
Some discussions on this Davydov model will be presented in Subsec.~\ref{subsec:davydov}.   

We mention that triaxial shapes have been studied extensively for medium-mass nuclei with weaker ellipsoidal deformations, as exemplified by works on $^{74,76}$Zn \cite{7476Zn}, $^{76}$Ge \cite{76Ge,76Geb}, $^{78}$Se \cite{78Se}, $^{76}$Kr \cite{yao_2014}, Kr isotopes \cite{rodriguez_2014}, and $^{80}$Zr \cite{rodriguez_2011}.   \textcolor{black}{It is pointed out once again that this article is focused on strongly deformed and clearly rotational nuclei as depicted in Fig.~\ref{fig:Ex2+}.}  In fact, none of 
%$^{74,76}$Zn, $^{76}$Ge, $^{78}$Se, $^{76}$Kr or $^{80}$Zr 
\textcolor{black}{the nuclei mentioned just above} appears in Fig.~\ref{fig:Ex2+}.
\textcolor{black}{An independent empirical analysis %based on experimental data 
\cite{andrejtscheff_1993} suggests consistent trends on the differences between a group of medium-mass weakly-deformed nuclei and another group of heavy strongly-deformed nuclei, as %which will be 
discussed %in some detail 
in subsec.~\ref{subsec:early_exp}.   }

%%%  Reinforced triaxiality in the $\gamma$ band  %%%%%%
\subsection{Governing triaxiality over the ground and $\gamma$ bands \label{sec:gamma band}}

The mean value of the deformation parameter $\gamma$ is quite close between the ground and $\gamma$ bands, with a minor difference of about 1$^{\circ}$.  This difference reflects a slight structure change from the ground band to the $\gamma$ band.   Its impact on the relative energy between these two bands is of a certain interest.  While the energy of the 0$^+_1$ state is calculated with many basis vectors (see Appendix~\ref{Ap_MCSM} for details), the first basis vector, denoted $\xi_0$ here ($\phi^{(1)}$  in eq.~(\ref{eq:mcsm_psi}) for 0$^+_1$), is most important among them: it indeed shows a large overlap probability with the final solution, 89\% after the projections and normalization.  
We next look at the 3$^+_1$ member of the $\gamma$ band.  This state has no counterpart in the ground band and should well represent features of the $\gamma$ band.  We first calculate its energy by using $\xi_0$ alone, similarly to the 0$^+_1$ state.   Although the obtained energy of 3$^+_1$ state is not too bad, its difference from the 0$^+_1$-state value is 1.27 MeV, which is notably larger than the 3$^+_1$ excitation energy by the present MCSM calculation shown in Fig.~\ref{fig:level}{\bf c}, that is 0.83 MeV.  However, instead of $\xi_0$, by using another state, $\xi_3$, determined by solely lowering the energy of the 3$^+_1$ state, the obtained 3$^+_1$ state comes down from 1.27 MeV to 0.76 MeV above the 0$^+_1$ state obtained from $\xi_0$.  This difference is apparently closer to the calculated 3$^+_1$ level energy (0.83 MeV in Fig.~\ref{fig:level}{\bf c}).  The overlap probability with the final solution becomes as high as 88\%.  Thus, the most optimum basis vector is slightly different between the 0$^+_1$ and 3$^+_1$ states.  
By including additional basis vectors into the actual MCSM calculation, the energies of the 0$^+_1$ and 3$^+_1$ states come down approximately in parallel, and their difference swiftly reaches the value shown in Fig.~\ref{fig:level}{\bf c}.  

The value of $\gamma$ is 8.5$^{\circ}$ and 9.8$^{\circ}$ for $\xi_0$ and $\xi_3$, respectively, being similar to the mean $\gamma$ values of the ground and $\gamma$ bands.  The analysis above indicates that this difference lowers the $\gamma$-band excitation energy by $\sim$0.5 MeV.  Once this lowering is incorporated, additional basis vectors do not substantially change the $\gamma$-band excitation energy, ending up with a 0.07 MeV shift in the final solution.      
These features suggest that the $\gamma$-band excitation energy is substantially lowered due to this enlargement (i.e., from 8.5$^{\circ}$ to 9.8$^{\circ}$) of the triaxial deformation.   This lowering will be discussed in Subsec.~\ref{sec:sketch} from a different viewpoint.
We stress that this effect is an outcome of $NN$ interactions contained in the MCSM calculation.

%%%%%%%%%%%%%%%%%%%%%%%%%%%%%%%%%%%%%%%%%%%%%%%%
%%%%  Section K
\section{underlying mechanism of the triaxiality, symmetry restoration} %and the variation of their effects between bands}
\label{sec:K}

\subsection{Implication of $K$ quantum number for rotational bands}

%%%  intrinsic state is introduced here !   first place in this paper   

We here begin to discuss by what underlying mechanism the triaxiality is favored.
The triaxial intrinsic state (i.e., body-fixed-frame state) $\phi$ is considered, and the state $\xi_0$ discussed in Sec.~\ref{sec:gamma band} is used as an illustrative example 
for it, while the argument is rather general.  
Assuming that various eigenstates can be generated from a single $\phi$ in a good approximation, we extract, from $\phi$, the state of definite $J$ and $M$, that are the total angular momentum and its z-projection in the laboratory frame.  This projection can be achieved, as usual, by rotating $\phi$ in the three-dimensional space with three Euler angles $\alpha, \beta$ and $\gamma$, and by integrating it with an appropriate weighting factor.  The process can be expressed by the following well-known equation (see \cite{ring_schuck_book} for instance),
%D^J_{MK}=exp(iM\alpha)d^J_{MK}(\beta)exp(iK\gamma)\\
%exp(i\alpha J_z)exp(i\beta J_y)exp(i\gamma J_z)|\phi\rangle
\begin{eqnarray}
&\Psi \bigl[\phi, J, M, K \bigr] \,=\,&(2J+1)/(8 \pi^2)    \int_0^{2\pi}   d\alpha \int_0^{\pi} d\beta \, {\rm sin}\beta \, \int_0^{2\pi} d\gamma \,\,\,\,\, \nonumber \\
&& \big\{ D^J_{M,K} (\alpha, \beta, \gamma) \big\}^*\, e^{i\alpha \hat{J}_z} \, e^{i\beta \hat{J}_y} \, e^{i\gamma \hat{J}_z} \, | \, \phi \rangle,
\label{eq:rot_phi}
\end{eqnarray}
where $D$ is the Wigner's function.  For the $D$ function and other related quantities, we adopt the convention of Edmonds \cite{edmonds}, which is also used in \cite{ring_schuck_book} with a summary in its Appendix A.  Note that $\gamma$ here is different from the deformation parameter $\gamma$, but no confusion is expected in this traditional usage of the character. (Note that the deformation parameter $\gamma$ will be mentioned in the unit of degrees, whereas the Euler angle $\gamma$ in radian.) 
Equation~(\ref{eq:rot_phi}) implies that the three-fold rotation of $\phi$ generates states with good ($J$, $M$) pairs.
One notices an additional index of $K$.  In fact, there can be different and independent states from the same $\phi$ for a given pair ($J$, $M$), and $K$ specifies them.
This $K$ arises from the $J_z$ rotation ($e^{i\gamma \hat{J}_z}$) with the weighting factor, 
\begin{equation}
D^J_{M,K} (\alpha, \beta, \gamma) \,=\, e^{iM\alpha} \, d^J_{M,K} (\beta) \, e^{iK\gamma } ,
\label{eq:Wigner}
\end{equation}
with $d^J_{M,K} (\beta)$ being the (small) $d$ function.     
This $K$ in eq.~(\ref{eq:Wigner}) is identified 
as the z-component $K$ for the rotation about the $z$ axis in the body-fixed frame (see the red circular arrow in the right drawing of Fig.~\ref{fig:image}{\bf d}). 
Although the $K$ quantum number is not conserved in a rigorous sense, it serves as an approximate quantum number at least for strongly deformed nuclei, as we shall see.  Moreover, the states of $J=0$ can be extracted only from $K=0$ components.   \textcolor{black}{We thus discuss the cases with definite integer $K$ values, to begin with.}
%The $J=2$ states have origins $-2 \le K \le 2$, for instance.

We first decompose $\phi$ according to $K$ value. 
This is straightforward because of the factorization in eq.~(\ref{eq:Wigner}):   
the integration with the Euler angle $\gamma$ is separated, and the projected state can be obtained by rotating $\phi$ about the $z$ axis %in the $xy$ plane %(i.e., about the $z$ axis (see Fig.~\ref{fig:image}) 
as, 
\begin{equation}
\Phi \bigl[\phi, K\bigr] \,=\, \frac{1}{{\mathcal N_K}} \frac{1}{2\pi}  \int_0^{2\pi} d\gamma \, e^{i\gamma (\hat{J}_z - K)} \, \phi ,
\label{eq:K-proj}
\end{equation}
where $\mathcal{N}_K$ is a normalization constant.

%%%%%%%%%%%%  FIGURE 6  %%%%%%%%%%%%%

\begin{figure*}[tb]
  \centering
  \includegraphics[width=14.5cm]{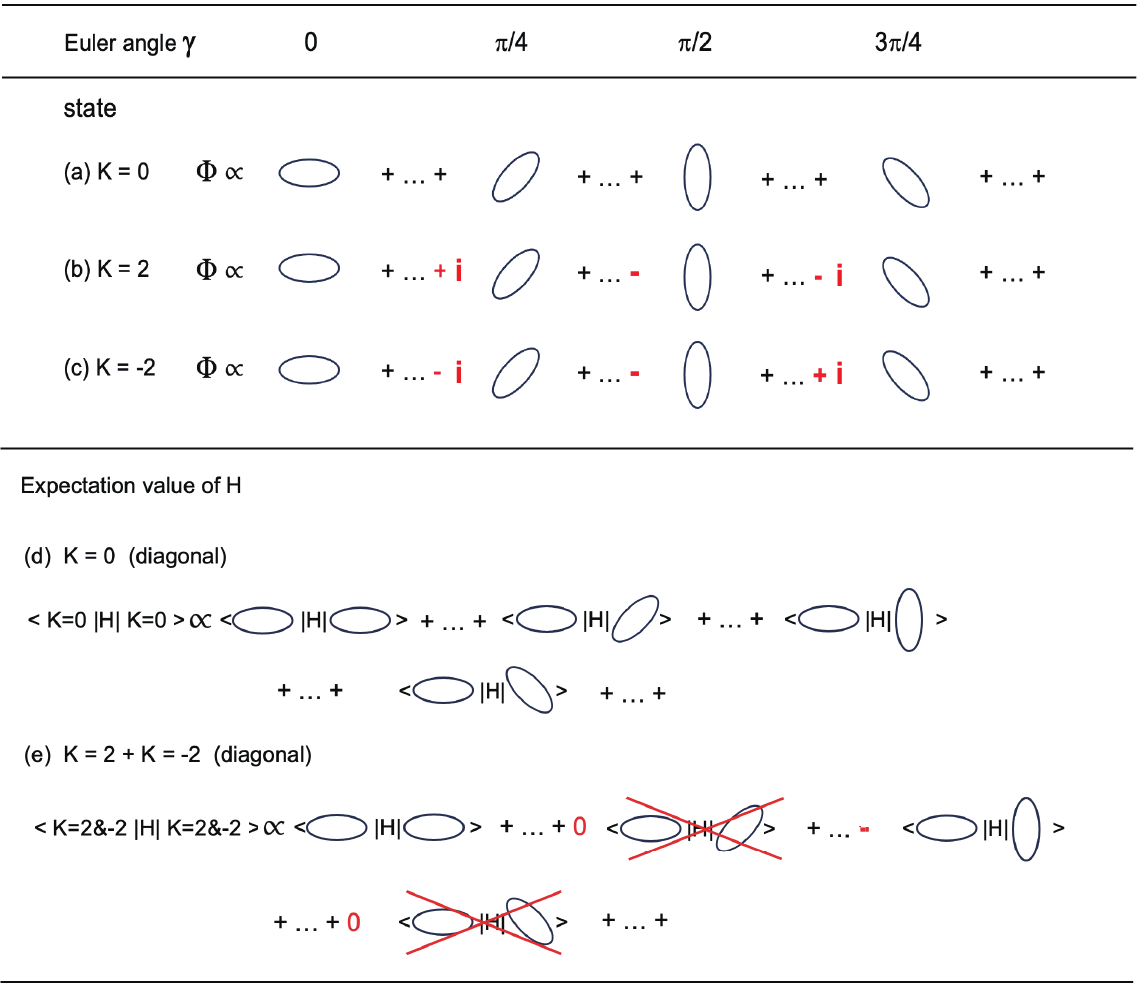}
    \caption{ {Schematic illustrations of K-projected states} }
  \label{fig:Kproj}  
\end{figure*}  

%%%%%%%%%%%%%%%%%%%%%%%%%%%%%%%%

\subsection{$K^P=0^+$ cases}
\label{K=0}

We begin with the $K=0$ state, which is particularly simple, because the contributions from all $\gamma$ angles are equally superposed:
\begin{equation}
\Phi \bigl[\phi, K=0\bigr] \,=\, \frac{1}{{\mathcal N_{0}}} \frac{1}{2\pi}  \int_0^{2\pi} d\gamma \, e^{i\gamma \hat{J}_z} \, \phi  . 
\label{eq:K=0state}
\end{equation}

The idea behind these complex mathematical expressions can be visualized in Fig.~\ref{fig:Kproj} (a), where the rotated states are explicitly displayed for discretely sampled angles $\gamma = 0, \pi/4, \pi/2, 3\pi/4$.   We see that all rotated states are uniformly superposed.  

The expectation value of the Hamiltonian $H$ with respect to the state in eq.~(\ref{eq:K=0state}) is given by 
\begin{eqnarray}
& h_{K=0} \,&=\, \langle \Phi \bigl[\phi, K=0\bigr] \, | H \, | \Phi \bigl[\phi, K=0\bigr] \rangle \nonumber \\
 & &=\, \frac{1}{|{\mathcal N}_0|^2} \frac{1}{{2\pi}} \, \int_0^{2\pi} d\gamma \, \langle \phi \, | \, H \, | \, e^{i\gamma \hat{J}_z} \, \phi \,\rangle . 
\label{eq:H K=0 ex}
\end{eqnarray}
This summation is visualized in Fig.~\ref{fig:Kproj} (d) again for selected discrete angles.   It is clear that all combinations are summed up with the same factor.  
The nuclear forces can connect $| \phi \,\rangle$ to $| e^{i\gamma \hat{J}_z} \, \phi  \,\rangle$, \textcolor{black}{the state rotated from $\phi$} by angle $\gamma$. 
%at $\gamma=0$ to the state obtained by rotating $\phi$ by non-vanishing $\gamma$.  
The integrand is largely negative (strong binding) at $\gamma$=0$^{\circ}$, and it is expected to continuously decrease in magnitude as $\gamma$ moves away from 0.  The integrated contributions are then considered to be also negative with realistic nuclear forces, giving additional binding energy to the nucleus (see Fig.~\ref{fig:Kproj} (d)).  Beyond $\gamma = \pi/2$, because of approximate symmetry between $\gamma$ and $\pi - \gamma$, the contribution will be bounced back to a good extent.   This feature is general, and its effect as additional binding energy is robust.  

We now evaluate how much this $K$ projection changes the energy, as the rest of the total effect of eq.~(\ref{eq:rot_phi}) will be discussed in the next section.  The $\xi_0$ state is taken as an example.  Its energy without the $K$ projection (i.e., ``unprojected'' energy) is -391.6 MeV.  (Note that parity projection lowers this energy by 1.2 MeV.)   Related quantities discussed hereafter are supposed to be projected onto the positive parity if not explicitly mentioned, except for ``unprojected'' quantities unless otherwise mentioned.   

The $K^P=0^+$ projection of $\xi_0$ lowers the energy down to -397.3 MeV.  
The energy obtained by projecting this $K^P=0^+$ state further onto $J^P=0^+$ turns out to be -399.3 MeV.
In other words, the $K^P$ projection lowers the energy by 5.7 MeV, while the J-projected energy is less than 2 MeV away.   
%The MCSM calculation produces this energy, -401.066 MeV, the $\xi_0$ state absorbs a good fraction of those correlations.   

We stress that any state with axial symmetry (i.e., deformation parameter $\gamma$=0$^{\circ}$) remains the same by the rotation about the $z$ axis, implying $| \phi \,\rangle = | e^{i\gamma \hat{J}_z} \, \phi  \,\rangle$ \cite{note_circle}.  
\textcolor{black}{Consequently, the couplings between the states of different orientations do not exist, and there is no way to generate additional binding energy by the present mechanism.} 
In the example case, if the state $\xi_0$ were not triaxial at all, the $K$ projection 
would yield no effect, meaning no binding energy gain.  Thus, the triaxiality \textcolor{black}{effect} may not be treated properly unless the $K$ projection is included.
In the present case, the $\xi_0$ state is taken from the MCSM calculation, where the K-projection is \textcolor{black}{automatically implemented for all possible $K$ values}, and thereby this problem does not occur.  \textcolor{black}{The $K$ value may become an approximately good quantum number due to some reasons.  The MCSM should  then generate eigensolutions of the Schr\"odinger equation including such features.}   However, if one starts from a many-body approach without $K$ projection, a false solution likely emerges.

In order to see how a triaxial intrinsic state appears, we carry out an HFB calculation for the same Hamiltonian.  Although the HFB approach may not provide very accurate solutions, we can grasp certain aspects relatively easily.  The HFB calculation is performed only with the projection for the particle numbers, and the projected energies shown below are calculated after the variation (like VBP) 
for parity, $K$ and $J$.

\begin{figure}[bt]
  \centering
%%%%  6.6cm was the limit to put it in the text  
  \includegraphics[width=7cm]{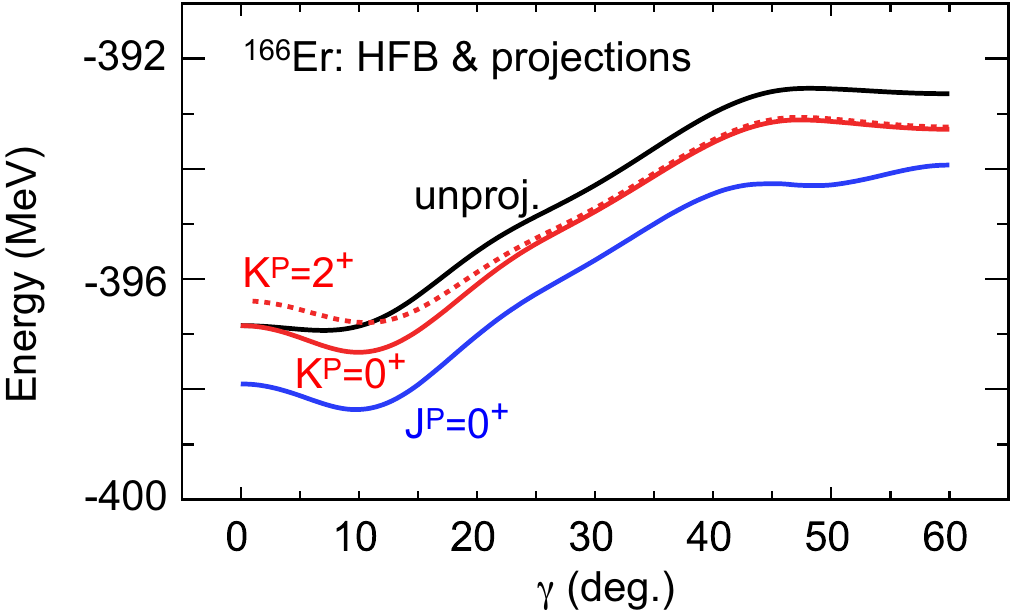}
    \caption{Energies of the ground state of $^{166}$Er given by HFB calculations as a function 
    of deformation parameter $\gamma$ (deg.) as a constraint.
    The energies are given by (black solid line) unprojected, (red solid line) K projected, and (blue 
    solid line) $J$  projected calculations.   
    The particle number is projected before variation, while parity, $K$ and $J$ are projected 
    after the variation.
    The $K^P$=$2^+$ energies are shown by red dotted line.
    Deformation parameter $\beta_2$ is fixed to the minimum point with actual value $\beta_2$=0.30.   
    } 
  \label{fig:hfb_proj_Er}  
\end{figure}

Figure~\ref{fig:hfb_proj_Er} shows the dependence of the calculated energies on the deformation parameter $\gamma$: 
its value is varied as a constraint to the HFB calculation, while another deformation parameter $\beta_2$ is fixed to the value of the minimum point.
% with $\gamma$=0$^\circ$.  
The unprojected energy only with the parity projection is basically flat from $\gamma$=0$^\circ$ up to $\gamma \sim$10$^\circ$.  The state projected onto $K^P$=0$^+$ captures additional binding through the mechanism presented above with Fig.~\ref{fig:Kproj} (d), and exhibits the energy minimum located around $\gamma$=10$^\circ$.  The additional binding gives a definite impact in drawing the shape of the nucleus, with the pronounced energy minimum around $\gamma$=10$^\circ$.  This $\gamma$ value is in good agreement with the one from the T-plot.  
Figure~\ref{fig:hfb_proj_Er} shows that the additional binding is saturated beyond $\gamma \sim$10$^\circ$, and the unprojected and K-projected energies are raised in parallel.  Figure~\ref{fig:hfb_proj_Er} also depicts that further projection onto $J^P$=0$^+$ basically lowers the $K$-projected energies by about the same amount up to $\gamma$=40$^\circ$, and that it does not change the location of the energy minimum.  Thus, for the location of the minimum, the $K$ projection is really crucial, 
and the induced shift of the $\gamma$ value to the energy minimum implies that the intrinsic structure and the restoration of rotational symmetry is not decoupled.

%%%%%%%%%%%%  FIGURE 7   %%%%%%%%%%%%%
% Fig7: gamma dependence

%%%%%%%%%%%%%%%%%%%%%%%%%%%%%%%%

\subsection{$K^P=2^+$ cases}
\label{K=2}

The case of $K^P=2^+$ is slightly more complicated, as both $K^P=+2^+$ and $K^P=-2^+$ must be considered because of time reversal symmetry in stationary states.  
As the parity projection is included unless otherwise stated, the parity is omitted in some cases, for the sake of brevity.  
Figure~\ref{fig:Kproj} (b) ((c)) depicts how the superposition of rotated $\phi$ occurs for $K=2$ ($K=-2$).   The multiplication factor becomes imaginary at Euler angle $\gamma = \pi/4$, and is negative at $\gamma = \pi/2$.  
%so as to fulfill time reversal symmetry, 
%and this combination will be denoted simply as $K^P=2^+$ for brevity.  
So, around $\gamma = \pi/2$, the multiplication factors of $K=\pm2$ projections show opposite signs to those of $K=0$.  

The expectation value of the Hamiltonian is evaluated similarly to the $K$=0 projection.   
%%% as (see Appendix \ref{Ap_Kproj} for more details).
We take equally-weighted linear combination of normalized $K$-projected states, $|K=2 \rangle$ and $|K=-2 \rangle$: $\tilde{\Phi} \bigl[\phi, K\bigr] \,=\,  \bigl( |K=2 \rangle + |K=-2 \rangle \bigr) /\sqrt{2}$ , and define
\begin{eqnarray}
& \tilde{h}_{K} \,&=\langle \tilde{\Phi} \bigl[\phi, K\bigr] \, | H \, | \tilde{\Phi} \bigl[\phi, K\bigr] \rangle 
\nonumber \\
 & &\propto %\frac{1}{|{\mathcal N}|^2} \, \frac{1}{{2\pi}}
      \, \int_0^{2\pi} d\gamma \, {\rm cos}(K\,\gamma) \, \langle \phi \, | H \, | \,     
     e^{i\gamma \hat{J}_z} \,  \phi \rangle.
\label{eq:H Kx}
\end{eqnarray}
The Hamiltonian does not couple a state of $K$=$K_1$ with a state of $K$=$K_2$ if $K_1 \ne K_2$.  Because of time-reversal symmetry, in the example of $K=\pm2$, the values of $\langle K=2\, | H | \, K=2 \rangle$ and $\langle K=-2 \,| H |\, K=-2 \rangle$ should be the same or approximately so, if $| \phi \rangle$ generates a dominant component of the state 
of interest.   This means 
%$\[ \langle K=2| + \langle K=-2| \} \,H\, \{ \rangle K=2| + \rangle K=-2| \} 
$\tilde{h}_{K} \sim \langle K=2 | H | K=2 \rangle \sim \langle K=-2 | H | K=-2 \rangle $.
Figure~\ref{fig:Kproj} (e) displays the contributions to $\tilde{h}_{K=2}$ from selected values of the Euler angle $\gamma$.  We find 
a vanished contribution at $\gamma=\pi/4$ (cf. $\cos(\pi/2)$=0) and a contribution opposite to the one for $K=0$ at $\gamma=\pi/2$ (cf. $\cos(\pi)$=-1).
This means that the $K=\pm2$ projections yield no additional binding energy from the contributions at $\gamma=\pi/4$, and that the contribution at $\gamma=\pi/2$ even diminishes the binding energy.   
Thus, the triaxiality generally decreases the binding energy gains of the $K=\pm2$ states compared to that of the $K=0$ state, locating $K=\pm2$ states above $K=0$ state.  
This effect is concretely discussed in the next section. %for the example with the $\xi_0$ state.
%1.5 MeV, which is nearly half of the additional energy gain of the $K^P=0^+$ state.  
%This high excitation energy is reduced by the $J^P$ projection (keeping K) and by increased triaxiality in $K^P=2^+$ state. 

This energy splitting is reduced in general, if the triaxiality becomes stronger.  
This is 
because for stronger triaxiality, $\langle \phi \, | H \, | \, e^{i\gamma \hat{J}_z} \,  \phi \rangle$ and  $\langle \phi \, |  \, e^{i\gamma \hat{J}_z} \,  \phi \rangle$ reduce faster as Euler angle $\gamma$ moves away from 0.  The faster reduction implies smaller differences of $K$=0 result from $K=\pm$2 results, as exemplified in Subsec.~\ref{sec:sketch}.
This is one of the reasons why the $\xi_3$ basis vector is better than $\xi_0$ basis vector as a single basis vector for calculating the energy of the $3^+_1$ (and also $2^+_2$) state of $^{166}$Er.

%%%%%%%%%%%%%%%%%%%%%%%%%%%%%%%%

\subsection{Relation to symmetry restoration and other remarks}
\label{symmetry restoration}

Coming back to general aspects, the $K$ projection of triaxial intrinsic state can be viewed from the symmetry restoration.  The triaxial intrinsic state has a particular orientation in the $xy$ plane, which violates the rotational symmetry about the $z$ axis. This symmetry is restored by rotating the intrinsic state about the $z$ axis, as shown in the previous two subsections, so that the $z$-component of the angular momentum takes an integer $K$.   As such $K$ value is conserved to a good approximation (exactly in some cases) for strongly deformed nuclei as discussed \textcolor{black}{in subsec.~\ref{subsec:Kmix},} it is a transparent and appropriate way to treat $K$ as a conserved quantum number.   We stress that this restoration mechanism is independent of actual nuclear forces.  In other words, the energy of such symmetry-restored state is a superposition of matrix elements of nuclear forces, and the weight factor is determined not by the forces but by the quantum-mechanical rotation procedure, in contrast to eigenvalue problems.  This feature makes the significance of the $K$ projection robust.

%We note that the MCSM calculation naturally includes possible $K$ mixings in the eigen solutions, independently of the amount of the mixing.

A systematic survey in terms of the Skyrme Hartree-Fock calculation indicated the dominance of the prolate nuclear shapes \cite{tajima_1996}.  \textcolor{black}{A recent Skyrme Hartree-Fock--Bogoliubov calculation \cite{scamps_2021} exhibits similar results, such as deformation parameter $\gamma$ =1.6$^{\circ}$ for $^{166}$Er, 5.6$^{\circ}$ for $^{164}$Dy, 1.2$^{\circ}$ for $^{154}$Sm, {\it etc}, which are well below the values obtained in the present study, already presented or to be presented. }
The universal finite-range liquid-drop model (FRLDM) similarly predicted no triaxial ground states for the deformed nuclei being discussed (see Fig. 2 of Ref. \cite{moller_2006} as mentioned earlier). 
These are examples of the conventional picture for the deformed nuclear shapes of heavy nuclei, where the properties of unprojected states were discussed.  Naturally, the triaxiality brought in by $K$ projection was absent.

The matrix element 
$\langle  \Phi \bigl[\phi, K=0\bigr] \, | H \, | \tilde{\Phi} \bigl[\phi, K=2\bigr] \rangle$ vanishes because of different $K$ values between bra and ket vectors.  When the total angular momentum $J$ is restored as discussed in the next section, this property is broken.   This breaking, however, is negligible as discussed \textcolor{black}{in subsec.~\ref{subsec:Kmix}}.    

Some details of the $K$ projection will be given in Appendix \ref{Ap_Kproj}. 

%%%%%%%%%%%%%%%%%%%%%%%%%%%%%%%%%%%%%%%%%%%%%%%
%%%%%%%%%%%%%%%%%%%%%%%%%%%%%%%%%%%%%%%%%%%%%%%

\section{A quantum theory of the rotation of multi-nucleon system \label{sec:J}}

After the discussions of the crucial consequence of the $K$ projection,
it is worth showing how the excitation energy at the rotor limit (see eq.~(\ref{eq:rot_energy}))
is obtained within the quantum many-body framework, without resorting to the quantization of the free rotational motion of a classical rigid body.

\textcolor{black}{Several attempts were made for this purpose in the past.   The developments by them can be more straightforwardly grasped in the view of the formulation presented below, and an overview  will be initiated in subsec.~\ref{subsec:remarks}.} 

\subsection{Formulation}

We here discuss the projection onto a good $J$ value.
The state $\phi_0$ is assumed to be a $K$=0 and normalized
intrinsic state, from which $J$-states are projected out.  
The state $\phi_0$ is actually the state discussed in the previous section.
The state $\phi_0$ is assumed to be of positive parity for simplicity, while the parity projection does not affect the following formulation.

Based on eq.~(\ref{eq:rot_phi}), 
we consider $J$-projected matrix elements and overlaps. 
Because of the rotational invariance of the Hamiltonian, $M$ can be taken to be 0
without losing the generality.  
The norm of the $J$-state component (with $M$=0) contained in $\phi_0$ is given by,
\begin{eqnarray}
&|{\mathcal N_J}|^2  \,&=\,\frac{2J+1}{8 \pi^2}\, \int_0^{2\pi} d\alpha \,  \int_0^{\pi} d({\rm cos}\beta) \, \int_0^{2\pi} d\gamma \,  \nonumber \\
& &   \,\,\,\,\,\,\,\,\,\,\, \langle \phi_0 \,|\, \big\{ D^J_{0,0} (\alpha, \beta, \gamma) \big\}^*
 e^{i\alpha \hat{J}_z} \, e^{i\beta \hat{J}_y} \, e^{i\gamma \hat{J}_z} \,  | \, \phi_0 \rangle \nonumber \\
& &= \, \frac{2J+1}{2} \, \int_0^{\pi} d({\rm cos}\beta) \, d^J_{0,0} (\beta) \, \langle \phi_0 \,|\, e^{i\beta \,\hat{J}_y} | \, \phi_0 \rangle , 
\label{eq:y_axis_0}
\end{eqnarray} 
where it is considered that $\phi_0$ is a $K$=0 state and the bra state is $\phi_0$ which is an  $M$=0 state.  

We introduce the norm kernel, 
\begin{equation}
n_y(\beta)  \,=\,  \langle \, \phi_0 \,|\, e^{i\beta \hat{J}_y} \,  | \, \phi_0 \rangle.
\label{eq:NJkernel}
\end{equation} 
Although this quantity is a complex number for a general state, here it is a real number because of $K, M=0$ for $|\,\phi_0 \rangle$.   
Equation~(\ref{eq:y_axis_0}) becomes
\begin{equation}
|{\mathcal N_J}|^2  \,= \, \frac{2J+1}{2} \, \int_0^{\pi} d({\rm cos}\beta) \, d^J_{0,0} (\beta) \, n_y(\beta) .
\label{eq:y_Nj}
\end{equation} 

Similarly, the expectation value of the Hamiltonian $H$ for a projected state is given by 
\begin{eqnarray}
& E_J \,&=\, %\langle \Phi \bigl[\phi, K=0\bigr] \, | H \, | \Phi \bigl[\phi, K=0\bigr] \rangle \nonumber \\
 \, \frac{2J+1}{2 |{\mathcal N}_J|^2}  \, \int_0^{\pi} d({\rm cos}\beta) \, \langle \phi_0 \, |  \, H \, d^J_{0,0} (\beta) \, e^{i\beta \hat{J}_y} \, | \, \phi_0 \,\rangle ,  \nonumber \\
& &=\, \frac{2J+1}{2 |{\mathcal N}_J|^2}  \, \int_0^{\pi} d({\rm cos}\beta) \, d^J_{0,0} (\beta) \, h_y(\beta) ,
\label{eq:H y}
\end{eqnarray}
where the energy kernel is introduced as, 
\begin{equation}
h_y(\beta)  \,=\,  \langle \, \phi_0 \,|\, H e^{i\beta \hat{J}_y} \,  | \, \phi_0 \rangle .
\label{eq:EJkernel}
\end{equation} 

The following identity is known 
\begin{equation}
d^J_{0,0} (\beta)  \,=\,  P_J ({\rm cos}\beta), 
\label{eq:dfunc}
\end{equation}
where $P_J ({\rm cos}\beta)$ stands for a Legendre \textcolor{black}{polynomial.} 

%The $P_J ({\rm cos}\beta)$ function satisfies
%\begin{equation} 
%\frac{d}{d({\rm cos}\beta)} \, \Big\{ (1 - {\rm cos}^2 \beta) \frac{d}{d({\rm cos}\beta)} P_J({\rm cos}\beta) \Big\} \, + J(J+1) P_J({\rm cos}\beta) = 0.
%\label{eq:PJeq}
%\end{equation}
%We now restrict ourselves to $\beta \approx 0$, because the values of $n_y(\beta)$ and $h_y(\beta)$ are reduced quickly as $\beta$ moves away from 0 as a consequence of strong deformation.  
%At the limit of $\beta \rightarrow$ 0, eq.~(\ref{eq:PJeq}) produces the derivative at $\beta=0$,
%\begin{equation} 
%\frac{d}{d({\rm cos}\beta)} \, P_J({\rm cos}\beta) \Big|_{\beta=0} \, = \, \frac{J(J+1)}{2},
%\label{eq:PJeq2}
%\end{equation}
%from which we obtain a linear approximation near $\beta=0$ 
%\begin{equation} 
%P_J({\rm cos}\beta) \, \approx \, 1 \,+\, F_J \, ({\rm cos}\beta - 1)\,\,\,\, {\rm for} \,\, \beta \approx 0 ,
%\label{eq:PJeq3}
%\end{equation}
%with
%\begin{equation} 
% F_J = J(J+1)/2.  
%\label{eq:FJ0}
%\end{equation}
%The function $d^J_{0,0} (\beta)  \,=\,  P_J ({\rm cos}\beta)$ in eqs.~(\ref{eq:y_Nj}) and (\ref{eq:H y}) is now approximated by this function.

\textcolor{black}{The Legendre polynomial $P_{\lambda} (z)$, as a function of $z$, is known to satisfy Legendre differential equation,
\begin{equation} 
\frac{d}{dz} \, \Big\{ (1 - z^2) \frac{d}{dz} P_{\lambda} (z) \Big\} \, + \lambda(\lambda+1) P_{\lambda} (z) = 0, 
\label{eq:Legendre1}
\end{equation}
where $z \in [-1, 1]$ and $\lambda=0, 1, 2, ..$.
We now look into $z\le1$ and $z\sim 1$.     %$z \precapprox 1$.
At the limit of $z \rightarrow$ 1, eq.~(\ref{eq:Legendre1}) produces the derivative at $z=1$,
\begin{equation} 
\frac{d}{dz} \, P_{\lambda} (z) \Big|_{z=1} \, = \, \frac{\lambda(\lambda+1)}{2},
\label{eq:Legendre2}
\end{equation}
where the so-called standardization condition $P_{\lambda} (1) = 1$ is used. 
The first two terms of the Legendre polynomial are written in the powers of $(z-1)$ as, 
\begin{equation} 
P_{\lambda} (z) \, = \, 1 \,+\, F_{\lambda} \, (z - 1)  \,+\,  ...,
\label{eq:Legendre3}
\end{equation}
with
\begin{equation} 
 F_{\lambda} = \lambda(\lambda+1)/2.  
\label{eq:Flam0}
\end{equation}
The two terms in eq.~(\ref{eq:Legendre3}) is a good approximation for $z \precapprox 1$. 
}

\textcolor{black}{
The values of $n_y(\beta)$ and $h_y(\beta)$ are reduced quickly as $\beta$ moves away from 0 as a consequence of strong deformation.  
With this situation, the d-function is approximated as 
\begin{equation} 
d^J_{0,0} (\beta)  \,=\,  P_J ({\rm cos}\beta) \, \approx \, 1 \,+\, F_J \, ({\rm cos}\beta - 1)\,\,\,\, {\rm for} \,\, \beta \approx 0 ,
\label{eq:PJeq3}
\end{equation}
with
\begin{equation} 
 F_J = J(J+1)/2.  
\label{eq:FJ0}
\end{equation}
As the integral is carried out with the variable ${\rm cos}\,\beta$ in eqs.~(\ref{eq:y_Nj}) and (\ref{eq:H y}), 
the d-function in eqs.~(\ref{eq:y_Nj}) and (\ref{eq:H y}) is naturally approximated by the function in eq.~(\ref{eq:PJeq3}), a polynomial of $({\rm cos}\,\beta - 1)$.
}

The range of $\beta$ runs from 0 to $\pi$.  Sizable contributions to the quantities in eqs.~(\ref{eq:y_Nj}) and (\ref{eq:H y}) are also expected for $\beta$ close to $\pi$, 
as the overlap is generally restored. For $\beta \sim \pi$, the linear and other approximations starting from $\beta = \pi$ back to smaller values work well also.  These contributions can be evaluated in the same way as those from $\beta \sim 0$.  Although they will not be explicitly discussed, they will be included in actual numerical results to be presented. 

We define
\begin{equation} 
n_k \, = \, \int d({\rm cos}\beta) \, n_y(\beta) \, ({\rm cos}\beta - 1)^k , \,\,\,\,  {\rm for } \, k=0, 1, 2, \dots.
\label{eq:ni}
\end{equation}
%and
%\begin{equation} 
%n_1 \, = \, \int d({\rm cos}\beta) \, n_y(\beta) \, ({\rm cos}\beta - 1).
%\label{eq:n1}
%\end{equation}
Likewise, similar quantities for the Hamiltonian are defined as
\begin{equation} 
e_k \, = \, \int d({\rm cos}\beta) \, h_y(\beta) , \, ({\rm cos}\beta - 1)^k , \,\,\,\, {\rm for } \, k=0, 1, 2, \dots.
\label{eq:hi}
\end{equation}
%and
%\begin{equation} 
%e_1 \, = \, \int d({\rm cos}\beta) \, h_y(\beta) \, ({\rm cos}\beta - 1).
%\label{eq:h1}
%\end{equation}
The projected energy of the state of $J$ is then given by
\begin{equation} 
E_J \, \approx \, \frac{e_0 \,+\, F_J e_1}{n_0 \,+\, F_J n_1},
\label{eq:EJ}
\end{equation}
which produces the excitation energies from the $J$=0 state,
\begin{equation} 
E_J \, - E_0 \,\approx \, \frac{F_J (e_1 n_0 \,-\,  e_0 n_1)}{(n_0 \,+\, F_J n_1) n_0} .
\label{eq:ExJ}
\end{equation}

\subsection{Contributions of leading and next-to-leading orders \label{subsec:K0_NLO}}

The quantities $n_0$ and $e_0$ are averages over the range of ${\rm cos} \beta$, and are relevant to the $J=0$ ground state as $P_{J=0}({\rm cos}\beta) = 1$.  As $n_y$ and $h_y$ are reduced quickly as $\beta$ moves away from $\beta=0$ (also from $\beta=\pi$), the magnitudes of $n_1$ and $e_1$ are smaller, respectively, than $n_0$ and $e_0$, due to the factor, ${\rm cos}\beta - 1$.
%We introduce $n_k$ and $e_k$, 
%\begin{equation} 
%n_k \, = \int d({\rm cos}\beta) \, n_y(\beta) \, ({\rm cos} \beta - 1)^k, \,\,\, {\rm for } \, k=2, 3, \dots ,
%\label{eq:n2}
%\end{equation}
%and
%\begin{equation} 
%e_k \, = \int d({\rm cos}\beta) \, h_y(\beta) \, ({\rm cos} \beta - 1)^k, \,\,\, {\rm for } \, k=2, 3, \dots .
%\label{eq:h2}
%\end{equation}
The magnitudes of $n_2$ and $e_2$ are smaller, respectively, than $|n_1|$ and $|e_1|$, and this trend may remain for certain higher $k$'s.
We then evaluate contributions of $n_k$'s and $e_k$'s (k=0, 1, 2) according to the order $k$, with the order of their product being the sum of the order of individual $n_k$ or $e_k$.
The leading-order (LO) contribution is given by $e_0 / n_0$ (see eq.~(\ref{eq:EJ})) for the states of all $J$ values to be considered.  It is also the exact $J$-projected energy of the $J$=0 state, $E_{J=0}$.  

The excitation energy of the state with $J$ (see eq.~(\ref{eq:ExJ})) is given in the next-to-leading order (NLO) of $({\rm cos}\beta - 1)$ by
\begin{eqnarray} 
E_x^{(k=1)} (J) \,= \, \frac{F_J (e_1 n_0 \,-\,  e_0 n_1)}{n_0^2} ,
\label{eq:ExJ2}
\end{eqnarray}
where the $F_J n_1/n_0$ term in the denominator of eq.~(\ref{eq:ExJ}) is dropped off because the numerator of eq.~(\ref{eq:ExJ}) is already of the NLO, and $n_0$ in the denominator is of the LO.   Eq.~(\ref{eq:ExJ2}) is rewritten as
\begin{eqnarray} 
E_x^{(1)} (J) \,= \, \frac{1}{2} J(J+1) \, \frac{e_0}{n_0} \, \Bigl\{ \,\frac{e_1}{e_0}\,-\,   \frac{n_1}{n_0}\,\Bigr\} .
\label{eq:ExJ3}
\end{eqnarray}
Note that at the LO, $E_x^{(0)} (J) \,= \, 0$, as stated above. 

The quantity $n_0$ is positive near $\beta$=0, where $n_1$ is negative.  As $h_y$ represents attractions from nuclear forces, $e_0$ is negative, and $e_1$  is positive.  As $\beta$ increases from zero, both $n_y(\beta)$ and $h_y(\beta)$ damp, but $h_y(\beta)$ damps more slowly than $n_y(\beta)$ does, because the Hamiltonian can connect two states oriented at more different angles.   This difference is expected to and indeed does result in an inequality $|e_1/e_0|  > |n_1/n_0|$, while $e_1/e_0$ and $n_1/n_0$ are both negative.   Combining all these features, the quantity after $J(J+1)$ in eq.~(\ref{eq:ExJ3}) becomes positive.
Thus, a perfect rotational excitation energy emerges.
Equation~(\ref{eq:ExJ3}) indicates that rotational excitation energies are generated by the Hamiltonian including nuclear forces.   This feature is robust and general.  For instance, the nuclear force does not have to be of any particular types and can include various terms even three-nucleon forces, for instance. 
What is needed is a strongly deformed state in the body-fixed frame shared by members of the band.

The present picture exhibits the ``quantum mechanical rotation'' of the nucleus.  The state observed in the laboratory is a superposition of this object pointing to different angular orientations.  The angular momentum, as a conserved quantum number, defines how the superposition occurs, and the multiplication factor for the $({\rm cos}\beta - 1)$ term indeed follows the $J(J+1)$ gradient.  This superposition provides us, in ideal cases, with the exact $J(J+1)$ excitation spectrum within a given rotational band.  

The origin of the rotational spectrum is the coupling between the same intrinsic state in two different angular orientations, originating in interactions between constituents (nucleons for atomic nuclei).  This is not the same as the increase of the kinetic energy of the free rotation of a rigid body in the conventional picture.  In the present view, the nuclear forces create a deformed intrinsic state, require the restoration of the rotational symmetry, and consequently produce the (nearly) $J(J+1)$ excitation spectrum, irrespectively of details of dynamics.

%%%%%%%%%%%%%%%%%%%%%%%%%%%%%%%%%
\subsection{Contributions of next-to-next-to-leading order}
\label{subsec:NNLO}
	
The Legendre function can be expanded for $\beta \approx 0$ up to $({\rm cos}\beta \,-\, 1)^2$ as
\begin{equation} 
P_J({\rm cos}\beta) \, \approx \, 1 \,+\, F_J \, ({\rm cos}\beta \,-\, 1) \,
+\, G_J \, ({\rm cos}\beta \,-\, 1)^2 \, .
\label{eq:PJ11}
\end{equation}
The coefficient $G_J$ is given as
\begin{equation} 
G_J = \frac{1}{16}\, J(J-1)(J+1)(J+2) = \frac{1}{16} \, \Big\{  \bigl(J(J+1)\bigr)^2 - 2J(J+1) \, \Big\},
\label{eq:PJ12}
\end{equation}
as proven by expanding $P_J(x)$ in terms of the polynomials of $(x-1)$.

The projected energy of the state of $J$ is then given by
\begin{equation} 
E_J \, = \, \frac{e_0 \,+\, F_J e_1 \,+\, G_J e_2}{n_0 \,+\, F_J n_1 \,+\, G_J n_2} ,
\label{eq:EJ2}
\end{equation}
with the excitation energies from the $J$=0 state,
\begin{equation} 
E_J \, - E_0 \,= \, \frac{F_J (e_1 n_0 \,-\,  e_0 n_1)\,+\,G_J (e_2 n_0 \,-\,  e_0 n_2)}{(n_0 \,+\, F_J n_1 \,+\, G_J n_2) n_0} .
\label{eq:ExJ2nd1}
\end{equation}
Expanding the denominator in terms of NLO term $n_1/n_0$ and next-to-next-to-leading order (NNLO) term $n_2/n_0$, we obtain the NNLO term from eq.~(\ref{eq:ExJ2nd1}),
\begin{eqnarray} 
&E_x^{(2)} (J) \,&= \, F_J^2 \, \frac{n_1}{n_0}  \,\Big\{ \frac{e_0 n_1}{n^2_0} \,-\,\frac{e_1}{n_0} \Big\} \, + \, G_J \, \frac{e_0}{n_0} \, \Bigl\{ \,\frac{e_2}{e_0}\,-\, \frac{n_2}{n_0}\, \Bigr\} , \nonumber \\
& &= \, - \, F_J \, E_x^{(1)} (J) \, \frac{n_1}{n_0} \,+ \, G_J \, \frac{e_0}{n_0} \, \Bigl\{ \,\frac{e_2}{e_0}\,-\,   \frac{n_2}{n_0} \,\Big\} .
\label{eq:ExJ2nd2}
\end{eqnarray}
The first term in the RHS is positive and raises the excitation energies, while the second term is likely negative and lowers the excitation energies. 

It is worth pointing that as the two terms of the RHS of eq.~(\ref{eq:ExJ2nd2}) cancel each other to a large extent, a large fraction of the remaining effects includes the $J(J+1)$ dependence because of $J^2(J+1)^2 - J(J-1)(J+1)(J+2) = 2 J(J+1)$.   

It is also emphasized that the effects at the NNLO are expressed by the two terms, $J(J+1)$ and $\{J(J+1)\}^2$.   This finding seems to be related to various empirical fits not only for nuclei but also for molecules \cite{atkins}.

%%%%%%%%%%%%%%%%%%%%%%%%%%%%%%%%%%%%%
\subsection{Verification Examples}

%%%%%%%%%%%%  FIGURE 8  %%%%%%%%%%%%%
% Fig 8: norm kernel

\begin{figure}[bt]
  \centering
  \includegraphics[width=6.5cm]{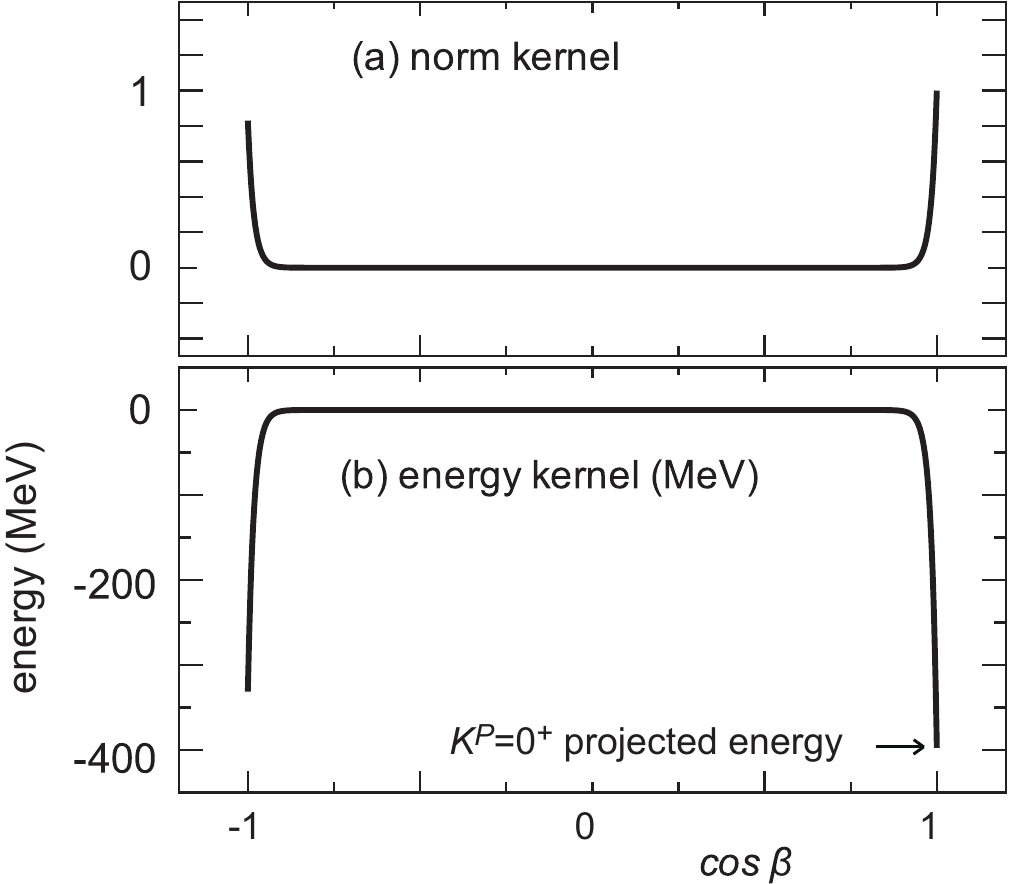}
     \caption{(a) norm kernel and (b) energy kernel as functions of the cos $\beta$.
    The case of $^{166}$Er is taken.  } 
  \label{fig:er_kernels}  
\end{figure}  

%%%%%%%%%%%%%%%%%%%%%%%%%%%%%%%%

Figure~\ref{fig:er_kernels} shows the norm and energy kernels, $n_y$ and $h_y$, as functions of cos$\beta$ for the $\xi_0$ state for $^{166}$Er.   The damping as cos$\beta$ deviates from 1 or -1 is shown to be fast as anticipated, while a closer look indicates that the damping is slower for the energy kernel in panel (b) than for the norm kernel in panel (a).
We point out that the energy at cos $\beta$=1 (i.e., $\beta$=0) in Fig.~\ref{fig:er_kernels} (b) is the energy obtained by the $K$=0 projection (-397.348 MeV) discussed in the previous section.  The energy comes further down for lower $J$ values (certainly for $J$=0) by restoring the quantum number, $J$. 

In this figure, the reflection symmetry with respect to cos~$\beta$=0 appears to a good extent, although this is not guaranteed with realistic $NN$ interactions.  It is pointed out that the figure is for one selected basis vector $\xi_0$ (see Subsec.~\ref{sec:gamma band}), but this symmetry is already fulfilled rather well, in accordance to the assumption of collective models.

Table~\ref{table:rot_ex} depicts the values of various quantities discussed in this section.
The values are obtained with the $\xi_0$ state introduced in Subsec.~\ref{sec:gamma band} for $^{166}$Er.  In later section, we will discuss the feature of $^{154}$Sm ($Z$=62, $N$=92).  Table~\ref{table:rot_ex} includes the values obtained for $^{154}$Sm, which will be 
discussed later.

%%%%%%%%%%%%%%   T a b l e   1     %%%%%%%%%%%%%%%%%%%
\begin{table}[tb]
 \caption{Examples of ground-state rotational bands. Energies are in MeV. }
 \label{table:rot_ex}
 \centering
 \begin{tabular}{l c c c c r}
 \toprule
 %  \hline
   quantity \textbackslash  nucleus  & \,\,\,\,\,$^{166}$Er & & \,\,\,\,\,$^{154}$Sm $0^+_1$ \\
 %\midrule
   \hline
   \hline
   $n_1$ / $n_0$   & \,\,-0.01731 & & -0.01654  \\
   $n_2$ / $n_0$ &  0.00061 & &  0.00056  \\
   $e_1$ / $e_0$ & -0.01736 & & -0.01661  \\
   $e_2$ / $e_0$ &  0.00061 & & 0.00056   \\
   \hline
   energy of $\phi_0$           & -397.35 & & -319.32 \\
   energy of J$^{\pi}$=0$^+$ state   & -399.30 & & -321.30 \\
   \hline
   Ex$^{(1)}$ (J=2$^+$)  & 0.0644  & &  0.0622  \\
   \,\,\,\,\,\,\,\,\,\,\,\, (J=4$^+$)  & 0.2147  & &  0.2072  \\
   \,\,\,\,\,\,\,\,\,\,\,\, (J=6$^+$)  & 0.4510   & &  0.4351  \\   
   \hline
   Ex$^{(2)}$ (J=2$^+$)  & 0.0011   &  &  0.0010  \\
   \,\,\,\,\,\,\,\,\,\,\,\, (J=4$^+$)  & 0.0037   & &  0.0029  \\
   \,\,\,\,\,\,\,\,\,\,\,\, (J=6$^+$)  & 0.0076   & &  0.0046  \\   
   \hline
    & NNLO &  exact proj. & NNLO & exact proj. \\
   \hline
   Ex$^{(T)}$ (J=2$^+$)  & 0.0655   & 0.0655 & 0.0631 & 0.0631 \\
   \,\,\,\,\,\,\,\,\,\,\,\, (J=4$^+$)  & 0.2184   & 0.2182 & 0.2100 & 0.2098 \\
   \,\,\,\,\,\,\,\,\,\,\,\, (J=6$^+$)  & 0.4586   & 0.4572 & 0.4396 & 0.4385 \\   
   \hline
   Ex(J=4$^+$) / Ex(J=2$^+$)  & 3.333  & &  3.326  \\
   Ex(J=6$^+$) / Ex(J=2$^+$)  & 6.997  & &  6.962  \\
   \bottomrule
   \end{tabular}
 \end{table}
 
Table~\ref{table:rot_ex} exhibits the rapid damping from $n_0$ to $n_2$ and that from $e_0$ to $e_2$.
The binding energy gain due to the couplings among different angles $\beta$ evaluated by the $J^{\pi}$ projection amounts to about 2 MeV, which is quite significant.
The next-to-leading order excitation energies are shown to dominate the excitation energies.
The excitation energies up to the NNLO are in good agreement with
the corresponding exact (or direct) projection.

It is of interest, in which situation $E_x(4^+_1)/E_x(2^+_1)$=3 may arise, as this is a possible border value for differentiating strongly deformed nuclei, as has been adopted in this paper.   
A similar question can be raised for $E_x(4^+_1)/E_x(2^+_1)$=2.5, which is more like the border to so-called gamma-soft nuclei, maybe combined with a proper criterion for other excited states.
By applying and extending the present ideas, we will be able to find solutions to those questions.

%%%%%%%%  FINITE   K  Values  %%%%%%%%%%% 
\subsection{Finite $K$ values}
 
The formulation for the $K$=0 rotational band can be generalized to 
rotational bands with finite $K$ values with certain formal complexities.
The final result of this subsection is shown in eq.~(\ref{eq:ExJ3K2}), which appears to be very simple and useful.  Even if one skips detailed derivation and jumps into eq.~(\ref{eq:ExJ3K2}), it does not prevent from understanding the rest of this paper.   We keep the derivation however, as it indicates some interesting and essential features of the collective motion. 

We mainly discuss the case of $K$=2, as other cases can be treated in the same way.
The wave function of the intrinsic state, from which relevant states are projected out, is supposed to have an invariance property with respect to angle-$\pi$ rotation about the $y$ axis, 
\begin{equation}
|\, \phi \, \rangle = e^{i \pi \hat{J}_y}  | \, \phi \, \rangle.
\label{eq:phi_pi}
\end{equation}
If not, it can be achieved by redefining $|\, \phi \, \rangle$ by $|\, \phi \, \rangle + e^{i \pi \hat{J}_y}  | \, \phi \, \rangle$ with a proper normalization.  Note that the $J$=0-projected state remains unchanged by this substitution, as this superposition is a part of the $J$-projection.  
  
As stated in Sec.~\ref{sec:K}, $K$=2 projection involves the projections onto $K$=2 and $K$=-2 subspaces.  The parity is always positive here and is not explicitly specified hereafter.   
We introduce $|\phi_K \rangle$ denoting the state projected from $\phi$ onto $K$ (see eq.~(\ref{eq:K-proj})):.
\begin{equation}
| \phi_K \rangle \,=\, \frac{1}{2\pi}  \int_0^{2\pi} d\gamma \, e^{i\gamma (\hat{J}_z - K)} \, | \phi \rangle\, .  %\,\,  K=\pm2 .
\label{eq:phiK}
\end{equation}
This is a part of the process shown in eq.~(\ref{eq:rot_phi}).
The $\pi$ rotation about the $y$-axis of $\phi_K$ is expressed as
\begin{eqnarray}
e^{i \pi \hat{J}_y} | \phi_K \rangle &=& \frac{1}{2\pi}  \int_0^{2\pi} d\gamma \, \Big\{ e^{i \pi \hat{J}_y}  e^{i\gamma (\hat{J}_z - K)} e^{-i \pi \hat{J}_y} \Big\} \Big\{ e^{i \pi \hat{J}_y}  \, | \phi \rangle \Big\} \nonumber \\
%&=& \frac{1}{2\pi}  \int_0^{2\pi} d\gamma \, \Big\{ e^{i\gamma (-\hat{J}_z - K)}  \Big\} \Big\{ e^{i \pi \hat{J}_y}  \, | \phi \rangle \Big\} \nonumber \\
&=& \frac{1}{2\pi}  \int_0^{2\pi} d\gamma \, e^{i\gamma (\hat{J}_z + K)}  \, |\phi \rangle 
= |\phi_{-K} \rangle , 
\label{eq:phiK_pi}
\end{eqnarray}
implying that $\phi_K$ and $\phi_{-K}$ are exchanged in this rotation.  

The state of $J$, $M(=J_z)$ projected from the $K$ component of $\phi$ is expressed, with the projection operator ${\mathcal P}_{J,M,K}$, as
\begin{eqnarray}
&&|\, J, \, M, \, K \rangle \, = \, {\mathcal P}_{J,M,K} \, | \, \phi \rangle,  
\label{eq:JMKstate}
\end{eqnarray}
with 
\begin{eqnarray}
{\mathcal P}_{J,M,K} \,
&=& \frac{2J+1}{8 \pi^2} \int_0^{2\pi} d\alpha \int_0^{\pi} d\beta \, {\rm sin}\beta \, \int_0^{2\pi} d\gamma \nonumber \\
&&\,\, \big\{ D^J_{M,K} (\alpha, \beta, \gamma) \big\}^* \, e^{i\alpha \hat{J}_z} \, e^{i\beta \hat{J}_y} \, e^{i\gamma \hat{J}_z} ,
\label{eq:ProjJMK}
\end{eqnarray}
where a Wigner function $D^J_{M,K} (\alpha, \beta, \gamma)$ is shown in eq.~(\ref{eq:Wigner}).
These two equations combined are essentially the same as eq.~(\ref{eq:rot_phi}), but the operator ${\mathcal P}_{J,M,K}$ is explicitly introduced for later purposes.  
The state in eq.~(\ref{eq:JMKstate}) is then rewritten as
\begin{eqnarray}
&&|\, J, \, M, \, K \rangle  \, \nonumber \\
&&= \frac{1}{2\pi} \int_0^{2\pi}   d\alpha \, e^{i\alpha (\hat{J}_z-M)} \, \frac{2J+1}{2}  \int_0^{\pi} d\beta \, {\rm sin}\beta \, d^J_{M,K} (\beta) \, e^{i\beta \hat{J}_y}   \nonumber \\
&&\,\,\,\,\,\, \frac{1}{2\pi} \int_0^{2\pi} d\gamma \, e^{i\gamma (\hat{J}_z-K)} \, | \, \phi \rangle , \nonumber \\
&&= \frac{1}{2\pi} \int_0^{2\pi} d\alpha \, e^{i\alpha (\hat{J}_z-M)} \, \frac{2J+1}{2} \int_0^{\pi} d\beta \, {\rm sin}\beta \, d^J_{M,K} (\beta) \, e^{i\beta \hat{J}_y}  \, | \, \phi_K \rangle . \nonumber \\
\label{eq:phk_Kproj}
\end{eqnarray}
We point out that the state $|\, J, \, M, \, K \rangle$ is obtained simply by the projection, and is not normalized.  All $|\, J, \, M, \, K \rangle$ states have the same ($J$, $M$) quantum numbers with different origins specified by $K$ values.
There is no orthogonality in general between $|\, J, \, M, \, K \rangle$ states with different $K$ values.   The integer $K$ can be interpreted as a kind of additional quantum number.  

It is now of interest to look into the relation between $| \, J, \, M, \, K \rangle$ and $| \, J, \, M, \, -K \rangle$. The latter is given by
\begin{eqnarray}
&&|\, J, \, M, \, -K \rangle  \, \nonumber \\
&&= \frac{1}{2\pi} \int_0^{2\pi}   d\alpha \, e^{i\alpha (\hat{J}_z-M)} \, \frac{2J+1}{2}  \int_0^{\pi} d\beta \, {\rm sin}\beta \, d^J_{M,-K} (\beta) \, e^{i\beta \hat{J}_y}   \nonumber \\
&&\,\,\,\,\,\, \frac{1}{2\pi} \int_0^{2\pi} d\gamma \, e^{i\gamma (\hat{J}_z+K)} \, | \, \phi \rangle , \nonumber \\
&&= \frac{1}{2\pi} \int_0^{2\pi} d\alpha \, e^{i\alpha (\hat{J}_z-M)} \, \frac{2J+1}{2} \int_0^{\pi} d\beta \, {\rm sin}\beta \, d^J_{M,-K} (\beta) \, e^{i\beta \hat{J}_y}  \, | \, \phi_{-K} \rangle .
\nonumber \\
\label{eq:-Lambda1}
\end{eqnarray}
The $d$ function in the integral can be transformed as
\begin{equation}
d^J_{M,-K} (\beta) \,=\, d^J_{-K,M} (-\beta) %\,=\, (-1)^{J-K} d^J_{K,M} (\pi-\beta) \nonumber \\
       \,=\, (-1)^{J-M} d^J_{M,K} (\pi-\beta).
\label{eq:-Lambda_d}
\end{equation} 
The integral with $\beta$ is then rewritten as
\begin{eqnarray}
&&\int_0^{\pi} d\beta \, {\rm sin}\beta \, (-1)^{J-M} d^J_{M,K} (\pi-\beta) \, e^{i\beta \hat{J}_y}  \, | \, \phi_{-K} \rangle , 
\label{eq:-Lambda2}
\end{eqnarray}
By substituting $(\pi-\beta)$ with $\beta'$, this integral becomes
\begin{eqnarray}
&&\int_0^{\pi} d\beta' \, {\rm sin}\beta' \, (-1)^{J-M} d^J_{M,K} (\beta') \, e^{i(\pi-\beta') \hat{J}_y}  \, | \, \phi_{-K} \rangle , \nonumber \\
&&=\int_0^{\pi} d\beta' \, {\rm sin}\beta' \, (-1)^{J-M} d^J_{M,K} (\beta') \, e^{-i\beta' \hat{J}_y}  \, | \, \phi_{K} \rangle , 
\label{eq:-Lambda2a}
\end{eqnarray}
where eq.~(\ref{eq:phiK_pi}) is used.  
The $\hat{J}_y$ operator comprises angular-momentum raising, $\hat{J}_+$, and lowering, $\hat{J}_-$, operators, and always changes the value of $J_z$ by one unit.  Both $e^{i\beta' \hat{J}_y}$ and $e^{-i\beta' \hat{J}_y}$ connect the state of $M$ to the state of $K$ and {\it vice versa}, because they contain, in their expanded forms, the $\hat{J}_y$ operator from the 0-th up to infinite powers.  The terms contributing to this connection includes the $|$K-M$|$-power of $\hat{J}_y$ multiplied by various even-power $\hat{J}_y$ terms.
In comparison between the contributions of $e^{i\beta' \hat{J}_y}$ and $e^{-i\beta' \hat{J}_y}$, the difference is represented by a single phase factor $(-1)^{K-M}$.  Thus, the integral in eq.~(\ref{eq:-Lambda2a}) is rewritten as
\begin{eqnarray}
\int_0^{\pi} d\beta' \, {\rm sin}\beta' \, (-1)^{J-K} \, d^J_{M,K} (\beta') \, e^{i\beta' \hat{J}_y}  \, | \, \phi_{K} \rangle .  
\label{eq:-Lambda3}
\end{eqnarray}
Putting this integral back to eq.~(\ref{eq:-Lambda1}), we obtain
\begin{equation}
|\, J, \, M, \, -K \rangle  \, = \, (-1)^{J-K} \, |\, J, \, M, \, K \rangle .
\label{eq:-Lambda4}
\end{equation} 
This equation  clearly indicates that the state $(-1)^{J-K} \, |\, J, \, M, \, -K \rangle$ is identical to $|\, J, \, M, \, K \rangle$.   
We stress that, for the ``additional (approximate) quantum number K'', $K< 0$ are redundant, and $K \ge 0$ are sufficient.  We shall refer only to $K \ge 0$ hereafter.

The present formulation is consistent with the conventional approach (see, for instance, \cite{ring_schuck_book}), with the state, 
\begin{equation}
\Omega^J_{M,K} \,\propto\, \Big\{ \, | \, J, \, M, \, K \rangle + (-1)^{(J+K)} \, | \, J, \, M, \, -K \rangle \, \Bigl\} ,
\label{eq:Phi_J}
\end{equation}
where $J$ and $K$ are assumed to be integers.  
As this state is built on a general intrinsic state without the symmetry in eq.~(\ref{eq:phi_pi}), 
the phase factor $(-1)^{(J+K)}$ is needed to make the state unchanged against the $\pi$-rotation about the $y$ axis in the body-fixed frame.
This phase factor $(-1)^{(J+K)}$ makes the expression in eq.~(\ref{eq:Phi_J}) invariant also with respect to the time reversal operation in the body-fixed frame.  The state of $J$ and $M$ is changed by the time-reversal operation, but its emerging mechanism defined in the body-fixed frame is supposed to be time-reversal invariant.  The same argument holds for the rotation about the $y$-axis in the body-fixed frame.  As mentioned already, this phase factor is not necessary in the present work.

In order to obtain the energy of the state $|\, J, \, M, \, K \rangle$, $M$=$K$ can be taken without losing the generality, because the energy does not depend on $M$.  The energy of $|\, J, \, K, \, K \rangle$ is then written as 
\begin{equation} 
E_{J,K} \,=\, \frac{\langle \phi |\, H \,{\mathcal P}_{J,K,K}\, | \phi \rangle}{\langle \phi |\, {\mathcal P}_{J,K,K}\,| \phi \rangle}.
\label{eq:energyPsi} 
\end{equation}
Note that the $K$ dependence originates in the $K$ projection of the state $\phi$ in the body-fixed frame, and this $K$ dependence naturally remains in the following arguments.   
We now calculate the numerator and denominator.  The latter is given by 
\begin{eqnarray}  
& & \langle \phi |\,  {\mathcal P}_{J,K,K} \,| \phi \rangle \nonumber \\
&=& 
\langle \phi | \, \frac{1}{2\pi} \int_0^{2\pi} d\alpha \, e^{i\alpha (\hat{J}_z-K)}  
\frac{2J+1}{2} \int_0^{\pi} d\beta \, {\rm sin}\beta \, d^J_{K,K} (\beta) \, e^{i\beta \hat{J}_y}  \, | \, \phi_K \rangle  \nonumber \\
&=& 
\frac{2J+1}{2}  \langle \phi_K | \int_0^{\pi} d\beta \, {\rm sin}\beta \, d^J_{K,K} (\beta) \, e^{i\beta \hat{J}_y}  \, | \, \phi_K \rangle .
\label{eq:phik_overlap}
\end{eqnarray} 
We generalize the definition of norm kernels from eq.~(\ref{eq:NJkernel}) to 
\begin{equation}
n^{(K)}_y(\beta)  \,=\,  \langle \, \phi_{K} \,|\,  e^{i\beta \hat{J}_y} \,    | \, \phi_{K} \rangle .
\label{eq:NJkernelK2}
\end{equation} 
The norm kernel for $K$=0 is then expressed as $n^{(0)}_y(\beta)$.  
The individual norm fraction is denoted by
\begin{equation}
n^{(K)}_J \,= \, \frac{2J+1}{2} \, \int_0^{\pi} d\beta \, {\rm sin}\beta \, d^J_{K,K} (\beta) \, n^{(K)}_y(\beta) .
\label{eq:y_nK2}
\end{equation}  

Likewise, we obtain the numerator in eq.~(\ref{eq:energyPsi}).   
For evaluating this quantity, we extend the energy kernels for $K$=0 to a more general form,
\begin{equation}
h^{(K)}_y(\beta)  \,=\, \langle \, \phi_{K} \,|\,  \, H \, e^{i\beta \hat{J}_y} \,  | \, \phi_{K} \rangle \, .
\label{eq:EJkernelK2}
\end{equation} 
The individual energy fraction is written as
\begin{equation}
e^{(K)}_J \,= \, \frac{2J+1}{2} \, \int_0^{\pi} d\beta \, {\rm sin}\beta \, d^J_{K,K} (\beta) \, h^{(K)}_y(\beta) .
\label{eq:y_eK2}
\end{equation}  

Equations~(\ref{eq:y_nK2}) and (\ref{eq:y_eK2}) are precise, but we handle those quantities in a different way.
For strongly deformed nuclei, norm and energy kernels are considered to be swiftly reduced as $\beta$ moves away from $\beta$=0 or $\beta=\pi$.  So, the behaviors of relevant integrands are discussed near $\beta=0$ and $\beta=\pi$.   
Utilizing the definition of the Wigner small $d$ function and the Jacobi function, we expand the following small $d$ functions near $\beta=0$ (i.e., cos$\beta$=1) as, 
\begin{equation}
d^J_{K,K} (\beta) = 1\, + F_{J,K} \,  ({\rm cos}\beta  -1) \, +  \,{\rm (higher \, order \, terms)} , 
\label{eq:d22}
\end{equation} 
with
\begin{equation}
F_{J,K} = \frac{1}{2} \, \bigl( \, J(J+1) - K^2 \, \bigr) . 
\end{equation}
Note that $F_J$ in eq.~(\ref{eq:FJ0}) implies $F_{J,K=0}$.   
On the other hand, near $\beta=\pi$, $d^J_{K,K} (\beta)$ with $K\ne0$ behaves in proportion to $\bigl(\,{\rm cos}(\beta-\pi) - 1\,\bigr)^K$ + higher-order terms, and thereby their 
contributions do not appear in the LO or NLO for $K\ge$ 2.  
Similarly to the $K$=0 case, norms are categorized in the hierarchy of the $k$-th power of the $({\rm cos}\beta - 1)$ term as, 
\begin{equation} 
\tilde{n}^{(K)}_k \, = \, \int d({\rm cos}\beta) \, n^{(K)}_y(\beta) \, ({\rm cos}\beta - 1)^k . \label{eq:nK2}
\end{equation}
where the tilde is placed in order to clearly distinguish from kernels.
The corresponding quantities for the energy, $\tilde{e}^{(K)}_k$, are similarly defined, 
by replacing $n^{(K)}_y(\beta)$ with  $h^{(K)}_y(\beta)$, as
\begin{equation} 
\tilde{e}^{(K)}_k \, = \, \int d({\rm cos}\beta) \, h^{(K)}_y(\beta) \, ({\rm cos}\beta - 1)^k . 
\label{eq:hK2}
\end{equation}
with $h^{(K)}_y(\beta)$ defined in eq.~(\ref{eq:EJkernelK2}).  
We will focus on the two lowest terms of $k$=0 (LO) and $k$=1 (NLO).

We obtain, up to the NLO,
\begin{equation} 
\langle \phi |\, {\mathcal P}_{J,K,K} \, |\phi \rangle = \tilde{n}^{(K)}_0 + \tilde{n}^{(K)}_1 \,F_{J,K} .
\end{equation}
The contributions to the energy are treated in the same way, 
and are represented by $\tilde{e}^{(K)}_0$ for the LO and by $\tilde{e}^{(K)}_1$ for the NLO.
We thus obtain up to the NLO,
\begin{equation} 
\langle \phi |\, H \,{\mathcal P}_{J,K,K} \, | \phi \rangle = \tilde{e}^{(K)}_0 + \tilde{e}^{(K)}_1 \, F_{J,K}.
\end{equation}

%%%%%%%%%%%%%%%%
Similarly to $K$=0 case, the contributions from a given order $k$ can be obtained, and 
$E_{J,K}$ in eq.~(\ref{eq:energyPsi}) is expressed for the LO and NLO combined as, 
\begin{equation} 
E^{(0+1)}_{J,K} \,= \, \frac{\tilde{e}_0}{\tilde{n}_0} \, 
                    +\, \frac{1}{2} \,\Big\{ J(J+1) - K^2  \Big\} \,  \frac{\tilde{e}_0}{\tilde{n}_0} \, \Big\{ \frac{\tilde{e}_1}{\tilde{e}_0}\,-\, \frac{\tilde{n}_1}{\tilde{n}_0} \Big\} ,
\label{eq:ExJ3K2}
\end{equation}  
where the factor $F_{J,K}$ is explicitly depicted, with $J=K, K+1, K+2, ...$. Here, $\tilde{n}_{0,1}$ and $\tilde{e}_{0,1}$ are defined in eqs.~(\ref{eq:nK2}) and (\ref{eq:hK2}), respectively.  The superscript $K$ for $\tilde{n}$'s and $\tilde{e}$'s is omitted here for brevity, while they are $K$ dependent.

This equation implies that rotational bands specified by $K$ are built on top of the $J=K$ state with the $J(J+1)-K^2$ dependence of the excitation energy.  This is exact at the NLO, and is consistent with the features shown for $K$=0 in Subsec.~\ref{subsec:K0_NLO}. 

It is remarkable that such a $\bigl( J(J+1)-K^2 \bigr)$ dependence is obtained within a quantum many-body scheme, without using the quantization of the classical rigid-body rotor with an intrinsic angular momentum $K$, as many textbooks show.  
It is noted that the ``moment of inertia'' of the ground band and that of other bands appear to be different in general.  
We note that the $K$ quantum number makes sense, as it is practically conserved in strongly deformed nuclei as discussed in subsec.~\ref{subsec:Kmix}.

We just note the contribution from the NNLO as
\begin{equation} 
E^{(2)}_{J,K} =  
  - (F_{J,K})^2 \,  \frac{\tilde{e}_0}{\tilde{n}_0} \, \frac{\tilde{n}_1}{\tilde{n}_0} \, \Big\{ \frac{\tilde{e}_1}{\tilde{e}_0}\,-\, \frac{\tilde{n}_1}{\tilde{n}_0} \Big\} 
  + G_{J,K} \,  \frac{\tilde{e}_0}{\tilde{n}_0} \, 
       \Big\{ \frac{\tilde{e}_2}{\tilde{e}_0}\,-\, \frac{\tilde{n}_2}{\tilde{n}_0} \Big\} ,                  
\label{eq:ExJ_NNLO}
\end{equation}  
where $G_{J,K}$ is defined as
\begin{equation}
G_{J,K} = \frac{1}{16}  \, \Big\{ \, \bigl( J(J+1)\bigr)^2 - 2(K^2 +1) J(J+1) +K^2 (K^2 +3)\, \Big\} .
\end{equation}
All terms in eq.~(\ref{eq:ExJ_NNLO}) are expressed again by polynomials of $J(J+1)$.

%%%%%  Sketch   %%%%%%%%%%
\subsection{Sketch for projected energy levels for $K$ and $J$ quantum numbers \label{sec:sketch}}

We here sketch the energies of the states projected for $K$ and $J$ quantum numbers.
The states $\xi_0$ and $\xi_3$ introduced in Subsec.~\ref{sec:gamma band} are taken.  
They are the most optimized single basis vectors in the MCSM calculation for $K$=0 and $K$=2 states, respectively.  As already indicated, 
the states $\xi_0$ and $\xi_3$ show $\gamma$=8.5$^{\circ}$ and 9.8$^{\circ}$, respectively.  

These two states are symmetrized so as to fulfill the condition in eq.~(\ref{eq:phi_pi}) (see the process shown below eq.~(\ref{eq:phi_pi})).  We hereafter denote such symmetrized states simply as $\xi_0$ and $\xi_3$, for the sake of brevity, except for the three unprojected energy levels as they are calculated from the original $\xi_0$ and $\xi_3$ states (i.e., no symmetry in eq.~(\ref{eq:phi_pi})). We indicate the magnitude of this symmetrization effect by noting that it lowers the energies of  $K^P$ projected states by $\sim$0.7 MeV for $\xi_0$.  

%%%%%%%%%%%%%%   F i g    9   %%%%%%%%%%%%%%%
\begin{figure*}[bt]
  \centering
  \includegraphics[width=13cm]{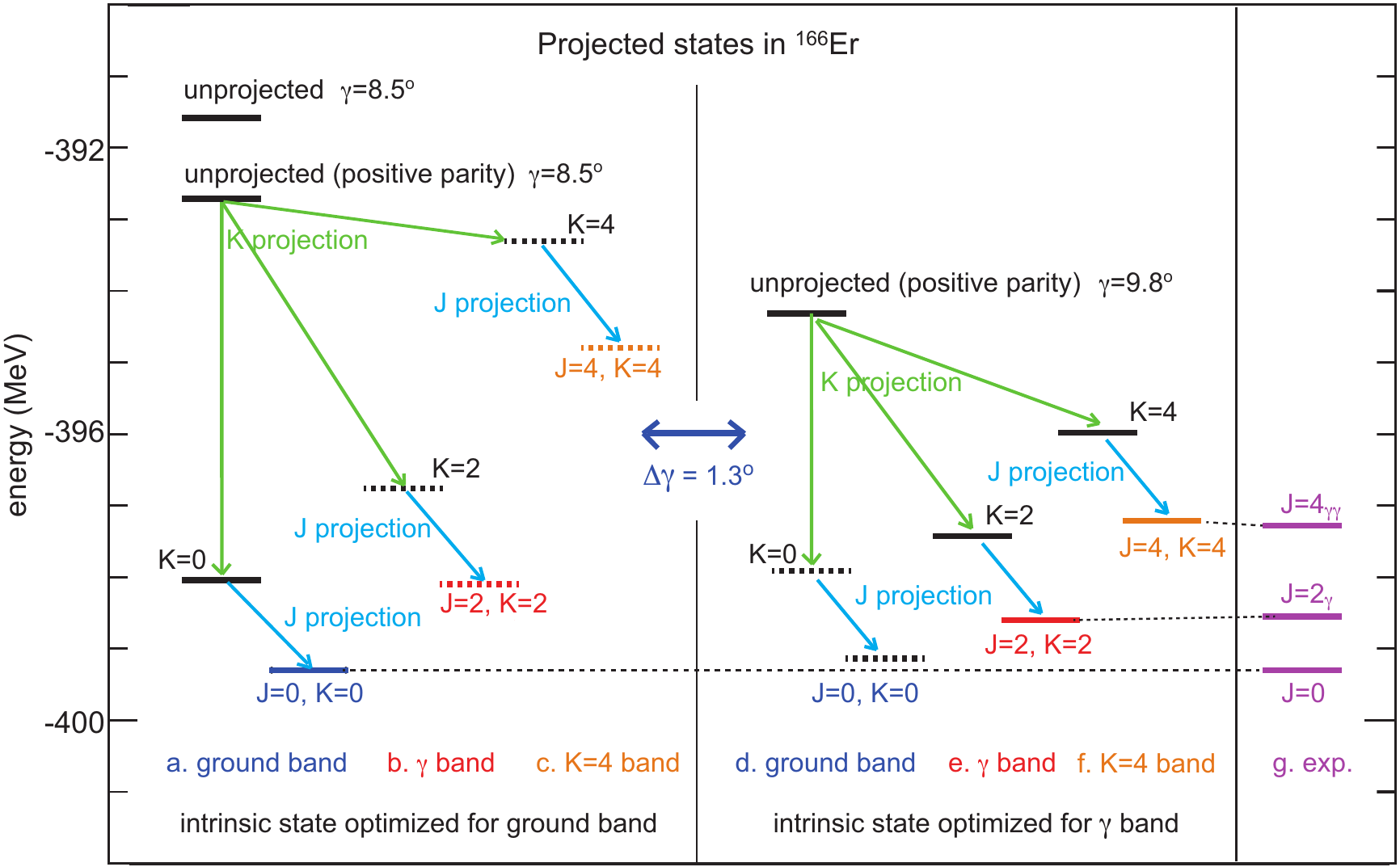}
\caption{Energies of projected states of $^{166}$Er and related experimental level energy.  All theoretical states are projected from the $\xi_0$ or $\xi_3$ state introduced in Subsec.~\ref{sec:gamma band}.  Parity is positive for all the states but unprojected state at the top. Columns {\bf a} and {\bf d} are for the ground band,  columns {\bf b} and {\bf e} are for the $\gamma$ band, and columns {\bf c} and {\bf f} are for the $K$=4 state.  Column {\bf g} is the observed excitation energies of 2$^+_{\gamma}$ and 4$^+_{\gamma\gamma}$ states.
%Dashed lines indicate upper state for a given $K$ or $J$=$K$ between two cases of $\gamma$=8.5$^{\circ}$  and 9.8$^{\circ}$.  
Between two panels {\bf a-c} ($\gamma$=8.5$^{\circ}$) and {\bf d-f} ($\gamma$=9.8$^{\circ}$), the lower energy level of a given quantum number is shown by a solid horizontal bar, while the higher one by the dashed horizontal bar.}  Green arrows imply energy gains due to $K$ projections, while blue arrows those due to $J$ projections.     
  \label{fig:KJproj_166Er}  
\end{figure*}  

The states $|\phi_K \rangle$ and $|\phi_{-K} \rangle$ now carry basically the same intrinsic structure.  Consequently, their energies $\langle \phi_K |\, H \,|\phi_K \rangle$ and $\langle \phi_{-K} |\, H \,|\phi_{-K} \rangle$ are equal.   Here, they will simply be denoted ``$K$=2 energy'', {\it etc.}   
The left part of Fig.~\ref{fig:KJproj_166Er} displays this energy projected from $\xi_0$ as well as the corresponding energies of $K$=0 and 4.  It is reminded that $\xi_0$ is the most important basis vector of the present CI calculation for the ground state of $^{166}$Er (see Subsec.~\ref{sec:gamma band}).
The unprojected energy (no parity projection either) is shown at the top of column {\bf a} of Fig.~\ref{fig:KJproj_166Er}.  All other states in Fig.~\ref{fig:KJproj_166Er} are projected to positive parity.  
The projected $K$=0 energy is lowered by more than 5 MeV from the unprojected energy, as shown in column {\bf a}.  The subsequent $J$=0 projection further lowers the energy by more than 1 MeV.  

Column {\bf b} depicts the energies of the states projected to $K$=2 located about 4 MeV below the unprojected state.  
The $K$=2 state is projected to $J$=2 with the energy $\sim$1.2 MeV above the $J$=0 level.  This $J$=2 level energy is too high compared to the observed excitation energy $\sim$ 0.8 MeV, which is shown in column {\bf g}.

The energies of the states projected from $\xi_3$ are displayed in panels {\bf d-f} of Fig.~\ref{fig:KJproj_166Er}.  
Between two panels {\bf a-c} and {\bf d-f}, the lower energy level of a given 
quantum number is shown by a solid horizontal bar, while the higher one is by the dashed bar.  So, the sequence of the solid bars represents the actual situation.
The $J$=$K$=2 level energy in column {\bf e} (solid red bar) indicates a good agreement with the experiment.  Thus, the increase of the deformation parameter $\gamma$ in the $K^P$=2$^+$ ($\gamma$) band gives a notable improvement.   The CI calculations with additional basis vectors provide more accurate results, while the basic trend of the level structure remains unchanged, as exhibited in Fig.~\ref{fig:tplot}.  

Figure~\ref{fig:KJproj_166Er} also displays that the excitation energies of $K$=2 and 4 states relative to the $K$=0 state are more compressed for the larger $\gamma$ value (columns {\bf d, e, f}) than those energies with the smaller $\gamma$ value (columns {\bf a, b, c}).  This illustrates the $\gamma$-induced level-energy compression discussed in Subsec.~\ref{K=2}. 

Figure~\ref{fig:KJproj_166Er} exhibits the energies of $K$=4 states, which will be discussed in Subsec.~\ref{sec:double_gamma} in detail.

%%%%%%%%%%%   Remarks on earlier approaches   %%%%%%%%%%%%%%%
\subsection{Remarks on earlier approaches}
\label{subsec:remarks}

 \textcolor{black}{The microscopic picture of nuclear rotation has been studied over decades from different angles, for example, \cite{peierls_yoccoz_1957,verhaar_1963a,verhaar_1963b,verhaar_1964,kamlah_1968,siemens_book}.  
Comprehensive reviews on low-spin members of rotational bands seem to be rather few, however.
Comparatively recent overviews can be found, for example, in \cite{satula_wyss_2005} as well as in the textbook \cite{ring_schuck_book}.  The former paper is mainly for high-spin states.  The latter article discusses rotational spectrum with the basic picture of A. Bohr: the rotation of deformed objects and the excitation energy as the kinetic energy of its free rotation. This point will be discussed in detail  in sec.~\ref{sec:summary}.}
  
 \textcolor{black}{We survey how the rotational features such as the $J(J+1)$ rule has been investigated in terms of quantum many-body problems in Appendix  \ref{Ap_rot_energy}.}
 
 \textcolor{black}{It is worth noting that comparatively recently, the rotational spectrum was studied in terms of the Effective Field Theory \cite{papenbrock_2011}.  It is not a many-body theory, but there are interesting viewpoints.  One of them can be the separation between the Nambu-Goldstone type rotation \cite{nambu_1960,goldstone_1961,goldstone_1962} and the vibrational excitation, as a separation due to different energy scales.  This separation corresponds to the separation between the J projection (yielding low excitation energy) and K projection (yielding high excitation energy) in the present perspective, although the relation is not direct. }

%%%%%%%%%%%%%%%%%%%%%%%%%%%%%%%%%%%%%%%%%% 

%%%   SECTION:  SECOND  MAJOR  ORIGIN 

\section{Second major source of triaxiality: nuclear forces} 
\label{sec:2nd source}

%%%%%%%%%%%%%%%%%%%%%%%%%%%%%%%%%%%
\subsection{Introduction}
\label{subsec:2nd source}

Section \ref{sec:K} showed that triaxial shapes arise as a consequence of the symmetry restoration represented by $K$ quantum number.  
This section exhibits its another source: nuclear forces.
 
%%%%%%%%%%   Figure  10    %%%%%%%%%%
\begin{figure}[tb]
  \centering
  \includegraphics[width=7cm]{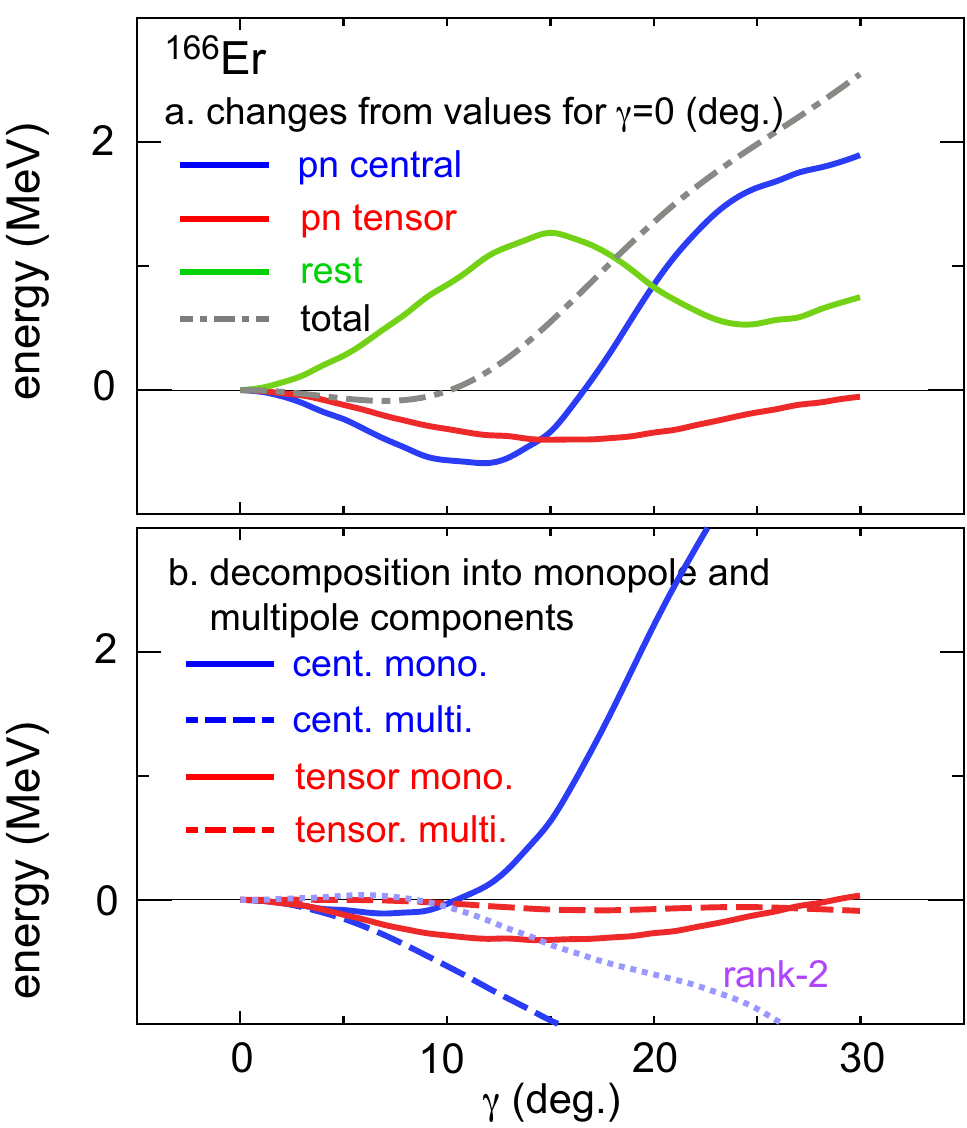}
    \caption{ Properties of the unprojected PES of $^{166}$Er as a function of the deformation parameter $\gamma$ as a constraint.  {\bf a.} Decomposition into contributions of proton-neutron central (blue) and tensor (red) forces and the rest (green) of the Hamiltonian. 
    {\bf b.} Decomposition of proton-neutron interaction effects into monopole and multipole components.    
    } 
  \label{fig:hfb_cto_166Er}  
\end{figure}  

Figure~\ref{fig:hfb_proj_Er} shows that the unprojected PES at $\gamma$=10$^{\circ}$ is as low as that at $\gamma$=0$^{\circ}$, with a shallow minimum between $\gamma$=0$^{\circ}$ and 10$^{\circ}$.  As this widely spread minimun becomes  the profound minimum after the $K$ projection, the understanding of the origin of this wide minimum is of great importance.
% for the clarification of the triaxiality.  

A  decomposition of the unprojected energy is carried out by calculating the expectation values of the \textcolor{black}{proton-neutron ($pn$)} central-force and tensor-force components of the Hamiltonian, with respect to the unprojected state specified by the deformation parameter $\gamma$ as a constraint.  The relative changes of their values from those at $\gamma$=0$^{\circ}$ are shown in Fig.~\ref{fig:hfb_cto_166Er}{\bf a} as functions of $\gamma$.  
The corresponding value of the rest of the Hamiltonian is also displayed.

Figure~\ref{fig:hfb_cto_166Er}{\bf a} indicates that the $pn$ central and tensor forces produce more binding energies for $\gamma$=0-15$^{\circ}$, while the rest of the Hamiltonian works oppositely, raising the energy in the whole region of Figure~\ref{fig:hfb_cto_166Er}{\bf a}.  We restrict ourselves to the region of $\gamma$=0-15$^{\circ}$ now.    The opposite trends between the $pn$ forces and the rest part is crucial in producing \textcolor{black}{a wide minimum around $\gamma$=7$^{\circ}$, which is a seed of the profound minimum around $\gamma$=10$^{\circ}$ after the $K$=0 projection as stated above.}

%%%%%%%%%%%%%%%%%%%%%%%%%%%%%%%%%%%%%%
\subsection{Monopole and multipole interactions: a brief note}
\label{subsec:monomulti}

%In order to explore underlying mechanisms for the variations shown in Fig.~\ref{fig:hfb_cto_166Er}{\bf a}, 
\textcolor{black}{
For further clarification, we decompose} the $pn$ central-force and tensor-force effects into their monopole and multipole contributions:  
A given two-body interaction, denoted by $\hat{v}$, \textcolor{black}{comprises} monopole and multipole interactions or components \cite{poves_1981}.  
When two nucleons are in single-particle orbitals $j$ and $j'$, 
\textcolor{black}{respectively, the monopole interaction implies an average effect of $\hat{v}$ over all possible angular-momentum couplings of the two nucleons.  The resulting interaction can be represented in terms of the occupation number operators.
For instance, the monopole interaction between a proton and a neutron is expressed as 
\begin{equation}
\hat{v}_{pn;mono} = \Sigma_{j,j'} \, v^{pn;mono}_{j,j'} \, \hat{n}^{(p)}_j \, \hat{n}^{(n)}_{j'} 
\label{eq:pn_mono}
\end{equation}
where $\hat{n}^{(p)}_j$ ($\hat{n}^{(n)}_{j'}$) denotes the number of protons in orbital $j$ (neutrons in $j'$) and $v^{pn;mono}_{j,j'}$ stands for coefficients called monopole matrix elements.
More details can be found, for instance, in a review \cite{otsuka_2020} or a brief lecture note \cite{otsuka_2022_emerging}.
Clearly, monopole(-interaction) contributions can linearly grow with the particle number in a particular orbital, and can be quite large, as} $\langle n^{(p,n)}_j \rangle$ can reach a value of $(2j+1)$. 
 
\textcolor{black}{
The rest of the interaction $\hat{v}$ is called the multipole interaction \cite{poves_1981}.
The multipole(-interaction) contribution does not have this linear dependence,} and is vanished if the orbital is fully-occupied.   Thus, the systematic trend of the multipole contributions differs from that of the monopole contributions.

A given $pn$ interaction can be expressed as
\begin{equation}
\hat{v}_{pn} = \Sigma_L \,\, \hat{v}^{(L)}_{pn} ,
\label{eq:pn_tot}
\end{equation}
with
\begin{equation}
\hat{v}^{(K)}_{pn}  \,=\, \Sigma_{j_p,j'_p,j_n,j'_n} \, f_{j_p,j'_p,j_n,j'_n, L} \, (\hat{o}^{(L)}_{j_p,j'_p} \, \hat{o}^{(L)}_{j_n,j'_n} ) ,
\label{eq:pn_L}
\end{equation}
where $L$ stands for angular momentum called rank, $f$ is coefficient, $( \,\, )$ means a scalar product, and $j_p$ and $j'_p$ ($j_n$ and $j'_n$) are proton (neutron) orbitals.   Here, $\hat{o}^{L}_{j,j'}$ implies the unit operator that annihilates a nucleon in the orbital $j'$ and creates a nucleon in the orbital $j$ with angular momentum $L$ transferred from $j'$ to $j$. 
(As the parity, $P=\pm$1, can be transferred, $L$ implicitly includes it.)
The $L$=0$^+$ part, $\hat{v}^{(L=0^+)}_{pn}$, is equal to the monopole interaction if the summation in eq.~(\ref{eq:pn_L}) is restricted to $j$=$j'$ including all possible additional quantum numbers. This condition is actually fulfilled with the present orbitals.  The $pn$ multipole interaction is given by $L\ne0^+$ terms in eq.~(\ref{eq:pn_tot}).  Among such terms, the term most relevant to the ellipsoidal shape deformation is $\hat{v}^{(L=2^+)}_{pn}$ term, a quadrupole term, as we shall discuss later.  
The multipole interaction is the origin of the shape deformation in general, because the monopole interaction effectively changes single-particle energies. %without mixing different configurations.  

%We note that the notation of ``rank $K$'' follows the standard convention, and this article is prepared so that it should not be confused with the quantum number $K$.  

%%%%%%%%%%%%%%%%%%%%%%%%%%%%%%%
\subsection{Central-force contribution}

\textcolor{black}{
Figure~\ref{fig:hfb_cto_166Er}{\bf b} shows the decomposition into monopole and multipole contributions. The central-force multipole contribution remains negative (more binding).}  Its monopole contribution is weakly negative for $\gamma < $10$^{\circ}$, but becomes repulsive for $\gamma >$10$^{\circ}$.  These properties are explained as follows.  As the triaxiality increases, the wave function becomes more complex in general.  \textcolor{black}{Wider types of correlations can then be involved or developed in wave functions, some of which can give rooms for more energy gain by central-force multipole interactions. This general argument is confirmed by Fig.~\ref{fig:hfb_cto_166Er}{\bf b} and other examples to be shown.  
Such interactions generally induce more occupations of some orbitals unfavored 
by monopole interaction.  This is considered to be a main reason for the rise of central-monopole contribution beyond $\gamma$=10$^{\circ}$.     
Up to $\gamma$=10$^{\circ}$, however, this rising trend is suppressed, because the nuclear system finds appropriate configurations so that monopole contributions can provide sufficient binding energy for growing $\gamma$.}  This effect is one example of the self-organization with monopole interaction in nuclei, which implies the optimization of configurations for a given triaxiality \cite{otsuka_2019}.  
Apparently, this optimization is not strong enough for $\gamma \gg$10$^{\circ}$.
 
%%%%%%%%%%   Figure 11  &  FIGURE 12  were here  %%%%%%%%%%

Figure~\ref{fig:hfb_cto_166Er}{\bf b} displays, by dotted line labeled ``rank-2'', 
the difference of the contribution of $\hat{v}^{(L=2^+)}_{pn}$ from its value for $\gamma$=0$^{\circ}$, as a function of $\gamma$.  \textcolor{black}{This difference stays nearly vanished up to 10$^{\circ}$.}  This looks peculiar at the first glance, and we now discuss how it arises.
Because of eq.~(\ref{eq:beta2_QQ}),    
$\beta_2^2$ approximately proportional to the expectation value of $(\hat{Q} \,\hat{Q})$ with respect to the unprojected intrinsic state.
The corresponding expectation value of $\hat{v}^{(K=2^+)}_{pn}$ is considered to be approximately  proportional to this expectation value, because of coherence between proton and neutron contributions.  
Thus, the expectation value of $\hat{v}^{(K=2^+)}_{pn}$ is expected to be $\propto \beta_2^2$, and stays constant as a function of $\gamma$ for $\beta_2$ fixed.   
Figure~\ref{fig:hfb_cto_166Er}{\bf b} confirms this feature.
%It is thus inferred that the $pn$ $L$=2$^+$ interaction may not notably move the triaxial minimum of unprojected state for a fixed $\beta_2$.  
The next natural higher multipole, $L$=4$^+$ (i.e., hexadecupole), appears to be the major driving force for larger triaxiality, as we shall see it in other examples.

\textcolor{black}{
The quadrupole ($L$=2$^+$) and hexadecupole ($L$=4$^+$) components of $pn$ central interaction do not show a large difference in their magnitude scales at the level of matrix elements of the states with one proton and one neutron. The expectation values of the former is, however, an order of magnitude larger than those of the latter, with respect to the states being discussed, because of stronger coherence effects.   It is of interest and importance that the hexadecupole interaction nevertheless produces crucial effects on the triaxiality in many cases.
}

\textcolor{black}{
The central interaction is expected to be strongly attractive for most of combinations between high-j orbitals, owing to large overlaps of their radial wave functions (see \cite{otsuka_2010}).  
Consequently, the total $pn$ monopole interaction is strongly attractive, if both proton and neutron orbitals are high-j orbitals.  A relevant feature was pointed out by Federman and Pittel \cite{federman_1977}, on the basis of the $^3$S$_1$ channel of the central interaction. }
A related feature was discussed in \cite{heyde_1985}, too.  

%%%%%%%%%%%%%%%%%%%%%%%%%%%%%%%%%%%%%%%%%%%%%%%
\subsection{Tensor-force contribution}

\textcolor{black}{
We now move on to the tensor-force contributions.
After the prediction of meson-mediated nuclear forces by Yukawa \cite{yukawa_1935}, the tensor force was formulated to the present form by Bethe \cite{bethe_1940}.    
We use the tensor force due to one $\pi$-meson exchange and one $\rho$-meson exchange processes.
The $pn$ interaction is presently given by the $V_{\rm MU}$ interaction as mentioned earlier, and the $V_{\rm MU}$ interaction includes this tensor force (see \cite{osterfeld_1992}) in addition to the central forces.  
The $NN$ interaction between nucleons in nuclei undergoes substantial renormalizations in general and is changed.   Although this is the case for central-force components, the tensor force, particularly its monopole interaction, remains rather unchanged by such in-medium renormalizations.  This unique property is referred to as the renormalization persistency \cite{ntsunoda_2011}.    Thus, the meson exchange processes remain the major origin of the tensor force in nuclei. It is amazing if such tensor force is directly connected to the formation of triaxial nuclear shape.}

Figure~\ref{fig:hfb_cto_166Er}{\bf b} includes the monopole and multipole contributions of the $pn$  tensor forces.   
The multipole contribution of the tensor force appears to be very minor.  Although its individual terms are not negligible in general, this minor contribution can be expected from the viewpoint of the high complexity of the tensor force resulting in cancellations among various contributions.  The present study provides actual cases of minor contributions of tensor multipole interaction.  

The monopole tensor contribution yields a steady substantial gain of the binding energy.  This feature is one of the crucial factors for the triaxial minimum, as discussed now.  
The monopole interaction of the tensor force generates an attraction between a proton in the $j_>=l+1/2$ orbital and a neutron in the $j'_<=l'-1/2$ orbital, where $j$ and $j'$ ($l$ and $l'$) denote the total (orbital) angular momenta, and $1/2$ implies the intrinsic spin of a nucleon \cite{otsuka_2005}.  Here, the symbols $j_>$ and $j_<$ are general notations to specify the particular couplings between orbital and spin angular momenta.  The attraction is quite strong if angular momenta $j$ and $j'$ are large \cite{otsuka_2005,otsuka_2020,otsuka_2022_emerging}.  This feature holds in exchanged combinations (proton $j_>$, neutron $j'_<$), too.  The tensor monopole interaction is, however, repulsive for ($j_>$,$j'_>$) or ($j_<$,$j'_<$) combination \cite{otsuka_2005}.  

\subsection{Triaxiality and nuclear-force components}
\label{subsec:triaxiality_forces}

\textcolor{black}{
Summarizing the discussions of this section, the triaxiality can be enlarged by the following mechanisms,\\
\indent (i) energy lowering of triaxial states due to high multipole, particularly hexadecupole, central interaction,\\
\indent (ii) for having (i), substantial inclusion of high-$j$ orbitals, such as $h_{11/2,9/2}$ and $i_{13/2,11/2}$, into active orbitals, \\
\indent (iii) energy lowering of these orbitals due to tensor and central interactions, as an example of quantum self-organization, because of too high ``bare'' energies of these orbitals.}

\textcolor{black}{
It is worth mentioning that the involvement of high-$j$ orbitals is essential, but low-$j$ orbitals also produce $Q_2$ and $Q_{-2}$ in response to the triaxiality.  In $^{166}$Er, these three are present, as an ideal example, while certain variations of the interplay, including the absence of some, are explored in the next subsection. } 

We just comment that triaxial states may show up without any of these features, although no such example has been found in heavy deformed nuclei being investigated.

It is also commented that the tensor force is known to be one of the major driving forces for various shape coexistence phenomena by lowering deformed intruder states , for instance, in $^{64,66,68}$Ni ($Z$=28, $N$=36,38,40) \cite{tsunoda_2014,otsuka_2016,leoni_2017,marginean_2020} or in $^{181-185}$Hg ($Z$=80, $N$=101-105) \cite{marsh_2018,sels_2019}, without significant triaxialities.

%%%%%%%%%%%%%%%%%%%%%%%%%%%%%%%%%%%%%%%%%%
\subsection{Explorations of triaxiality in neighboring nuclei}
\label{subsec:triaxial_other}

%%%%%%%%%%   Figure  11 (earlier 13)   %%%%%%%%%%
\begin{figure*}[tb]
  \centering
  \includegraphics[width=13cm]{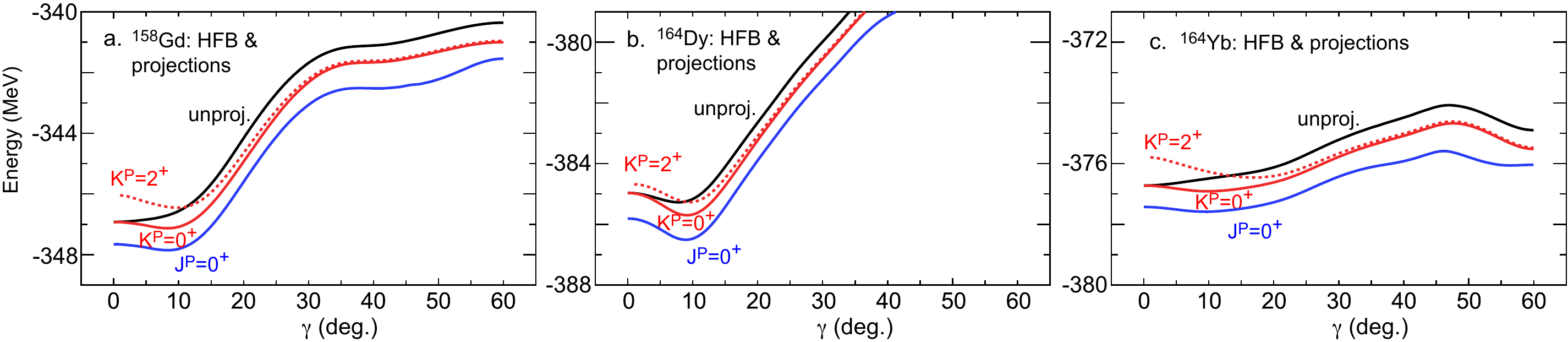}
    \caption{ PES of a.$^{158}$Gd ($Z$=64, $N$=94), b.$^{164}$Dy ($Z$=66, $N$=98) and c.$^{164}$Yb ($Z$=70, $N$=94) obtained by constrained HFB calculation.  See the caption of Fig.~\ref{fig:hfb_proj_Er}.
   }
  \label{fig:GdDyYb_hfb}  
\end{figure*}  
%%%%%%%%%%%%%%%%%%%%%%%%%%%%%%%

It is now of interest to explore other nuclei, \textcolor{black}{located near $^{166}$Er ($Z$=68,$N$=98) in the Segr\`e chart.} We consider three samples, $^{158}$Gd ($Z$=64, $N$=94), $^{164}$Dy ($Z$=66, $N$=98) and $^{164}$Yb ($Z$=70, $N$=94).  
The triaxiality of these nuclei is confirmed the QVSM calculation for the ground and low-lying states, as shown in Sec.~\ref{sec:nuclei_around}.  Figure~\ref{fig:GdDyYb_hfb} shows the PES obtained 
by the constrained HFB calculation in the same manner as in Fig.~\ref{fig:hfb_proj_Er}.  

%%%%%%%%%%   Figure 12  (earlier 14)    %%%%%%%%%%
\begin{figure*}[tb]
  \centering
  \includegraphics[width=13cm]{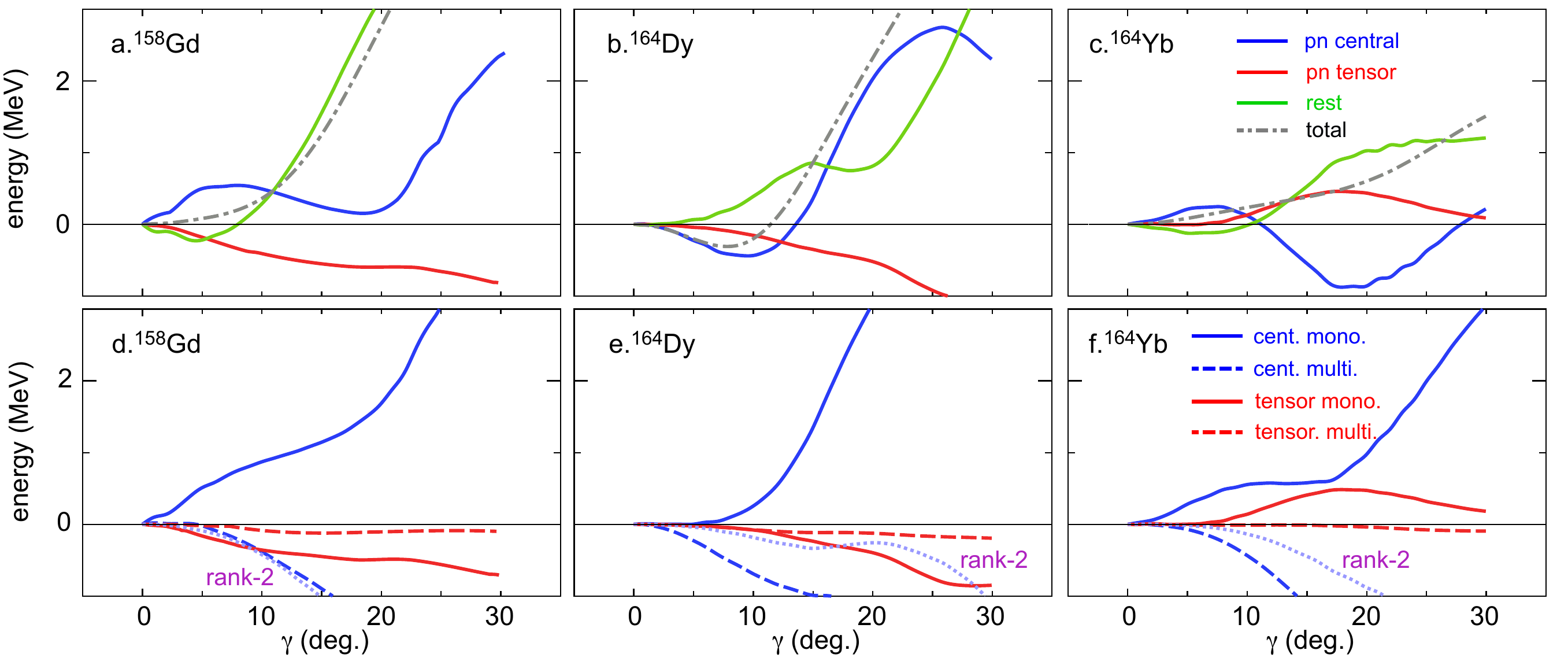}  
    \caption{ Properties of the PES of (a.,d.) $^{158}$Gd, (b.,e.) $^{164}$Dy and (c.,f.) $^{164}$Yb as a function of $\gamma$ as a constraint.  See the caption of Fig.~\ref{fig:hfb_cto_166Er}  {\bf a.} (upper panels) and ({\bf b.} (lower panels).
    } 
  \label{fig:cto_mm_GdDyYb}  
\end{figure*}  
%%%%%%%%%%%%%%%%%%%%%%%%%%%%%%%

In Fig.~\ref{fig:GdDyYb_hfb}, the unprojected PES shows no triaxial minimum for $^{158}$Gd or $^{164}$Yb, but the slope is slow enough so that a triaxial minimum appears in the $K^P$=0$^+$ PES.   The unprojected PES of $^{164}$Dy exhibits a minimum, and it becomes quite deep after $K^P$=0$^+$ projection, being deeper than the one for $^{166}$Er.  The unprojected PES of $^{164}$Yb is not far from being flat for $\gamma<$20$^{\circ}$, which suggests that the structure of this nucleus may be somewhat closer to that of so-called ``gamma-soft nuclei''.  For $^{164}$Yb, the minimum arises after $K^P$=0$^+$ projection, although it is shallower than the other two or $^{166}$Er.  In all the cases, the minimum in the $J^P$=0$^+$ projected PES is obvious, and will be confirmed by the QVSM calculations in Sec.~\ref{sec:nuclei_around}. 
 
%Deeper understanding can be conducted by looking into Fig.~\ref{fig:cto_mm_GdDyYb}.
Figure~\ref{fig:cto_mm_GdDyYb}{\bf a-c} display the same types of plots as the ones in Fig.~\ref{fig:hfb_cto_166Er}{\bf a}.  The contribution of the $pn$ tensor interaction in $^{158}$Gd and $^{164}$Dy is as substantial as in $^{166}$Er, but it is vanished or repulsive for $^{164}$Yb.  It is reminded that the tensor monopole interaction can be repulsive depending on the orbital combination.   Other terms also produce low PES values for $^{164}$Yb up to $\gamma \sim$10$^{\circ}$, giving a  certain gamma-soft feature.  

%The pn central force favors a triaxial minimum in $^{164}$Dy but does not for $^{158}$Gd.
%The triaxiality of  $^{158}$Gd is owing largely to the pn tensor force and somewhat to the rest term.  

Figure~\ref{fig:cto_mm_GdDyYb}{\bf d-f} display the analysis %by decomposing $pn$-force effects into monopole and multipole contributions, 
like Fig.~\ref{fig:hfb_cto_166Er}{\bf b}.  
\textcolor{black}{
Up to $\gamma \sim$10$^{\circ}$, the multipole component of the central interaction is crucial, while its rank-2 part shows a small variation around zero similarly to Fig.~\ref{fig:hfb_cto_166Er}{\bf b}.  The monopole tensor interaction is important for $^{158}$Gd and $^{164}$Dy, while remains neutral for $^{164}$Yb.  As a whole, the $pn$ tensor monopole and $pn$ central hexadecupole interactions are shown to play crucial roles for keeping unprojected PES low enough, up to $\gamma \sim$10$^{\circ}$.} 
%for the emergence of substantial triaxiality in low-lying states.

%%%%%%%%%%   Figure  13  (earlier 15)    %%%%%%%%%%

\begin{figure}[tb]
  \centering
  \includegraphics[width=8.6cm]{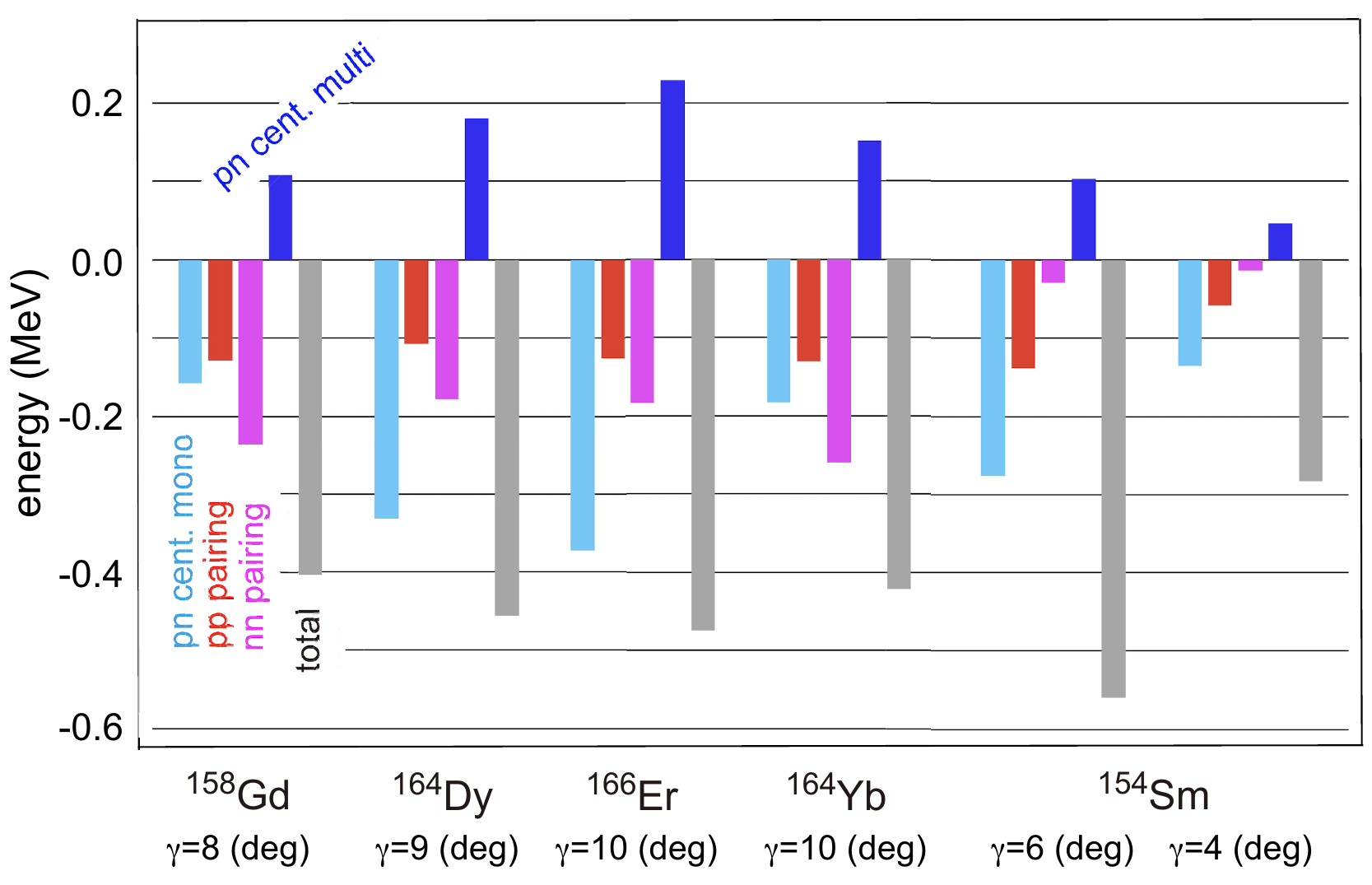}
    \caption{Contributions of selected components of the interaction for the energy change from the unprojected state to the $K$=0 projected state for $^{158}$Er, $^{164}$Dy, $^{166}$Er and $^{164}$Yb, obtained by utilizing the constrained HFB calculation.   The selected components comprise the $pn$ central monopole interaction (light blue), the pp (red) and the nn (pink) pairing interactions, and the $pn$ central multipole interaction (dark blue), as indicated in the figure.  The total change is also displayed (gray).  The $\gamma$ taken for this analysis is shown.  The same analysis is presented for $^{154}$Sm with two $\gamma$ values.} 
\label{fig:K0shift}  
\end{figure}  

%%%%%%%%%%%%%%%%%%%%%%%%%%%%%%%

\subsection{Contributions to K=0 lowering}

\textcolor{black}{
The $K$ projection is shown, in sec.~\ref{sec:K}, to be crucial in producing the triaxiality.  After having discussions on the interaction components, we now discuss what parts of the interaction produce larger contributions.}
Figure~\ref{fig:K0shift} shows major contributions such as (i) $pn$ central monopole contribution, 
 (ii) $pp$ pairing contribution, (iii) $nn$ pairing contribution, (iv) $pn$ central multipole contribution as well as the total contribution.  %, for $^{158}$Gd, $^{164}$Dy, $^{166}$Er and $^{164}$Yb.  
The pairing interaction refers to $pp$ and $nn$ interactions coupled to $J^P$=0$^+$, after the removal of monopole components which is rather small.  It is very natural that the pairing contributions increase the binding energy after the $K$ projection, as they can attractively connect wave functions oriented in different angles.  
We point out that the pairing contributions to the unprojected states become less attractive as
$\gamma$ increases from 0$^{\circ}$, because the $xy$-plane cross section becomes more deformed.  This tendency remains in the $K^P$=0$^+$ projected states as a function of $\gamma$, but the repulsive tendency is weakened. %as $\gamma$ grows, lowering the energies from the unprojected values.  
This is an interesting aspect of the pairing interaction because the pairing interaction contributes to the stabilization of the triaxial minimum of $K^P$=0$^+$ projected states.  It can be taken into account when the pair transfer reaction is studied: pair correlations can be somewhat stronger than in unprojected states.  

The $pn$ multipole central contribution produces less binding energy for $\gamma$ larger after the $K^P$=0$^+$ projection.  This is interpreted as a result of the effective smearing of deformed shape in the $xy$ plane by this projection.  In contrast to this, the $pn$ monopole central contribution becomes more attractive, because the occupation of more favored orbitals can increase.  Although the $K^P$=0$^+$ projected state may give an impression of slightly weaker deformation (or correlation) energy, the basic building block remains to be the unprojected state.

We point out that the pairing effect is considerably constant among $^{158}$Gd, $^{164}$Dy, $^{166}$Er and $^{164}$Yb.  The same quantities for $^{154}$Sm are also shown, and will be discussed in Sec.~\ref{sec:154Sm}.
 
%%%%%%%%%%%%%%%%%%%%%%%%%%%%%%%%%%%%%%%%%%%%%%%%

\section{Interim Summary for the triaxiality and the rotation \label{sec:7}}

\subsection{Small and medium triaxialities}

The discussions presented above indicate that there are two major underlying mechanisms for triaxiality.  
%Both originate in the $NN$ interaction, but the effects emerge in different ways.  
One of them is due to the restoration of broken rotational symmetry caused by $\gamma \ne$ 0. 
%in the body-fixed $xy$ plane.  
The restoration is materialized by the $K$ projection.  This is expected to appear virtually in all cases, as the restoration gives additional binding energies to the projected states with $\gamma \ne$ 0.   For $\gamma$= 0, no additional binding energy of this kind arises by definition \cite{note_circle}.  If there is this mechanism alone, the resulting triaxiality does not seem to be so large, 
as we will see a concrete example in Sec.~\ref{sec:154Sm}.  This triaxiality can then be referred to as  \textcolor{black}{``triaxiality created/enhanced by symmetry restoration''.  Because this mechanism produces small values of $\gamma$, \textcolor{black}{the resulting triaxiality is called as ``\textcolor{black}{small trixiality}" hereafter.}  The small triaxiality} may exist in eigenstates in many CI calculations, but may not show up in unprojected calculations, i.e., what standard DFT approaches normally do.  The $K$ projection unveils it.  
We stress that the present $K$-projection effect occurs for any isolated deformed composite objects including non-nuclear cases, because the symmetry restoration is universal.
It is an intriguing question whether or not this mechanism is related to the bending of triatomic molecules and polymers.

The other origin is specific components of nuclear forces.    
In the example of $^{166}$Er, a pronounced energy minimum arises around $\gamma=  9^\circ$ in the $K^P$ and $J^P$ projected PESs (see Fig.~\ref{fig:hfb_proj_Er}).  
This kind of triaxiality is not only due to the symmetry restoration but  also due to the flat bottom of the unprojected PES in  Fig.~\ref{fig:hfb_proj_Er}.  Thus, there is the second mechanism for the triaxiality, which is a direct consequence of  specific interactions.  This triaxiality can be referred to as ``triaxiality induced by specific-interactions''.  
This mechanism produces, combined with the enhancement by the symmetry-restoration, medium values of $\gamma$, for instance $\gamma \approx$9$^\circ$ for $^{166}$Er.  The resultant (total) triaxiality can then be called as ``medium trixiality".   Such specific interactions include two primary contributors: $pn$ tensor monopole interaction and $pn$ central multipole interaction.  Regarding the latter, the rank-2 (i.e., quadrupole) interaction is crucial for the deformation (i.e., large $\beta_2$), but is now shown not to generate the triaxiality within the same $\beta_2$ value, at least near the actual minimum. Thus, the primary contributor to the latter is ascribed to higher rank components, predominantly the rank-4 (i.e., hexadecupole) component of the $pn$ central multipole interaction.
    
The $pn$ tensor monopole and the $pn$ central multipole (excluding rank 2) interactions can produce substantial effects in many nuclei; in some cases both do explicitly, but one of them may do so in others cases.  Because of this general relevance, the medium triaxiality arises in a large number of deformed nuclei (as further presented in Sec.~\ref{sec:nuclei_around}), and produces additional binding energy of the order of magnitude of 1 MeV.   \textcolor{black}{The medium triaxiality occurs with 5$^{\circ} \lesssim \gamma \lesssim 15^{\circ}$, for examples investigated so far.  This also implies $\gamma \lesssim 5^{\circ}$ for small triaxiality.  Its lower bound will be discussed in Sec.~\ref{sec:154Sm}.}  

There are thus a variety of origins of the \textcolor{black}{medium} triaxiality, and they can act coherently.    This finding is an intriguing outcome of this study, and may have an impact on future studies on heavier nuclei and fission processes \cite{leander,nature_review_SHE,sobiczewski_2007,kowal_SHE_2010,algerian_2017}.   
Other components of the interaction may also contribute depending on cases.   Note that neither of the two primary contributors was included in the traditional Pairing + QQ model.

%It is worth mentioning that $pn$ rank-2 and rank-4 interactions lower excitation energies of $K$=2 and 4 states in Fig.\ref{fig:KJproj_166Er}.   This is remarkable because only pp and nn quadrupole  ($J^P=2^+$) pairing interactions among other components provide substantial lowering of the 2$^+_{\gamma}$ and 4$^+_{\gamma\gamma}$ bands.   This is another feature pertinent to the $pn$ hexadecupole (rank-4) interaction.    
%The QVSM calculation shows consistent trends in the obtained 2$^+_{\gamma}$ eigenstate.
%We will discuss, in Secs.~\ref{sec:nuclei_around} and \ref{sec:154Sm}, how the two origins of the triaxiality work in more nuclei.

We mention that the energy-minimum point of the $K$ projected PES and that of the $J$ projected PES  depict about the same value of $\gamma$, as an indicator of a stable triaxiality within a given rotational band labeled by $K$ values.

\subsection{Self-organization involving monopole interactions}

We now comment on the monopole interaction.  Although it directly favors no specific shape, 
the story does not stop there.  The effects of monopole interaction largely vary as occupation patterns change.  Different shapes are associated with different occupation patterns.  Monopole interactions can thus produce different binding energies for different shapes.  Moreover, the unique linearity of the monopole effect as functions of occupation numbers can make such effects quite substantial (see Subsec.~\ref{subsec:monomulti}).  
Because the monopole interaction effectively changes single-particle energies, the single-particle environment can thus be tailored for certain shapes if circumstances allow, and this can yield significant consequences.  This mechanism is referred to, in general, as (the quantum mechanical version of) the self-organization \cite{otsuka_2019,otsuka_2022_emerging}, and can be considered as a meeting point between single-particle and collective aspects of nuclear structure.   In actual cases, the final shape may emerge from a pool of possible shapes with more or less similar deformation (or correlation) energies.  As the binding energy includes both monopole and multipole contributions, the monopole-interaction effect can be crucial for ``choosing'' the actual shape.   It is also noted that monopole and multipole effects can be cooperative with positive feedback between them in the self-organization.  In the case of $^{166}$Er, certain configurations favoring triaxiality indeed gain more binding energy due to this mechanism. 

\textcolor{black}{The medium triaxiality is often enhanced by this self-organization mechanism.  It, however, occurs with proper numbers of nucleons in the relevant large-$j$ orbitals. 
If they are too few or too many, it produces no or different effects.}  
The near-prolate shapes in Hg isotopes \cite{marsh_2018} turn out to be an example of too many neutrons in the 1$i_{13/2}$ orbit for triaxial shapes. 

%%%%%%%%%%%%%%%%%%%%%%%%%%%%%%%%%%%%%%%%%%%%%%%%%%%
%%%  K  m i x i n g   %%%%%%%%%%

\subsection{Practical conservation of K quantum number in strongly deformed nuclei}
\label{subsec:Kmix}

\textcolor{black}{
The mixing of states of different K quantum numbers, called K mixing, is discussed here.
The eigenstates of angular momentum $J$ are considered, and the parity, being positive, is omitted for brevity.
The MCSM basis vectors, $\phi^{(i)}, i$=1, 2, ..., are assumed to have been obtained by the MCSM calculation.  
The K-projected states with $K$=-$J$, -$J$+1, ..., $J$-1, $J$ are obtained in the form of  eq.~(\ref{eq:K-proj}), and are denoted as $\phi^{(i)}_K, i$ = 1, 2, ...  
Following eq.~(\ref{eq:phk_Kproj}), a laboratory-frame state is generated from  $\phi^{(i)}$, as, 
\begin{eqnarray}
&&| J, M, K, \phi^{(i)} \rangle  \, \nonumber \\
&&=\frac{2J+1}{4\pi} \int_0^{2\pi} d\alpha e^{i\alpha (\hat{J}_z-M)} \int_0^{\pi} d\beta \, {\rm sin}\beta \,d^J_{M,K} (\beta) e^{i\beta \hat{J}_y}  \phi^{(i)}_K. 
\label{eq:phk_Kproj-r}
\end{eqnarray}
The Hamiltonian is diagonalized in a Hilbert space spanned by basis vectors in eq.~(\ref{eq:phk_Kproj-r}) for $i$=1, 2, ... and $K$=-$J$, ..., $J$.  Non-orthogonality among these basis vectors can be properly handled.  
Because the K quantum is defined for the given $z$ axis, 
the basis vector $| J, M, K, \phi^{(i)} \rangle$ changes if the $z$-axis is varied, but the final results of the MCSM calculation does not change, as all $K$ values are taken.
In order to see the physics insight, however, the longest ellipsoidal axis ($z$ axis in the body-fixed frame) is aligned to the same direction for all $\phi^{(i)}_K$'s.  This is closer to the original image of the intrinsic state than arbitrarily oriented states.
The MCSM calculation gives amplitudes for K-projected basis vectors.  If we re-orient basis vectors $\phi^{(i)}, i$=1, 2, ... as stated above, new amplitudes can be obtained from these amplitudes by using the rotation operator.  After the longest axes of all $\phi^{(i)}_K$'s are aligned to the same direction, this direction is defined as the $z$ axis, and various $K$ components are obtained.  The MCSM eigenstates is a superposition of all such components.  Up to this point, just different conventions are used, and calculated quantities in the laboratory frame remain unchanged.  For more concrete mathematical expressions, see Appendix \ref{Ap_MCSM}. 
}

\textcolor{black}{
We here introduce a truncation that $K$ value is restricted to a specific value.   For $J^P$=0$^+$ state, only $K$=0$^+$ contributes.  This truncation is perfect by definition.  For $J^P$=2$^+$ state, however, a similar property does not hold.  We calculate the overlap probability between the eigenstate and the state obtained with a specific $K$ value, where the superposition is made only with the components of this $K$, for instance $K$=2$^+$ for $J^P$=2$^+$.  The result is shown in Table~\ref{table:K_prob}.  The first row represents the $J^P$=0$^+$ state of $^{166}$Er, and the probability is 1 as stated.  The second row is for the $J^P$=2$^+_1$ state with the probability=0.999 for $K$=0$^+$, but the probability becomes less than 0.001 for $K$=2$^+$.  The third row indicate the same quantities with the probabilities 0.000 and 1.000, respectively.  Table~\ref{table:K_prob} clearly indicates that the $J^P$=2$^+_{1,2}$ state is almost perfectly of $K$=0$^+$ and 2$^+$, respectively.
Table~\ref{table:K_prob} shows the probabilities of various $K$ components for some MCSM eigenstates, suggesting that the K quantum number is practically conserved, at least up to $\gamma \sim 15^{\circ}$.
The process for obtaining the results in Table~\ref{table:K_prob} is explained in more detail in Appendix \ref{Ap_MCSM}. 
}
 
\textcolor{black}{
The validity of the alignment of the longest axes taken so far is examined by loosening the alignment 
%by $\sim$1 $^{\circ}$ 
into random directions for various basis vectors.  The overlap probability decreases in all cases linvestigated so far, which is naturally expected.
The energy of the K truncated wave functions show very good agreements with the corresponding original eigenvalues, for naturally major components, being consistent with the pattern of the overlap probabilities.
If an odd integer is taken for $K$, the overlap probability is very small. 
}    
 
%%%%%%%%%%%%%%   T a b l e   2     %%%%%%%%%%%%%%%%%%%
\begin{table}[tb]
 \caption{Examples of overlap probabilities of K=$0^+$, $2^+$ and K=$4^+$ components with selected low-lying eigenstates. }
 \label{table:K_prob}
 \centering
 \begin{tabular}{l c c c c c c r}
 \toprule
 %  \hline
    nucleus  & \,\,eigenstate\,&\,\,$\langle \gamma \rangle ^{\circ}$ \,\, &\,\,K=$0^+$ &  \,\,\,\,\,K=$2^+$  &  \,\,\,\,\,K=$4^+$\\
%%      quantity \textbackslash  nucleus  & \,\,eigenstate\,&\,\,K=$0^+$ &  \,\,\,\,\,K=$2^+$ \\
 %\midrule
   \hline
   \hline
   $^{166}$Er & \, $0^+_1$  \, & \, 8.2 \, & \,\,1.000 &    &    \\
                      & \, $2^+_1$  \, & \, 8.2 \, &  \,\,0.999 &  \,0.000  &    \\
                      & \, $2^+_2$  \, & \, 9.1 \, &  \,\,0.000 &  \,1.000  &    \\
                      & \, $3^+_1$  \, & \, 9.1 \, & \,\,0.000 &  \,0.999  &    \\
                      & \, $4^+_1$  \, & \, 8.2 \, & \,\,0.998 &  \,0.001  &  \,0.000  \\
                      & \, $4^+_3$  \, & \, 9.5 \, & \,\,0.000 &  \,0.000  &  \,1.000  \\
   $^{158}$Gd & \, $0^+_1$ \, & \, 5.9 \, & \,\,1.000 &    &    \\
                      & \, $2^+_1$  \, & \, 6.0 \, & \,\,0.997 &  \,0.000  &    \\
                      & \, $2^+_2$  \, & \, 8.0 \, & \,\,0.000 &  \,1.000  &   \\
                      & \, $3^+_1$  \, & \, 8.0 \, & \,\,0.000 &  \,0.999  &    \\
                      & \, $4^+_1$  \, & \, 6.0 \, & \,\,0.989 &  \,0.000  &  \,0.000  \\
   $^{164}$Dy & \, $0^+_1$  \, & \, 7.3 \, & \,\,1.000 &    &   \\
                      & \, $2^+_1$  \, & \, 7.3 \, & \,\,0.997 &  \,0.000  &    \\
                      & \, $2^+_2$  \, & \, 8.1 \, & \,\,0.000 &  \,1.000  &    \\
                      & \, $3^+_1$  \, & \, 8.1 \, & \,\,0.000 &  \,0.999  &    \\
                      & \, $4^+_1$  \, & \, 7.4 \, & \,\,0.990 &  \,0.001  &  \,0.000  \\
                      & \, $4^+_3$  \, & \, 8.2 \, & \,\,0.000 &  \,0.000  &  \,1.000  \\
   $^{164}$Yb & \, $0^+_1$ \, & \, 8.7 \, & \,\,1.000 &    &   \\
                      & \, $2^+_1$  \, & \, 8.7 \, & \,\,0.999 &  \,0.000  &    \\
                      & \, $2^+_2$  \, & \, 14.2 \, & \,\,0.000 &  \,1.000  &    \\
                      & \, $3^+_1$  \, & \, 14.2 \, & \,\,0.000 &  \,1.000  &    \\
                      & \, $4^+_1$  \, & \, 8.7 \, & \,\,0.995 &  \,0.003  &  \,0.000  \\
                      & \, $4^+_3$  \, & \, 15.1 \, & \,\,0.005 &  \,0.000  &  \,0.994  \\
  $^{154}$Sm & \, $0^+_1$ \, & \, 3.7 \, & \,\,1.000 &    &   \\
                      & \, $2^+_1$  \, & \, 3.7 \, & \,\,0.984 &  \,0.000  &    \\
                      & \, $0^+_2$  \, & \, 12.6 \, & \,\,1.000 &    &   \\
                      & \, $2^+_2$  \, & \, 12.6 \, & \,\,0.991 &  \,0.006  &    \\
                      & \, $2^+_3$  \, & \, 13.8 \, & \,\,0.003 &  \,0.993  &    \\
                      & \, $3^+_1$  \, & \, 13.8 \, & \,\,0.004 &  \,0.995  &    \\
                      & \, $2^+_4$  \, & \, 5.9 \,   & \,\,0.000 &  \,0.999  &    \\
        \bottomrule
   \end{tabular}
 \end{table}

\begin{figure}[tb]
  \centering
  \includegraphics[width=5.5cm]{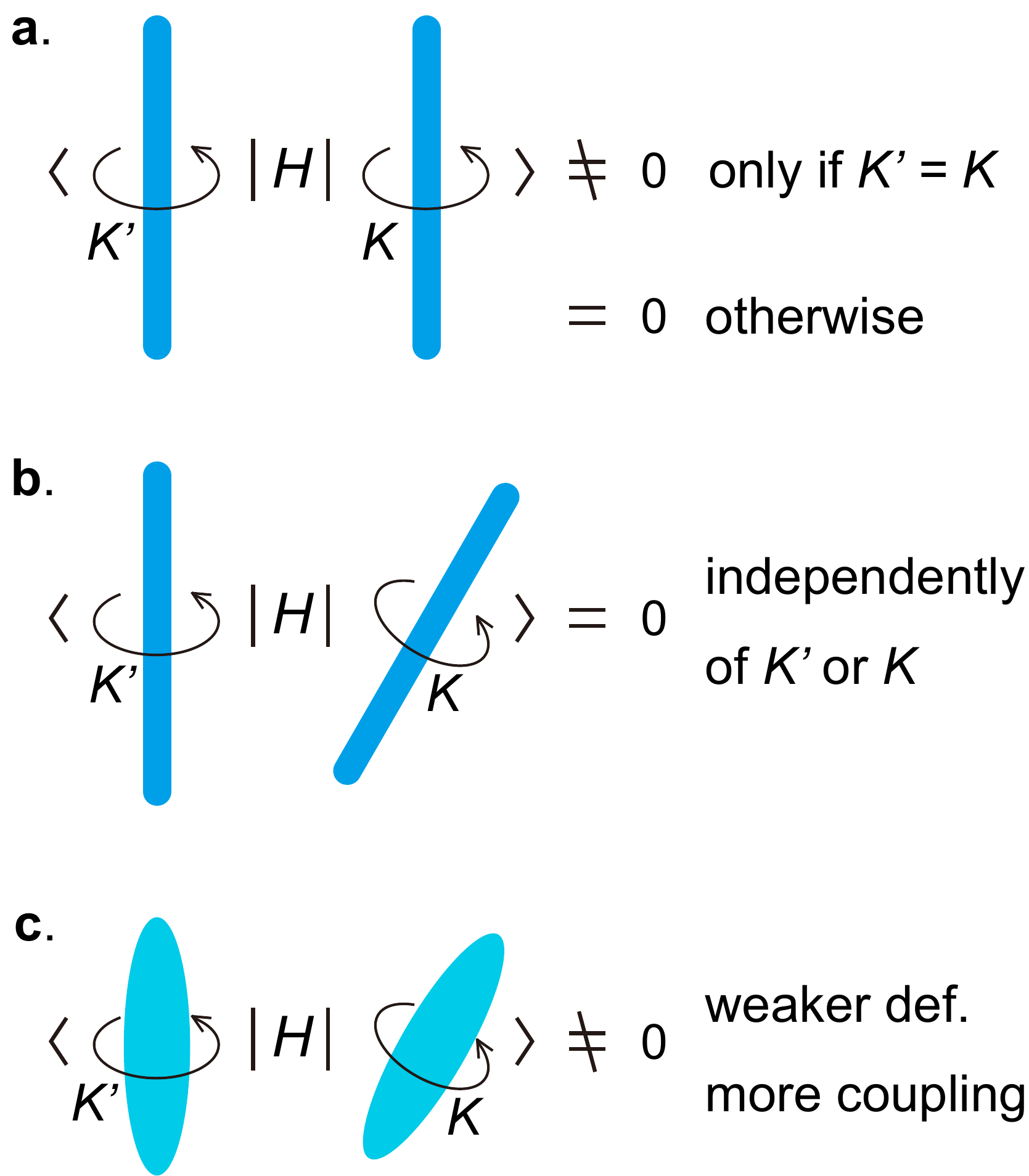}
     \caption{ {\bf a, b} Schematic illustration of the conservation of K quantum number at the limit of strong deformation represented by rods representing triaxial intrinsic states projected onto $K$ or $K'$.  {\bf c} Schematic illustration of the K mixing in weakly deformed nuclei expressed by ellipsoids.  See the caption of {\bf a, b}.
   }
  \label{fig:Kconsv}  
\end{figure}

\textcolor{black}{
Besides numerical investigations, we present intuitive explanations about the suppression of K mixing.   The suppression occurs due to small magnitudes of matrix elements connecting states of different K values as well as large energy differences between such states.  The former is expressed by matrix element,
\begin{equation}
\langle \, J, \, M, \, K', \,\psi \, | \,H \, |\, J, \, M, \, K, \, \psi \,  \rangle, 
\label{eq:Kmix}
\end{equation}
with $\psi$ denoting a triaxial intrinsic state.  
The calculation of this quantity includes the integration of the right-hand side of eq.~(\ref{eq:phk_Kproj-r})  for both bra and ket in eq.~(\ref{eq:Kmix}).      
%The variables (Euler angles) ($\beta$, $\alpha$) and ($\beta'$, $\alpha'$) appear in the integration.  }
Among integrations with Euler angles, we first focus on the matrix element of $H$ between K-projected bra ($\psi_{K'}$) and ket ($\psi_K$) states (similarly to the discussion in the previous paragraph), as visualized in Fig.~\ref{fig:Kconsv}: rods represent $\psi_{K'}$ in bra and $\psi_K$ (or its rotated one) in ket. 
Figures~\ref{fig:Kconsv} {\bf a, b} show the cases with sufficiently strong deformations. 
}  

\textcolor{black}{
As the rod is supposed to be triaxial, it is ``rotating'' about its own longest (or $z$) axis with angular-momentum component $K$ or $K'$, in this simple visualization.  As $K$ and $K'$ are defined in the body-fixed frame, $K'$ in {\bf b} refers to the intrinsic $z$ axis of the leaning rod.   
Only the relative configurations between bra and ket states matter because the Hamiltonian is rotationally invariant.  Thus, the two cases in panels {\bf a, b} basically represent all possible cases, at least qualitatively.  A strong deformation, symbolized by the rod, results in a rapid decrease of relevant overlap and Hamiltonian matrix element between two states, as relative Euler angles between the longest axes of two rods increase from zero.  In the limiting case of sudden decrease, the matrix element of the Hamiltonian vanishes if two rods point to orientations different between bra and ket, as shown in panel {\bf b}.   In contrast, in panel {\bf a},  the rod is in the same orientation for bra and ket.  This is a necessary condition, in order that the Hamiltonian couples bra and ket states.  However, if $K \ne K'$, the coupling does not occur because the Hamiltonian needs $K=K'$ for non-vanishing effects.  Thus, the trends in Fig.~\ref{fig:Kconsv} {\bf a, b} suggest that the K quantum number is practically conserved with sufficiently strong deformation.} 

%%%%%%%%%%   Figure  14  (no earlier figure)    %%%%%%%%%%
%%%%%%%%%%%%%%%%%%%%%%%%%%%%%%%

\textcolor{black}{
Actual situations in Table~\ref{table:K_prob} appear to be consistent with this general feature.  
The energy difference between different K values is of the order of magnitude $\sim$1 MeV (see sec.~\ref{sec:K} and Appendix~\ref{Ap_Kproj}).  Table~\ref{table:K_prob} then implies, based on perturbation estimate, that the mixing matrix elements are substantially smaller than 0.1 MeV.
Note that $\beta_2$ is about 0.3 in the cases shown in Table~\ref{table:K_prob}, and  strongly deformed nuclei with $A \sim 160$ are known to be characterized by $\beta_2 \gtrsim 0.25$. 
%We examined the K mixing for $^{166}$Er with the first MCSM basis vector, $\xi_0$, introduced in sec. \ref{sec:gamma band}.  The $\xi_0$ basis vector produces five 2$^+$ states with $K$=-2, .., 2, and the lowest one in energy corresponds to the 2$^+_1$ state.  The probability of the $K=0$ component in this 2$^+_1$ state appears to be 99.992   \%.  The effect of $K$$\ne$0 components on the excitation energy is calculated to be, in the order of magnitude, as small as  the contribution of the NNLO term shown in sec.~\ref{subsec:NNLO}.  
We thus expect that the K mixing is practically negligible for strongly deformed heavy nuclei being discussed. 
This is consistent with empirical assignment of K values for low-lying bands of deformed heavy nuclei, as illustrated, for instance, in \cite{bohr_mottelson_book2,ring_schuck_book}. }

It is further commented that practical conservation of K quantum number occurs with the Bohr Hamiltonian up to $\gamma \sim$10$^{\circ}$ \cite{tsunoda_2021}, 
\textcolor{black}{and that this conservation  
may have certain resemblances with the partial dynamical symmetry \cite{leviatan_2011}.} 

%\textcolor{black}{A brief sketch is presented in the next subsection for the cases of weaker deformations.}

%%%%%%%%%%%%%%%%%%%%%%%%%%%%%%%%%%%%%%%%%%%%%%%%%%%%%%

\subsection{Large triaxialities in weakly deformed nuclei}
\label{subsec:large}

\textcolor{black}{
If the deformation is weak ($\beta_2<$0.25), the situation becomes more like panel {\bf c}: bra- and ket-vector states have sizable overlap, even if one of them is tilted from the other.  %They are then connected by the Hamiltonian.  
The $K'$ defined in the body-fixed frame spread over various values in the laboratory frame, including $K$.  The bra-ket coupling by the Hamiltonian becomes finite.  Thus, the K mixing occurs, more substantially with weaker deformation.     
As already emphasized at the beginning of subsec.~\ref{subsec:Kmix}, the MCSM automatically handles such changes between strong and weak deformations, and includes K mixing if it happens.}
\textcolor{black}{
Other many-body theories and models are also designed so as to pursue this task in their own ways (see Appendix \ref{Ap_large} for more details).}

\textcolor{black}{
As well-known, weaker deformation often results in $E_x(4^+_1)/E_x(2^+_1) < 3.00$, implying that they are outside the present study.  Many of such nuclei are observed to be gamma-soft nuclei with the triaxiality being so strong that $\gamma$ can be near 30$^{\circ}$.   Such situation may be called ``large triaxiality'', where the $2^+_2$ level energy often moves down around or below the $4^+_1$ one, as qualitatively understood also from the argument for Fig.~\ref{fig:KJproj_166Er}.   Large triaxiality in gamma-soft nuclei, e.g., in $^{66}$Zn \cite{66Zn}, $^{74,76}$Zn \cite{7476Zn}, $^{76}$Ge \cite{76Ge,76Geb}, $^{78}$Se \cite{78Se}, $^{72}$Kr \cite{wimmer_2020}, $^{76}$Kr \cite{yao_2014}, $^{70-98}$Kr \cite{rodriguez_2014}, $^{80}$Zr \cite{rodriguez_2011}, is of great interest also as an extension of the present work, including certain K mixing.}

\textcolor{black}{Large triaxiality and $\gamma$-softness have close relations to the Wilet-Jean $\gamma$-unstable model \cite{wilets_1956}, and to the O(6) limit \cite{ibm_O6} of Interacting Boson Model \cite{ibm_book}.  The $\gamma$-instability of the O(6) limit is shown to produce the energy minimum at $\gamma$=30$^{\circ}$ after the $J^P$=$0^+$ projection \cite{otsuka_1987}. %, as we discussed through the fermionic CI calculation.  
Thus, this subject has a wide related area to be further investigated. } 

%%%%%%%%%%%%%%%%%%%%%%%%%%%%%%%%%%%%%%%%%%%%%%%%
%%%%%%    Related properties and models    %%%%%%%%%%%%%%%
 
\section{Related properties and models \label{sec:8}} 

\subsection{$\gamma$ vibration model} 
\label{subsec:gamma_vib}

The picture described above is confronted with the interpretation in terms of the $\gamma$ vibration raised \cite{bohr_1952} and stressed \cite{bohr_mottelson_book2,bohr_nobel} by A. Bohr.  The $\gamma$ vibration is a vibrational distortion of the axially-symmetric circular cross section (view B) of Fig.~\ref{fig:image}{\bf f}, leading to the intuitive image in Fig.~\ref{fig:level}{\bf a}, where the 2$^+_2$ state carries one $\gamma$ phonon.  This interpretation contradicts the present picture, as this cross section is not a circle but an ellipse as displayed in Fig.~\ref{fig:image}{\bf f}.   The excitation from the ground state to the 2$^+_2$ state is considered, presently, to be a rotation of the same entity in the quantum mechanical sense (see Fig.~\ref{fig:level}{\bf b}), with the change of $K$ quantum number from 0 to 2.  

It is reminded that the terminology, $\gamma$ band, is used in the present work also, just for referring to the rotational band built on the lowest 2$^+$ state with $K$=2, which naturally excludes the 2$^+_1$ state.  The $\gamma$ band in the present study has nothing to do with the $\gamma$ vibration.   The band head of the $\gamma$ band is sometimes called the 2$^+_{\gamma}$ state in literature and also in the present work.  
We just report that the present CI calculation shows no evidence or hint of a $\gamma$ vibration thus far. 

Microscopic investigation on the presence of the $\gamma$ vibration, without explicitly assuming an axially-symmetric ground state, has been very difficult.   As a related work, in \cite{delaroche_2010}, this question was indirectly argued, among other issues, from a broader microscopic manner with the Gogny interaction, a well-accepted mean-field model.  In this approach \cite{delaroche_2010}, the Bohr Hamiltonian was derived microscopically and was diagonalized.  It successfully described the excitation energies of the first 2$^+$ states of many nuclei.  The vast majority of second 2$^+$ states were considered as $\gamma$ vibrations, but their excitation energies are systematically overestimated \cite{delaroche_2010}, casting a mystery to date.  

A Relativistic Hartree-Bogoliubov (RHB) calculation by Li {\it et al.} \cite{li_2010} shows the PES for $^{166}$Er with the energy minimum at $\gamma = 6^{\circ}$.  This deviation from the prolate minimum is assessed as ``the calculated PES is soft in the $\gamma$ direction but, as shown below, not soft enough ...'' (excerpt from \cite{li_2010}), and a softer PES works against the $\gamma$-phonon picture.  The calculated value of B(E2; 2$^+_2\rightarrow0^+_1$) is smaller by about a factor of 1/2 than the experimental one, as another indication of the necessity of a softer PES.  The triaxial deformation seems to be suggested also from the description of the so-called double $\gamma$-phonon state (to be discussed in the next subsection), with a statement ``the potential surface might have a minimum for $\gamma \ne0$'' (excerpt from \cite{li_2010}).  Although the RHB calculation does not include the tensor force because of the missing Fock term, the scheme may produce reasonable results because the adopted interaction was fitted to a large number of deformed nuclei.  If the tensor force could have been included somehow, the energy minimum would have moved to a point with stronger triaxiality.  The properties described in \cite{li_2010} and related works, e.g. \cite{yang_2021}, thus appear to be in a basically consistent passage towards the present study, despite some apparent differences, 
\textcolor{black}{{\it e.g.}, no $K$ projection.}

%Further survey for the $\gamma$ vibration will be of high interest.  
From the experimental side, low-energy vibrational modes from the deformed ellipsoid were extensively reviewed 
with existing data by Sharpey-Schafer {\it et al.} \cite{Sharpey-Schafer_2019}, casting a doubt over the $\gamma$ vibration picture for the second 2$^+$ states, as described in the next subsection. 

%%%%%  beta vibration
%%%\subsection{$\beta$ vibration}
It is noted that the so-called $\beta$ vibration was proposed by Aage Bohr together with the $\gamma$ vibration \cite{bohr_mottelson_book2,bohr_1952}.   Although it is one of the traditional pictures, the existence of the $\beta$-vibrational mode has been widely investigated as reviewed, for instance, by Garrett {\it et al.} \cite{garrett_2018} and by Sharpey-Schafer {\it et al.}  \cite{Sharpey-Schafer_2019}, indicating the negative likelihood of this mode.  Although this is an important and somewhat related subject, it lies outside the scope of this study, and will not be discussed in this article.

%%%%   DOUBLE gamma
\subsection{Double $\gamma$ phonon state \label{sec:double_gamma}}

With the $\gamma$ phonon, the double $\gamma$-phonon (i.e., $\gamma\gamma$) state can be considered, and we now discuss such $\gamma\gamma$ $J^P$=4$^+$ state.  
A simple estimation of its level energy is twice that of the $\gamma$-phonon $J^P$=2$^+$ state, resulting in 2$\times$0.79 MeV=1.58 MeV for $^{166}$Er.  Its experimental candidate state has been identified based on $E2$ decay to the 2$^+_2$ state, which should be relatively strong due to one-phonon annihilation.  The observed excitation energy, 2.03 MeV (see \cite{ensdf,fahlander_1996,garrett_1997,tsunoda_2021}), however, appears to be substantially higher than the value estimated above.  
This discrepancy has attracted attention over decades. 

The $\gamma\gamma$-phonon state was a focus of the review paper by Sharpey-Schafer {\it et al.} \cite{Sharpey-Schafer_2019}, pointing out that there are candidates for this state among experimentally observed states, but that certain measured properties do not match characteristics of $\gamma\gamma$-phonon states, such as excitation energies and transfer reaction cross sections.  This argument was first made for the nucleus $^{190}$Os, which was once believed to be the best candidate for the $\gamma$-phonon picture.  A similar argument was presented in the case of $^{166}$Er \cite{Sharpey-Schafer_2019}.  We can quote ``There is now doubt about the identification of the first excited $K^{\pi}=2_{\gamma}^+$ bands in all deformed nuclei as arising from $\gamma$ shape vibrations of the nuclear mean field.''  \cite{Sharpey-Schafer_2019}, and it is clear that the present study endorses this doubt from theoretical side. 

The present theoretical 4$^+_3$ state is primarily the $K$=4 member of almost the same triaxial shape as that of the ground state.  The triaxiality is slightly enlarged for $K$ larger (see Figs.~\ref{fig:tplot}{\bf c} and \ref{fig:KJproj_166Er}).
Its excitation energy is 2.22 MeV by the QVSM calculation, being closer to the observed value, 2.03 MeV \cite{ensdf,fahlander_1996,garrett_1997}.  The calculated value $B(E2; 4^+\rightarrow 2^+_2)$=10.4 W.u. is also in agreement with the experimental value 8 $\pm$ 3 W.u. \cite{ensdf}.  Thus, the calculated 4$^+_3$ state appears to correspond to the 4$^+$ state assigned traditionally as the $\gamma\gamma$-phonon state.  The difficulty of too high excitation energy then disappears, after the $\gamma\gamma$-phonon assignment is replaced by the $K$=4 triaxial rotation.   Note that in the present QVSM calculation, other 4$^+$ states may exist below the present 4$^+_3$ state, with shapes far different, but they are skipped because of limited computational resources.

Figure~\ref{fig:KJproj_166Er} depicts essential features of $K$=4 states from the viewpoint of the projected energies.  
The comparison between the left (columns {\bf a-c}) and right (columns {\bf d-f}) parts of Fig.~\ref{fig:KJproj_166Er} suggests that as $\gamma$ increases from 8.5$^{\circ}$ to 9.8$^{\circ}$, the energy gain by $K$-projection becomes smaller for $K$=0 and 2, whereas larger for $K$=4.  Consequently, the change $\Delta \gamma$=1.3$^{\circ}$ reduces excitation energies of $K$=4.   
These features, visible in Fig.~\ref{fig:KJproj_166Er}, result in a salient agreement between the $J$=$K$=4 level in column {\bf f} and the experimental $J$=$4^+_{\gamma\gamma}$ level in column {\bf g}.    This trend remains in the QVSM calculation with more basis vectors.

%An identical trend is seen in the right part of Fig.~\ref{fig:KJproj_166Er}.

%%%%%  Davydov model
\subsection{Davydov model \label{subsec:davydov}}
The triaxial nuclear shape was intensively discussed by Davydov and his collaborators already around 1958 \cite{davydov1,davydov2}.  They proposed a rigid-triaxial-rotor model (Davydov model) that is a special case of the Bohr Hamiltonian \cite{bohr_mottelson_book2,bohr_1952,bohr_1953,yamazaki_1963,fortunato_2005,rohozinski_2009}.  Their systematic studies were an indispensable achievement in the study of nuclear structure.  For $\gamma=$9$^{\circ}$, the ratio $B(E2;2^+_2\rightarrow 0^+_1)/B(E2;2^+_1\rightarrow 0^+_1)$ is calculated to be 0.0233 in the Davydov model \cite{tsunoda_2021}, reproducing the experimental value (0.0238). %and the presently calculated value (0.0288)}. 
%(see Fig.~\ref{fig:level}{\bf c}).  
As electromagnetic properties directly scan the shapes, the degree of the triaxiality shown by the Davydov model appears to be consistent with the present scheme.   

Excitation energies are another story, however.  
Experimentally $E_x(2^+_2)$/$E_x(2^+_1)$= 9.8 has been measured.  The Davydov model yields this ratio $\sim$20 for $\gamma=$9$^{\circ}$, and requires $\gamma\sim$13$^{\circ}$ to reproduce this experimental value.   Thus, the Davydov model depicts an internal inconsistency within the model.  This is probably because the Davydov model describes the structure of the nucleus in terms of the free rotation of a rigid body with a fixed triaxial shape, and because the excitation energies are due to increased kinetic rotational energies.  This view clearly differs from the present picture of deformed nuclei and their energies, where the excitation energies are due to the total Hamiltonian including $NN$ interactions, and a small change of the triaxiality is a natural consequence.   The present QVSM calculation, which incorporates such features, reproduces the observed $\gamma$-band excitation energy, as presented in Figs.~\ref{fig:level} and \ref{fig:KJproj_166Er}.  The Davydov model is, on the other hand, too rigid and simple to handle such effects, and fails in predicting excitation energies.  
We emphasize, on the other hand, that the concept of triaxial shapes, one of the two main messages by Davydov and his collaborators \cite{davydov1,davydov2}, is definitely appropriate from the present view, and deserves high appreciation.
 
We here depict a related feature about the rigid-triaxial-rotor model based on the present work.  
 
Figure~\ref{fig:KJproj_166Er} shows the excitation energies measured from the $K$=0 state for  $\gamma$=8.5$^{\circ}$ and 9.8$^{\circ}$.  
They are calculated based on the right-hand side of eq.~(\ref{eq:H Kx}).  By utilizing the expansion, ${\rm cos}(K\gamma) \sim 1- (1/2) K^2 \gamma^2 + ...$, the integral can be expanded by a power series, $K^0$, $K^2$ and higher-power terms.  By handling the norm in a similar way, and by expanding the normalized matrix elements in terms of the powers of $K^2$, the energies are expressed by a polynomial comprising $K^0$, $K^2$, and higher-power terms.  The $K^0$ term obviously gives the energy of the $K$=0 state.  The $K^2$ and higher-power terms provide the excitation energies.  For a certain triaxiality, the energy kernel in eq.~(\ref{eq:H Kx}) damps more quickly as $\gamma$ increases, and higher-power terms become negligibly small.  We here restrict ourselves to the limit consisting of the $K^0$ and $K^2$ terms; the former is nothing but the energy of the $K$=0 state, and the latter gives the excitation energies proportional to $K^2$.  

This $K^2$ dependence can be examined by the ratio $\{E(K$=4)-$E(K$=0)\}/\{$E(K$=2)-$E(K$=0)\}, that is four in the limit.  Its actual value by the full $K$ projection calculation is 
3.69 and 3.93 for left ($\gamma$=8.5$^{\circ}$) and right ($\gamma$=9.8$^{\circ}$) parts of Fig.~\ref{fig:KJproj_166Er}, respectively.   A similar trend is found in the other sequence of $J$=$K$ being 0-2-4, with slightly smaller values of the ratio.  The $K^2$ dependence of the excitation energy thus holds to a rather good extent. 
If $\gamma$ is fixed, the present case somewhat corresponds to the rigid triaxial rotor.  

Figure~\ref{fig:KJproj_166Er} indicates that 
the $K^2$ term with a fixed value of $\gamma$ turns out to produce too high $K$=2 and 4 excitation energies.  Figure~\ref{fig:KJproj_166Er} suggests that the change of $\gamma$=8.5$^{\circ}$ for $K$=0 to $\gamma$=9.8$^{\circ}$ for $K$=2 and 4 substantially lower the excitation energies, which become closer to experimental values, as stated in the previous subsection.  Thus, even a small variation of $\gamma$ can be crucial.  
This variation is not an adjustable parameter, however, and it is a consequence of the quantum many-body calculation with a Hamiltonian.    
We comment that in the rigid rotor modeling (see for instance \cite{jehangir_2018}), the $K^2$ term may have been interpreted as rotational kinetic energy in the $xy$ plane, and the $K$=4 state appears at 4 times excitation energy of $K$=2 state \cite{jehangir_2018}.  This 
was argued to be an evidence against the triaxiality, but what to be excluded is the excessively perfect rigidity of the triaxiality.

%%%%%  Staggering %%%%%%%%%%%%%%%%
\subsection{Staggering}
Triaxial shape deformation has been discussed also from the viewpoint of the staggering of level energies, an effect due to odd or even integer of the angular momentum $J$.  
The signature of the staggering is quantified for a given band by \cite{casten_book,zamfir_1991}
\begin{equation}
S(J,J-1,J-2) = \frac{ \{E(J)-E(J-1)\}-\{E(J-1)-E(J-2)\} }{ E_x(2^+_1) } ,
\label{eq:S}
\end{equation}
where $E(J)$ denotes the energy of spin-$J$ member of the band, and $Ex(2^+_1)$ is used for the normalization purpose of the energy scale.  The actual value of $S$ thus obtained from experimental energies of $^{166}$Er turned out to be 0.290 and 0.269 for $J$=4 and 6, respectively \cite{ensdf}.  These values seem to be shown also in Fig. 3 of \cite{zamfir_1991}.  They turned out to be 0.326 and 0.302, after replacing $E_x(2^+_1)$ with the corresponding quantity obtained from the level-energy difference between the 6$^+$ and 2$^+$ members of the $\gamma$ band.  This value is close enough to the ideal value $S$=1/3, suggesting no notable degree of staggering nor $K$ mixing, \textcolor{black}{in a perfect agreement with the consequence of subsec.~\ref{subsec:Kmix}.}
%Similar properties are seen in heavy deformed nuclei to be discussed.  

%%%%%  Deformed Shell Model   %%%%%%%%%%%%%%%
\subsection{Deformed Shell Model and other approaches}
Quite a few calculations were performed by Sun {\it et al.} with the deformed shell model for the description of the structure of many heavy nuclei, including triaxial shapes \cite{sun_2000,sun_2002}.   The value of $\gamma$ was assumed somehow {\it a priori}, and basis vectors are generated for the deformed single-particle field.  They were used for the diagonalization of the Hamiltonian.  The actual values of $\gamma$ appear to be larger than the ones obtained in the present work, and the connection between the two approaches is an open question.

Some other approaches to triaxial shapes were reviewed by Frauendorf in \cite{frauendorf_2015} including the Tidal Wave approach \cite{frauendorf_2011} and Triaxial Projected Shell Model \cite{sheikh_1999}.   The latter seems to have the same root as the deformed shell model.    

%%%%%%%%%%%%  FIGURE 15   (earlier 16)  %%%%%%%%%%%%%
\begin{figure}[bt]
  \centering
  \includegraphics[width=8.7cm]{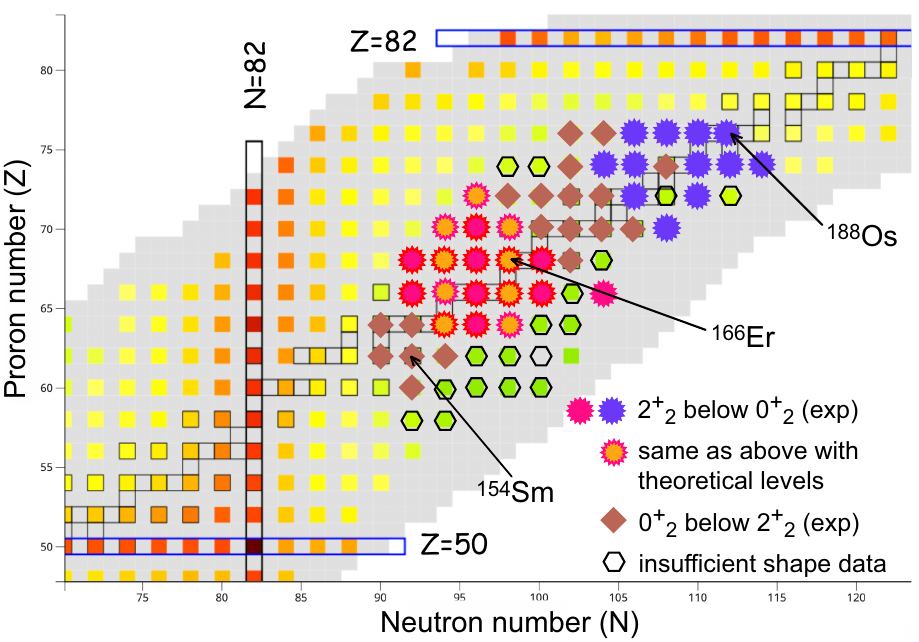}
    \caption{ {Nuclear chart (part) and deformed nuclei with substantial triaxialities (red and purple star symbols).}   Brown diamonds imply the nuclei with the 0$^+_2$ level below the 2$^+_2$ level.  
   If not covered by stars or diamonds, red squares indicate the lowest 2$^+$ states with higher level energies ($>$ 1 MeV), while green squares stand for those with lower 2$^+$ level energies ($<$ 0.1 MeV).  The squares with colors in between indicate intermediate cases.   The background is given by NuDat 3.0 \cite{nudat3}.  The nuclei with theoretical descriptions shown in Fig.~\ref{fig:levels} are red stars with orange inside.
    } 
  \label{fig:chart}  
\end{figure}    
%%%%%%%%%%%%%%%%%%%%%%%%%%%%%%%%%%%%%

%%%%%%%%%%%%%%%%%%%%%%%%%%%%%%%%%%%%%%%%%%%%%%%%
\section{Nuclei around $^{166}$Er in the Segr\`e chart}
\label{sec:nuclei_around}

\subsection{Trends}

We next apply the present study to some nuclei around $^{166}$Er in the Segr\`e (nuclear) chart.

One of the experimentally measurable signatures for a substantial triaxiality can be the observed 2$^+_2$ level below the observed 0$^+_2$ level, and we adopt it.  Figure~\ref{fig:chart} displays a part of the Segr\`e (nuclear) chart (even-even nuclei with $Z$=50-82 and $N$=82-122), where red star symbols (including those with orange inner part) indicate 18 deformed nuclei (3.00 $\le E_x(4^+_1)/E_x(2^+_1)\le$ 3.33) with this signature \cite{ensdf,nudat3}.   These 18 nuclei are traditionally supposed to have prolate ground states, as already stated.  The purple star symbols form a separate group, some of which may have been considered, in the past, to be triaxial.  In fact, one of them, $^{188}$Os, was considered, theoretically, to have $\gamma$=30$^{\circ}$ \cite{hayashi_1984}.  
Missing star symbols do not necessarily imply prolate-like shapes but rather mean insufficient experimental data for judgment. 

%%%%%%%%%%%%  FIGURE 16  (earlier 17)  %%%%%%%%%%%%%
%%%  levels of 9 nuclei  including 162,166Er and 164Dy 
\begin{figure*}[tb]
  \centering
  \includegraphics[width=18cm]{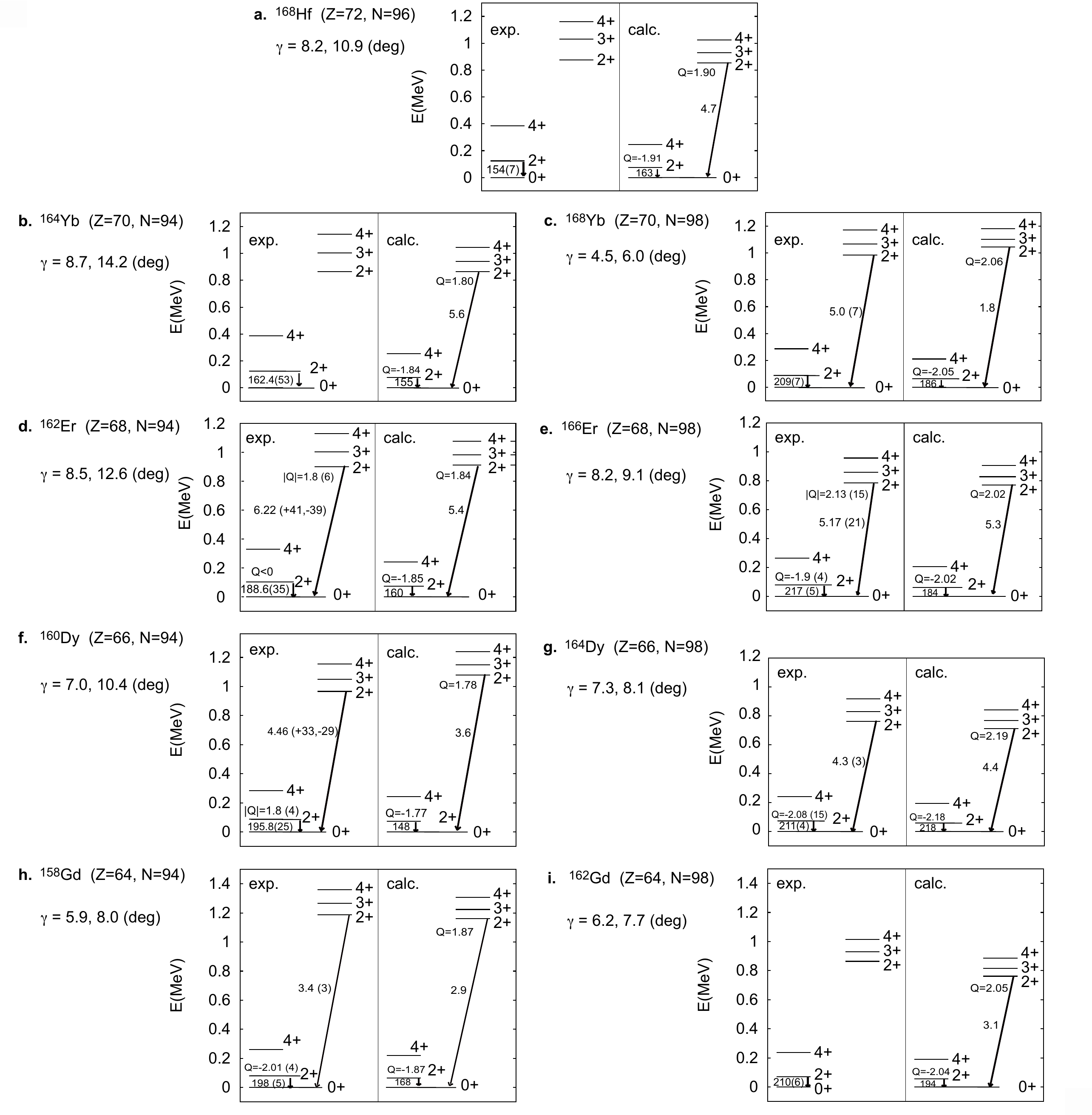}
    \caption{ Lowest level energies of {\bf a} $^{168}$Hf, {\bf b} $^{164}$Yb, {\bf c} $^{168}$Yb, {\bf d} $^{162}$Er,  {\bf e} $^{166}$Er, {\bf f}  $^{160}$Dy, {\bf g}  $^{164}$Dy, {\bf h} $^{158}$Gd, and {\bf i} $^{162}$Gd.
Theoretical and experimental \cite{ensdf} values of $B(E2;2^+_1\rightarrow 0^+_1)$ and $B(E2;2^+_2\rightarrow 0^+_1)$ (see arrows) are shown in W.u., if available.  The spectroscopic electric quadrupole moment is shown (see "Q") in the unit of $e$ barn for the $2^+_1$ and $2^+_2$ states: all theoretical values and available experimental values \cite{ensdf}.  
The mean values of the deformation parameter $\gamma$ of the 0$^+_1$ and 2$^+_2$ states are  shown for each nucleus below the nucleus name.}
  \label{fig:levels}
\end{figure*}  
%%%%%%  162Er   164Dy   158Gd   %%%%%%%%%%%%%%%

The CI calculations discussed in this article require huge computer resources, and thereby we have performed them for some selected ones. % as displayed in Fig.~\ref{fig:chart}.
%$^{162}$Er, $^{164}$Dy, $^{158}$Gd and $^{170}$Er besides $^{166}$Er.  
Figure~\ref{fig:levels} {\bf a-i} show primary results of such calculations in comparison to the corresponding experimental values.  
These panels show the level energies of low-lying states, the values of $B(E2;2^+_1\rightarrow 0^+_1)$, $B(E2;2^+_2\rightarrow 0^+_1)$ (in W.u.), and spectroscopic electric quadrupole moments (in $e$ barn).   Considering also that the same Hamiltonian is used for these nuclei, the agreement to experiment \cite{ensdf} is rather good.  For instance, the 2$^+_2$, 3$^+_1$ and 4$^+_2$ states of $^{162}$Er (panel {\bf d}) are shifted upwards by the almost same amounts between theory and experiment, compared to those of $^{166}$Er (panel {\bf e}).   A further shift is seen in $^{158}$Gd similarly between experiment and theory (panels {\bf h} and {\bf e}).  All shown levels remain quite unchanged between $^{164}$Dy and $^{166}$Er (panels {\bf g} and {\bf e}), but the $B(E2;2^+_2\rightarrow 0^+_1)$ value differs between these two nuclei also similarly in both theory and experiment.  
Thus, a good overall agreement is obtained between theory and experiment, except for the underestimation of $B(E2;2^+_2\rightarrow 0^+_1)$ value of $^{168}$Yb.  
As the changes of the $B(E2;2^+_2\rightarrow 0^+_1)$ value by up to a factor of two from nucleus to nucleus is reproduced rather well in the present calculation, the discrepancy for $^{168}$Yb needs to be remedied by improving the Hamiltonian 
in a future project with a wider coverage of nuclei.   
The $N$=96 isotones, $^{160}$Gd, $^{162}$Dy, $^{164}$Er and $^{166}$Yb, are described similarly well but are not included in Fig.~\ref{fig:levels} for brevity.
The nuclei shown in Fig.~\ref{fig:levels} and these four fulfill the experimental criteria of the substantial triaxiality mentioned above.

The T-plots for the nuclei, $^{158-162}$Gd, $^{160-164}$Dy, $^{162-166}$Er and $^{164-168}$Yb, $^{168}$Hf, exhibit large values of deformation parameter $\gamma$, mostly being above around 6$^{\circ}$.  The values are close to (though somewhat smaller than) those of the PES minima by the HFB calculation displayed in Figs.~\ref{fig:hfb_proj_Er} and \ref{fig:GdDyYb_hfb}.  
These nuclei thus exhibit \textcolor{black}{medium} triaxiality in the present QVSM calculation as well as in the present HFB calculation.  In other words, the \textcolor{black}{medium} triaxiality emerges in a wider region of the Segr\`e (nuclear) chart, and shows a large overlap with the region of substantial triaxiality designated by the experimental criterion stated above.     

%%%%%%%%%%  Figure 17  (earlier 18)   %%%%%%%%%%%%%
\begin{figure}[!tb]
  \centering \hspace{4mm}
  \includegraphics[width=6.8cm]{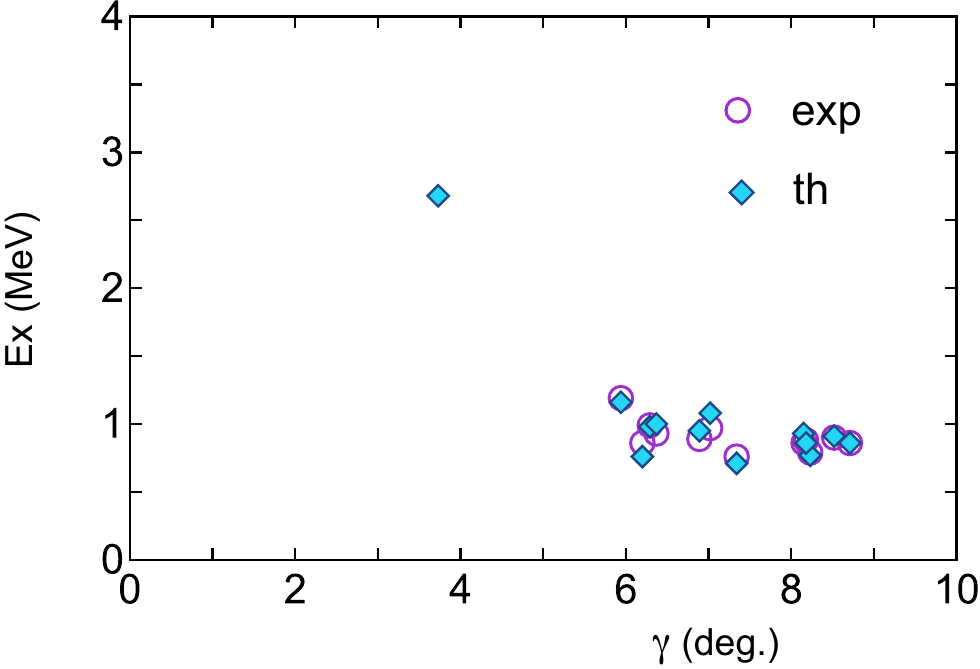}
    \caption{ Systematics of experimental (open circle) and theoretical (blue diamond)  
    2$^+_{\gamma}$ excitation energy as a function of the deformation parameter $\gamma$, 
    except for the far left blue diamond, which represents the $2^+_{g\gamma}$ state of 
    $^{154}$Sm to be discussed in Subsec.~\ref{subsec:154Sm_Tplots}.   
    The symbols represent, from right to left, $^{164}$Yb ($\gamma$=8.7$^{\circ}$), $^{162}$Er 
    (8.5$^{\circ}$), $^{166}$Er (8.2$^{\circ}$), $^{168}$Hf (8.2$^{\circ}$), $^{164}$Er 
    (8.2$^{\circ}$), $^{164}$Dy (7.3$^{\circ}$), $^{160}$Dy (7.0$^{\circ}$), $^{162}$Dy 
    (6.9$^{\circ}$), $^{166}$Yb (6.4$^{\circ}$), $^{160}$Gd (6.3$^{\circ}$), $^{162}$Gd 
    (6.2$^{\circ}$), $^{158}$Gd (5.9$^{\circ}$), $^{154}$Sm (3.7$^{\circ}$), respectively.  }
  \label{fig:2gam_ex}  
\end{figure}  
%%%%%%%%%%%%%%%%%%%%%%%%%%%%%%%

%%%%%%%%%%  Figure 18  (earlier 19)   %%%%%%%%%%%%%
\begin{figure}[!tb]
  \centering
  \includegraphics[width=7cm]{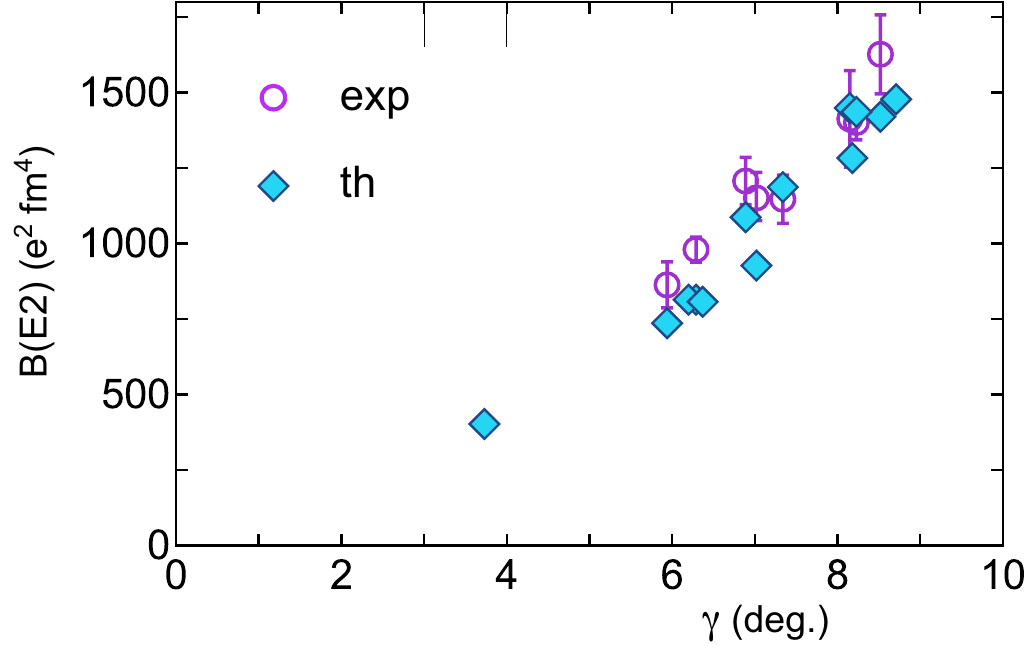}
    \caption{ Systematics of experimental (open circle) and theoretical (blue diamond) 
    B(E2; 0$^+_1 \rightarrow$ 2$^+_{\gamma}$) values as a function of the deformation 
    parameter $\gamma$.  
   See the caption of Fig.~\ref{fig:2gam_ex}. 
   The far left blue diamond represents B(E2; 0$^+_1 \rightarrow$ 2$^+_{g\gamma}$) value of 
   $^{154}$Sm (see Subsec.~\ref{subsec:154Sm_Tplots}).}   
  \label{fig:2gam_E2}  
\end{figure}  
%%%%%%%%%%%%%%%%%%%%%%%%%%%%%%% 
 
As these nuclei are located in the middle of the deformed rare-earth region in the Segr\`e chart and are traditionally supposed, \textcolor{black}{{\it e.g.} \cite{bohr_1952,bohr_1953,bohr_mottelson_book2,rowe_book,deshalit_book,ring_schuck_book,bohr_nobel,kumar_baranger1968,bes_sorensen1969,moller_1995,tajima_1996,moller_2006,delaroche_2010}} to have prolate ground states and $\gamma$ vibrational 2$^+$ states \textcolor{black}{(despite some claims for dominant triaxilaities \cite{Sharpey-Schafer_2019,sun_2002} or for modest triaxialities in scattered nuclei as stated already, {\it e.g.} \cite{li_2010,yang_2021,scamps_2021}),} the present picture gives a strong impact to the overall understanding of nuclear shape, with a suggestion of the \textcolor{black}{medium} triaxiality prevailing in a number of heavy deformed nuclei, combined with their robust mechanisms.  
%The prevailing triaxiality includes both the small triaxiality and the medium triaxiality.

It is of interest how the excitation energies and B(E2) values can be plotted as functions of the $\gamma$ value of the 0$^+_1$ state, where the $\gamma$ value is extracted from QVSM calculations filtered by good agreement to experiment.
Figure~\ref{fig:2gam_ex} displays such a plot of $2^+_{\gamma}$ level energies for $^{158,160,162}$Gd, $^{160,162,164}$Dy, $^{162,164,166}$Er, $^{164,166}$Yb and $^{168}$Hf.  
Likewise, Fig.~\ref{fig:2gam_E2} displays the $B(E2; 0^+_1 \rightarrow 2^+_{\gamma})$ values for the same nuclei.  The systematic trends emerge surprisingly well \textcolor{black}{in comparison to experimental data, and suggest that the degree of triaxiality exhibited by the present $\gamma$ value is related directly to the structure evolution of these nuclei.}  These two figures show the values for $^{154}$Sm, and we will discuss them in Sec.~\ref{sec:154Sm}.

If one moves away from these nuclei in the Segr\`e (nuclear) chart, the triaxiality generally changes from the ranges shown in Figs.~\ref{fig:2gam_ex} and \ref{fig:2gam_E2}.  This is an interesting project to be pursued in the future, and we will touch upon the structure of $^{154}$Sm in Sec.~\ref{sec:154Sm}, as an entry point to it.  
%%%%%%%%%%%%%%%%%%%%%%%%%%%%%%%

%%%   Re-visit to early experiments 
\subsection{Re-visit to early experiments} 
\label{subsec:early_exp}

Quite a few experiments were conducted decades ago for some nuclei of current interest, yielding extensive data by various probes. Examples are found in \cite{cline_1986} for $^{168}$Er, in 
\cite{fahlander_1992} for $^{166}$Er, in \cite{werner_2005} for $^{158}$Gd and 
$^{164}$Dy, where the value of the deformation parameter $\gamma$ was deduced from the data of Multiple Coulomb Excitation experiments {\it \`a la} Cline \cite{cline_1986,kotlinski_1990}, with some variations, based on the Kumar invariant approach \cite{kumar_1972}.  
The typical indication is represented by excerpts from \cite{cline_1986}, ``The asymmetric rigid rotor using $\gamma$=9 $^\circ$ reproduces the data well ...'' and ``The individual E2 matrix elements and the rotational invariants for the ground and $\gamma$ bands in $^{168}$Er all are consistent with rotation of a quadrupole deformed rotor with asymmetry centroid of $\gamma$ $\approx$ 9$^\circ$, ...''.   Despite such clear experimental message, there was no statement to suggest a triaxial ground state instead of prolate one, which might be due to the paradigm of the ``preponderance of axially symmetric shapes'' \cite{kumar_baranger1968}.  
The observed features of $^{168}$Er resemble those of $^{166}$Er, where the QVSM calculation points to $\gamma$ $\sim$ 9$^\circ$ as discussed so far.   
The value of $\gamma$ is reported as $\gamma$$\approx$10$^\circ$ for $^{166}$Er \cite{fahlander_1992}, $\gamma$=6 (2)$^\circ$ for $^{158}$Gd and $\gamma$=7 (4)$^\circ$ for $^{164}$Dy \cite{werner_2005}.  The present QVSM calculation shows $\gamma$=5.9$^\circ$ 
for $^{158}$Gd and $\gamma$=7.3$^\circ$ for $^{164}$Dy, in a salient agreement with the values deduced from the experiment.      
Thus, with unbiased eyes, there were experimental evidences supporting the present view 
that these nuclei depict \textcolor{black}{medium} triaxiality.  
\textcolor{black}{A related analysis was reported by the systematic survey of experimental data available at that time  \cite{andrejtscheff_1993}, which depicts features consistent with the present study}.

As other existing experimental data are also consistent with the present calculation, \textcolor{black}{medium} triaxiality appears to occur in a large number of nuclei in the region of the Segr\'e chart being discussed.  Note that this is additional to the basic (modest) triaxiality expected in all deformed nuclei.  The idea of ``preponderance of axially symmetric shapes''  \cite{kumar_baranger1968} was supported by microscopic calculations in the 1950s \cite{kumar_baranger1968,bes_sorensen1969}.   Such calculations did not include either the tensor interaction or the $pn$ central high-multipole interaction, as literally shown by the name of the approach, the Pairing+QQ model \cite{kumar_baranger1968,bes_sorensen1969}.  The tensor force effect had been practically overlooked for decades also in the systematic study of the shell structure \cite{otsuka_2020}.   A similar history may have been repeated for the shapes, 
leading to a possible misevaluation of precious experimental data. 
 
%%%%%%%%%%%%%%%%%%%%%%%%%%%%%%%
%%%%%%%%%%  154 Sm   %%%%%%%%%%%
\section{Structure of $^{154}$Sm}
\label{sec:154Sm}

We now move on to the structure of $^{154}$Sm, which has been considered to have the ground state in a (axially-symmetric) prolate shape and exhibit $\beta$ and $\gamma$ vibrational excitations.  
The structure of this nucleus was discussed by an MCSM approach prior to this work \cite{otsuka_2019}, with the statement ``the T-plot circles for the ground state are concentrated around $\beta_2$=0.28 and $\gamma$=0$^{\circ}$, a prolate shape''.  This was our understanding at that time.  Note that the picture presented in \cite{otsuka_2019} for side bands was already different from the traditional one.  This nucleus has now been studied by an advanced version of the MCSM, i.e., QVSM \cite{shimizu_2021}, and the results are presented in \cite{tsunoda_2023}. 
We expand this calculation with some insights shown below. 
We focus on the triaxiality of $^{154}$Sm in this article, partly because many other important aspects were discussed in \cite{otsuka_2019}.   

%%%%%%%%%%%%%%%  Figure 19    (earlier 20)  %%%%%%%%%%%%%%%%%
%%%  levels of 154Sm 
\begin{figure}[tb]
  \centering
  \includegraphics[width=7cm]{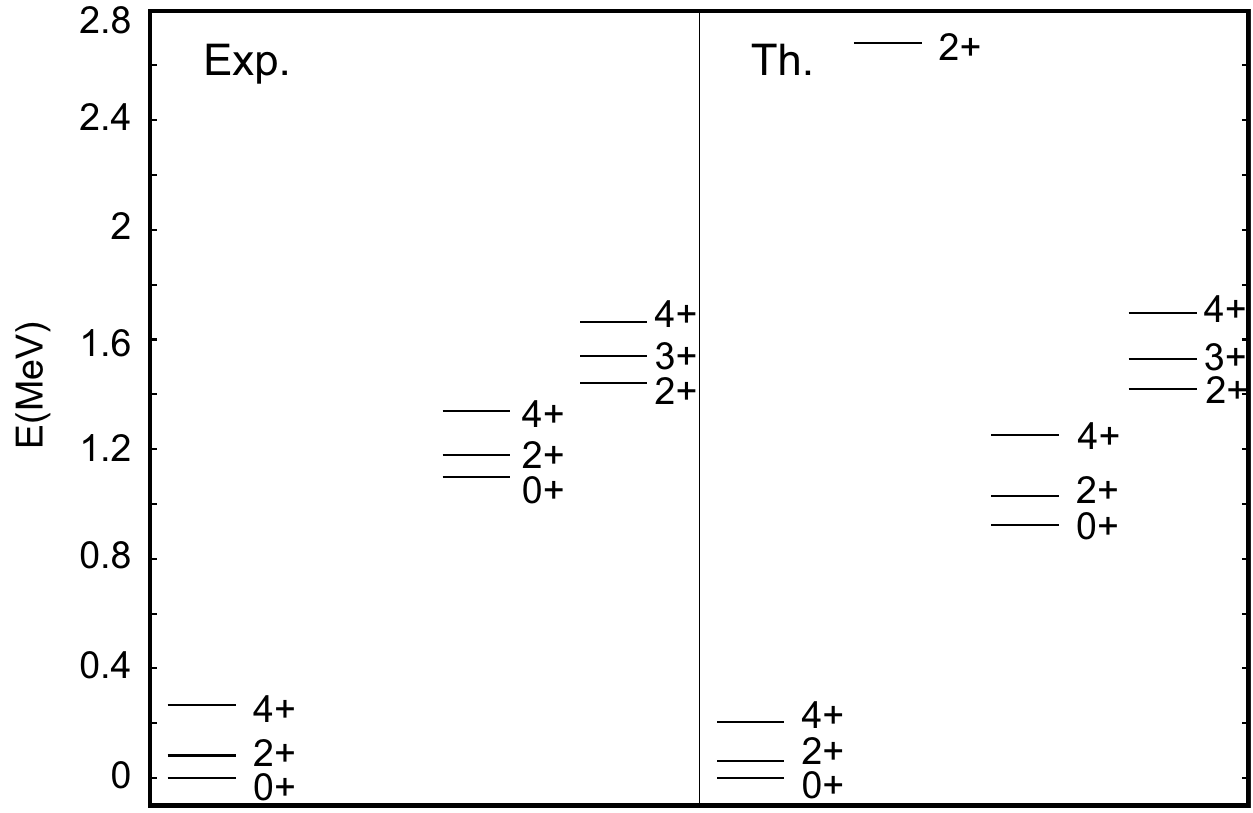}
    \caption{ Lowest level energies of $^{154}$Sm compared to experimental ones \cite{ensdf}.  }
  \label{fig:154Sm_levels}
\end{figure}  

%%%%%%%%%%%%%%%%%%%%%%%%%%%%%%%%%%%%%%%%%%
\subsection{Energies \label{subsec:154Sm_energies}}

Figure~\ref{fig:154Sm_levels} shows level energies of low-lying states of $^{154}$Sm with additional states than in \cite{tsunoda_2023}.  The level energies of the ground band, the so-called $\beta$ band (on top of the $0^+_2$ level) and the $\gamma$ band    
(on top of the $2^+_3$ level) are well described.  
%We note that there is a band built on another $0+$ state in the experiment, but it is not discussed here because of very different character.
 
%%%%%%%%%%%%  FIGURE  20   (earlier 21)  %%%%%%%%%%%%%
% Fig: gamma dependence for 154Sm

\begin{figure}[tb]
  \centering
  \includegraphics[width=6.5cm]{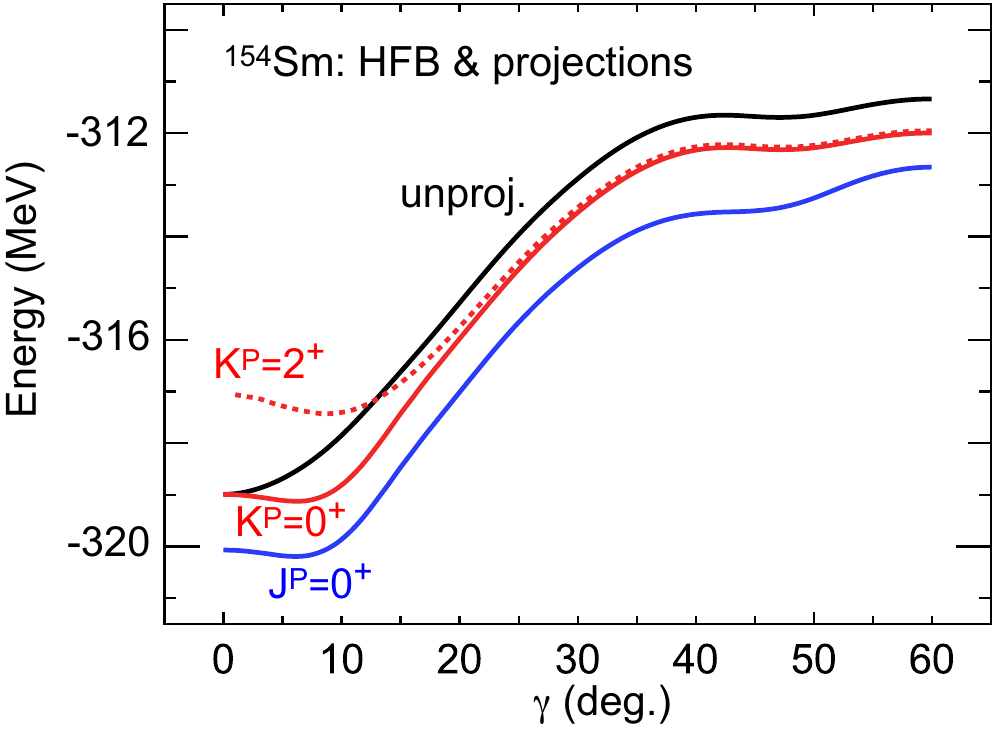}
    \caption{Energies of the ground state of $^{154}$Sm given by HFB calculations.
    See the caption of Fig.~\ref{fig:hfb_proj_Er}. 
    } 
  \label{fig:hfb_proj_Sm}  
\end{figure}  

%%%%%%%%%%%%%%%%%%%%%%%%%%%%%%%%

Such a salient agreement brings us to an analysis using the HFB calculation similar to the one shown in Fig.~\ref{fig:hfb_proj_Er}.   The outcome is displayed in Fig.~\ref{fig:hfb_proj_Sm}. 
The unprojected energy starts to rise at $\gamma$=0$^\circ$, and keeps rising 
with a rate, $\sim$5 MeV from $\gamma$=10$^\circ$ to 30$^\circ$, which is about 
twice faster than the corresponding rate of $^{166}$Er in Fig.~\ref{fig:hfb_proj_Er} ($\sim$3 MeV from $\gamma$=10$^\circ$ to 30$^\circ$).  Nevertheless, the $K$ projection locates its minimum around $\gamma$=6$^\circ$, as a consequence of the underlying mechanism described in Sec. \ref{sec:K}.  The $J$ projection does not cause any major change besides an overall shift of the energy downwards.  Note that the correlations incorporated by the QVSM calculation place the T-plot circles with mean value $\gamma \sim$3.7$^\circ$.  It has been seen for $^{166}$Er and other nuclei that additional correlations incorporated by QVSM calculations reduce the HFB $\gamma$ values by 1-2$^{\circ}$, and the same trend appears here.  

The unprojected PES in Fig.~\ref{fig:hfb_proj_Sm} monotonically increases by about 4 MeV as $\gamma$ changes from 0$^{\circ}$ to 20$^{\circ}$.  This is comparable to the corresponding trends of other approaches such as Gogny Hartree-Fock Bogoliubov approximation \cite{Robledo_2008}, Relativistic Energy-Density Functional method \cite{Li_2009} and relativistic Hartree-Bogoliubov model \cite{Niksic_2010}.  In this sense, the present work appears to exhibit rather consistent features up to the unprojected PES.  However, the $K$ projection was not activated in those approaches, leading to no triaxial ground states.

\begin{figure}[tb]
  \centering
  \includegraphics[width=6.5cm]{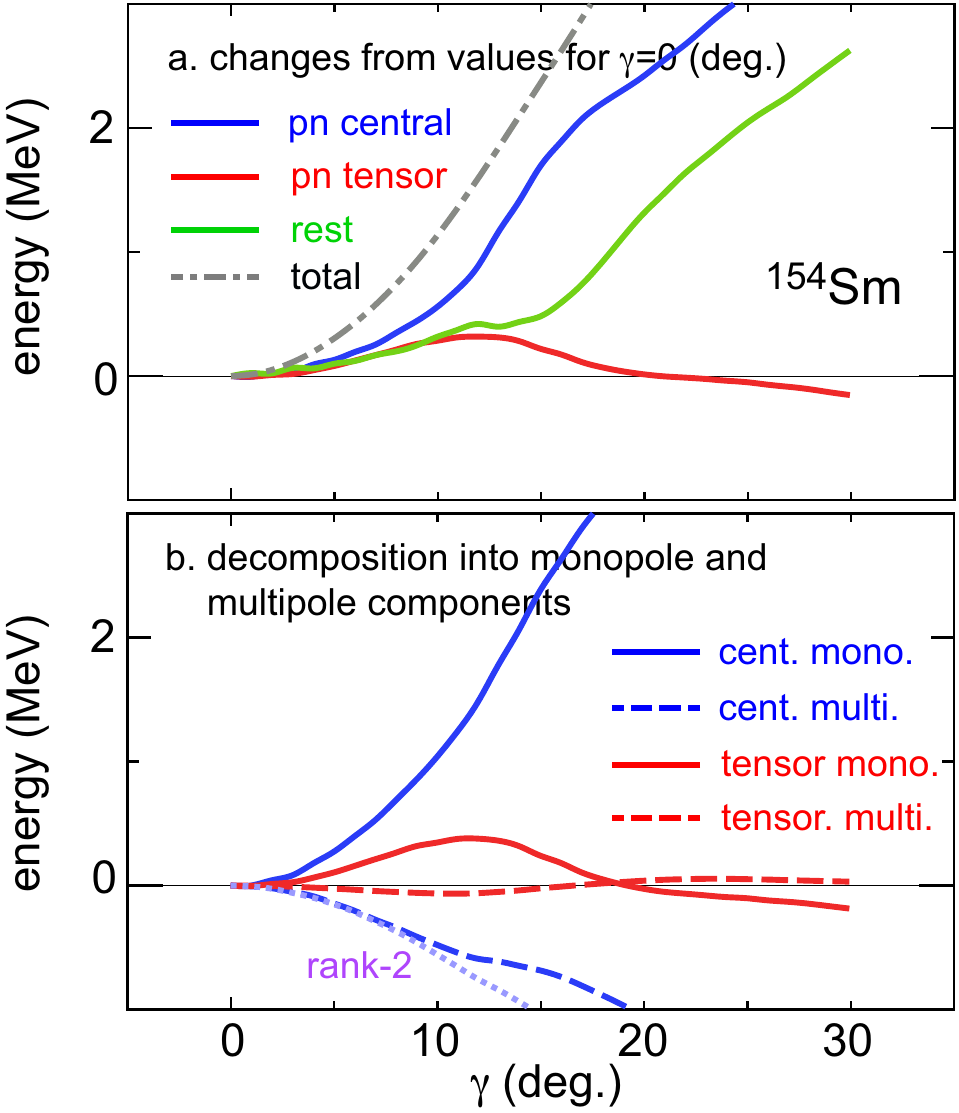}
    \caption{Properties of the PES of $^{154}$Sm as a function of $\gamma$ as a constraint.  See the caption of Fig.~\ref{fig:hfb_cto_166Er}  {\bf a.} (upper panel) and ({\bf b.} (lower panel).
    } 
  \label{fig:hfb_comb_Sm}  
\end{figure}

The analysis similar to the one for $^{166}$Er shown in Fig.~\ref{fig:hfb_cto_166Er} is carried out as shown in Fig.~\ref{fig:hfb_comb_Sm}.   Figure~\ref{fig:hfb_comb_Sm}{\bf a} indicates that all three components of the interaction work against triaxiality until $\gamma$ becomes about 20$^{\circ}$.  This may be a typical phenomenon in nuclei comparatively close to prolate shapes.  
Figure~\ref{fig:hfb_comb_Sm}{\bf b} presents more anatomy: the central-force multipole interaction favors triaxiality, but its effect is canceled out by strong opposition from the central monopole interaction.  The tensor monopole interaction does not help either.  
The rank-2 central-multipole contribution does not stay near 0, most likely due to a rapid structure change by imposed $\gamma$ values, as consistently shown by the fast rise of the unprojected PES.
%The rank-2 central-multipole contribution remains close to 0 up to $\gamma \sim$7$^{\circ}$.

The unprojected PES thus looks against the emergence of the triaxiality, and the triaxiality seen in Fig.~\ref{fig:hfb_proj_Sm} is perfectly due to the projection onto $K$=0, i.e., the symmetry restoration.

Figure~\ref{fig:K0shift} includes two cases of $^{154}$Sm with $\gamma$=6$^{\circ}$ and 4$^{\circ}$ with binding-energy gain of 0.56 and 0.28 MeV, respectively, supporting the occurrence of the \textcolor{black}{small trixiality.}  The gain is smaller for $\gamma$=4$^{\circ}$, as expected,  but remains finite.
Thus, $^{154}$Sm exhibits how the \textcolor{black}{small trixiality} arises, which is different from the \textcolor{black}{medium} triaxiality discussed in previous sections.   There must be many nuclei showing the \textcolor{black}{small trixiality} due to the same underlying mechanism.

%%%%%%%%%%%%  FIGURE  21  (earlier 22)   %%%%%%%%%%%%%
% Fig: gamma dependence for 154Sm

%%%%%%%%%%%%%%%%%%%%%%%%%%%%%%%%

%%%%%%%%%%%%  FIGURE  22   (earlier 23)   %%%%%%%%%%%%%
%%%%%%%%%%  Tplots %%%%%%%%%%%%%
\begin{figure*}[tb]
  \centering
  \includegraphics[width=17.5cm]{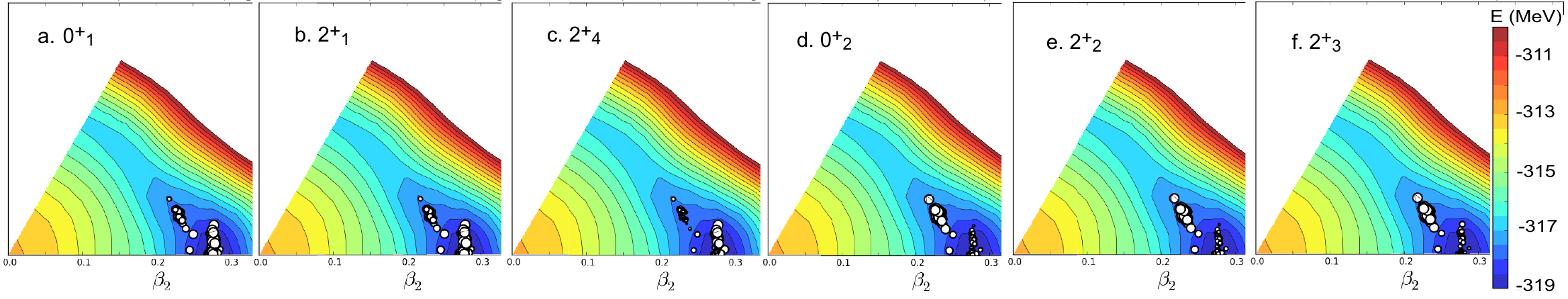}
    \caption{ T-plots for the $0^+_{1,2}$ and $2^+_{1,2,3,4}$ states of $^{154}$Sm.    } 
  \label{fig:154Sm_Tplot}  
\end{figure*}  
%%%%%%%%%%%%%%%%%%%%%%%%%%%%%%%

\subsection{Tplots \label{subsec:154Sm_Tplots}}

Figure~\ref{fig:154Sm_Tplot} shows Tplots for the $0^+_{1,2}$ and $2^+_{1,2,3,4}$ states.
One finds two major patterns: one is shared by the $0^+_{1}$, $2^+_{1}$ and $2^+_{4}$ states, and the other by the $0^+_{2}$, $2^+_{2}$ and $2^+_{3}$ states.  The latter group shows T-plot circles at rather large values of $\gamma$, indicating that the states of the latter group are of 
\textcolor{black}{medium} triaxiality with an average value of $\gamma \sim$ 13$^{\circ}$.  
So, there is a shape coexistence for $^{154}$Sm with the \textcolor{black}{small and medium} triaxialities.  Although this phenomenon was pointed out in \cite{otsuka_2019}, the in-depth explanation was missing.

The ground band built on the $0^+_{1}$ state was considered in \cite{otsuka_2019}, based on human sight, to be of $\gamma \sim$ 0$^{\circ}$, without high-resolution inspection.  The objective extraction of the mean $\gamma$ value has been implemented in the present study \cite{otsuka_2022}, and the obtained result appears to be $\gamma \sim$ 3.7$^{\circ}$.  Very recently, an experimental value of $\gamma$, 5.0 $\pm$1.4$^{\circ}$, is obtained for $^{154}$Sm by the experiment on its Giant Dipole Resonance \cite{kleemann_2024}.  

The obtained small but finite $\gamma$ value is consistent with the HFB analysis above, and produces a $K^P$=2$^+$ state as well as $K^P$=0$^+$ state.  The $K^P$=2$^+$ $2^+$ level is calculated, in the present QVSM calculation, to be at the excitation energy $\sim$ 2.7 MeV with a similar T-plot pattern to the  $0^+_{1}$ and $2^+_{1}$ states.  We emphasize that this state emerges because of the triaxiality of the ground state, and denote it by $2^+_{g\gamma}$, meaning the ``$2^+_{\gamma}$'' state within the family of the ground state having  similar triaxialities.  This $2^+_{g\gamma}$ state is predicted to be excited from the $0^+_{1}$ state with B(E2) $\sim$ 400 e$^2$fm$^4$.   
  
Figures~\ref{fig:2gam_ex} and \ref{fig:2gam_E2} include the values for the $2^+_{g\gamma}$ state of $^{154}$Sm, which are only predictions.  These values appear to be on smooth extrapolations.   It is of great interest to see them experimentally in the future.

%%%%%%%%%%%%  FIGURE  23  (earlier 24)   %%%%%%%%%%%%%
%%%%%%%%%%%%  FIGURE gamma Nd Sm  %%%%%%%%%%%%%
\begin{figure}[tbh]
  \centering
  \includegraphics[width=8.7cm]{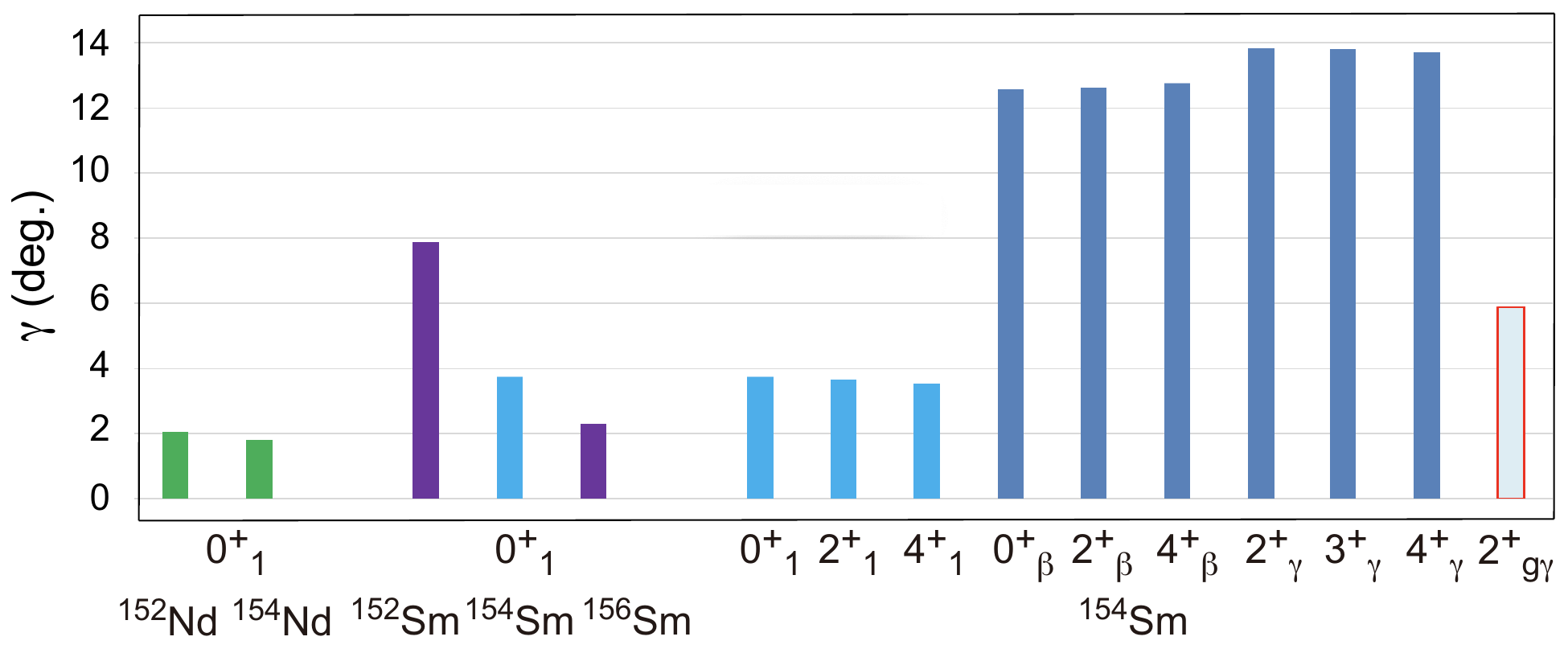}
    \caption{Deformation parameter $\gamma$ (degrees) for the 0$^+_1$ states of
    $^{152,154}$Nd and $^{152,154,156}$Sm nuclei.  For $^{154}$Sm, the value for the members of other bands are shown also.  The $\beta$ and $\gamma$ bands are triaxial states with large values of the deformation parameter $\gamma$.  The far right column is for the $\gamma$-band head built on the ground state.
    } 
  \label{fig:gamma_NdSm}  
\end{figure}  

%%%%%%%%%%%%%%%%%%%%%%%%%%%%%%%%

\subsection{Deformation parameter $\gamma$ \label{subsec:154Sm_gamma}}

We briefly survey the triaxiality of some nuclei surrounding $^{154}$Sm in the Segr\`e chart.  The values of the deformation parameter $\gamma$ is displayed in Fig.~\ref{fig:gamma_NdSm} for certain low-lying states of $^{154}$Sm, as well as for the 0$^+_1$ states of $^{152,154}$Nd and those of $ ^{152,154,156}$Sm.   The $\gamma$ value is almost the same between $^{152}$Nd and $^{154}$Nd, likely suggesting a stable modest triaxiality of $\gamma\sim$2$^{\circ}$ in this region of the Segr\`e chart.  A consistent tendency is found in the chain of the $^{152,154,156}$Sm isotopes: $\gamma$ is as large as about 8$^{\circ}$ for comparatively less deformed nucleus $^{152}$Sm, and becomes closer to 2$^{\circ}$ in strongly deformed nucleus $^{156}$Sm.  Further studies on heavier isotopes are of interest.

The $\gamma$ value of various states of $^{154}$Sm is also shown in Fig.~\ref{fig:gamma_NdSm}.  The states are labeled by $\beta$ for the $0^+_2$ band, while the 
$\gamma$ band means the lowest band built on a $2^+$ state.  The $\gamma$ value is almost constant within each band.  The $\gamma$ value is larger by $\sim$1$^{\circ}$ for the $\gamma$ band than for the $0^+_2$ band, similarly to the $\gamma$-value difference between the $0^+_1$ and $2^+_2$ bands of $^{166}$Er, 
\textcolor{black}{a stretching effect discussed in Subsec.~\ref{sec:sketch}.}  The last column corresponds to the $2^+_{g\gamma}$ state discussed just above (Subsec. \ref{subsec:154Sm_Tplots}).  It belongs to the family of the 0$^+_1$ state and depicts a similar T-plot to the one for the 0$^+_1$ state (see Fig.~\ref{fig:154Sm_Tplot}).  The $\gamma$ value of the $2^+_{g\gamma}$ state is larger by $\sim$2$^{\circ}$ than that for the 0$^+_1$ state.  This shift can be understood 
\textcolor{black}{in terms of the same kind of the stretching as the one mentioned above with a larger change.}   It does not seem to change the basic picture of the $2^+_{g\gamma}$ state as the $K$=2 member of the ground-state structure. 

%%%%%%%%%%%%%%%%%%%%%%%%%%%%%%%%%%%%%%%%%%%%%%%%
%%%%%%     S u m m a r y        %%%%%%  
\section{Summary and prospects \label{sec:summary}}  

\textcolor{black}{This article shows the wide appearance of triaxial shapes in heavy deformed nuclei with examples of  $^{166}$Er, $^{154}$Sm and some other heavy nuclei.   
The rotational bands in these nuclei are shown, in a fully quantum many-body picture (i.e., without 
\textcolor{black}{the quantization of the free rotation} of rigid rotor), to arise from such triaxial ellipsoidal shapes, with the J(J+1) - K$^2$ rule of the level energies with K quantum numbers practically conserved.  This rule is re-derived by using polynomial properties of the Wigner d-functions, with results extended in a line from earlier works. }
\textcolor{black}{
Prevailing triaxial shapes are unveiled in deformed heavy nuclei.  The empirically known rotational-band pattern is obtained with the classification by K quantum number, while the internal structure is different from conventional picture {\it \`a la} A. Bohr. 
%The prevailing triaxial shapes thus appear, with empirically known rotational-band pattern classified by K quantum number, while the internal structure is different from conventional picture {\it \`a la} A. Bohr. 
} 
%This seems to be in a contrast to the preponderance picture of prolate shapes, which was stressed by Aage Bohr, and has been a textbook item.  

%, but it will be superseded, in future textbooks, by the present novel concept.    
Two origins of the triaxiality are indicated.  Once a triaxial shape arises in the $xy$ plane (see Fig.~\ref{fig:image}), it breaks the symmetry of the rotation in this plane.  This symmetry 
is restored by projecting onto a $K$ quantum number as the intermediate step before the projection onto a good $J$ value.  The $J$ projection generally causes $K$ mixing, but it is quite weak if the deformation is strong enough, as in heavy deformed nuclei discussed in this paper. Realistic nuclear forces lower the $K$=0 state below $K$$\ne$0 states.  Because of this binding-energy gain, the energy minimum is 
\textcolor{black}{moved to a larger} %located at a finite 
$\gamma$ value.  This mechanism due to the symmetry restoration is clearly robust.  It occurs in virtually all cases of strong deformation, including $^{166}$Er and $^{154}$Sm.  The triaxiality by this mechanism is called \textcolor{black}{``triaxiality created/enhanced by symmetry restoration'' or simply ``small triaxiality''.}

The triaxiality due to this symmetry restoration 
%not only computationally appears in the result of the mathematical procedure, but also 
demonstrates how important is the interplay between intrinsic degrees of freedom (shape deformation parameters) and extrinsic features (e.g. $J$ and $K$ values characterizing bands and levels).  For example, by projecting on  $K^P$=0$^+$, the \textcolor{black}{potential energy surface (PES)} as a function of $\gamma$ is changed from the unprojected one, \textcolor{black}{yielding substantial additional binding energies for finite $\gamma$ values, pushing the PES minimum to larger triaxiality.}  As another example, \textcolor{black}{different bands exhibit different $\gamma$ values, which can be interpreted as a stretching of triaxial deformation as $K$ increases.}
%Although this stretching remains less than 20\% of the ground-state $\gamma$ value, it results in physical quantities in} good agreement with experiment.  
%If the \textcolor{black}{medium} triaxiality occurs, it is generally seen that the $\gamma$ value increases from the ground band to the $\gamma$ band and the $\gamma\gamma$ state. 
%This is a kind of stretching effect with $K$ value due to nuclear forces.  
The difference is as ``small'' as $\Delta \gamma$=1.3$^{\circ}$ between $K^P$=0$^+$ and 2$^+$ (see Fig.~\ref{fig:KJproj_166Er}), but it is related to a significant lowering of energies.  We need to be careful about the gauge of $\gamma$, and $\gamma \sim$9$^{\circ}$ should not be considered as a small number without a particular significance.  
%A notable stretching \textcolor{black}{does not occur} as $J$ increases, probably due to more stable equilibrium. 

Another origin of the triaxiality is the specific components of nuclear forces.  
The monopole component of tensor force and the high multipole component of central forces are crucially important as well as the involvement of  high-$j$ orbitals, such as $h_{11/2,9/2}$ and $i_{13/2,11/2}$.    
\textcolor{black}{These factors combined} are shown to bring about ``\textcolor{black}{medium} triaxiality'' with $\gamma \sim 9^{\circ}$ in $^{166}$Er, and that with $\gamma \sim$ 6$^{\circ}$-13$^{\circ}$ in some other nuclei, as far as investigated so far.  
Due to the renormalization persistency \cite{ntsunoda_2011}, the tensor force in nuclei is closely related to the one-meson exchange processes, leading to a direct connection between such an ``elementary-particle'' process and nuclear shapes. The $\pi$-meson exchange is treated separately also in the modern theory of nuclear forces \cite{weinberg_1990} as materialized as the chiral Effective Field Theory \cite{machleidt_2011} of the Quantum Chromo Dynamics. 

As a separate but related feature, the $\gamma$ value extracted from the CI calculation depicts clean trends, for instance, with the B(E2;2$^+_{\gamma}\rightarrow$0$^+_1$) (see Fig.~\ref{fig:2gam_E2}).  Figure ~\ref{fig:2gam_E2} implies that the parameter $\gamma$ is one of the dominating variables, and endorses the $K$ rotation as the basic mode.  

The constrained Hartree-Fock-Bogoliubov (HFB) analysis was conducted to show the contributions of 
certain components of nuclear forces as functions of $\gamma$: tensor monopole and central multipole interactions are quite important for \textcolor{black}{medium} triaxiality.  Interestingly, the quadrupole interaction does not produce additional binding energy for triaxial states, as long as $\beta_2$ is fixed. 
The effects of higher rank interactions, particularly hexadecupole interaction, can enlarge triaxiality and those effects can be enhanced by triaxiality, yielding more binding energies for larger triaxiality.    This may lead us to a plausible connection between triaxiality and fission, mediated by hexadecupole properties.   It is noted that higher rank interactions are naturally included in the present work, as the $pn$ central interaction is given by a Gaussian finite-range interaction (V$_{\rm MU}$) with the strengths fixed so as to be consistent with microscopic interactions for a number of nuclei, from $sd$-shell nuclei up to Hg isotopes \cite{otsuka_2010,marsh_2018}.  
\textcolor{black}{
The V$_{\rm MU}$ interaction seems to exhibit properties similar to a Chiral EFT interaction, at least for certain level energies \cite{gade_2024}.}

The superheavy nuclei are one of major objectives of nuclear physics \cite{nature_review_SHE,kowal_SHE_2010}.   The triaxiality is certainly more relevant to superheavy nuclei, as the deformation will be more significant source of their binding energy.  

%The concept of the \textcolor{black}{medium} triaxiality is applicable to a number of nuclei.
%, also because there are several origins in nuclear forces.   

%%%%%%%%%%%%%%    F i g   2 4      (earlier 25)  %%%%%%%%%%%%%%%%
%%%%%%%%%%  /Users/takaotsuka/Desktop/スクリーンショット 2024-02-18 22.25.42.pngFigure 10  (old Extended FIGURE 2)  %%%%%%%%%%%%%
\begin{figure*}[!tb]
  \centering
  \includegraphics[width=14cm]{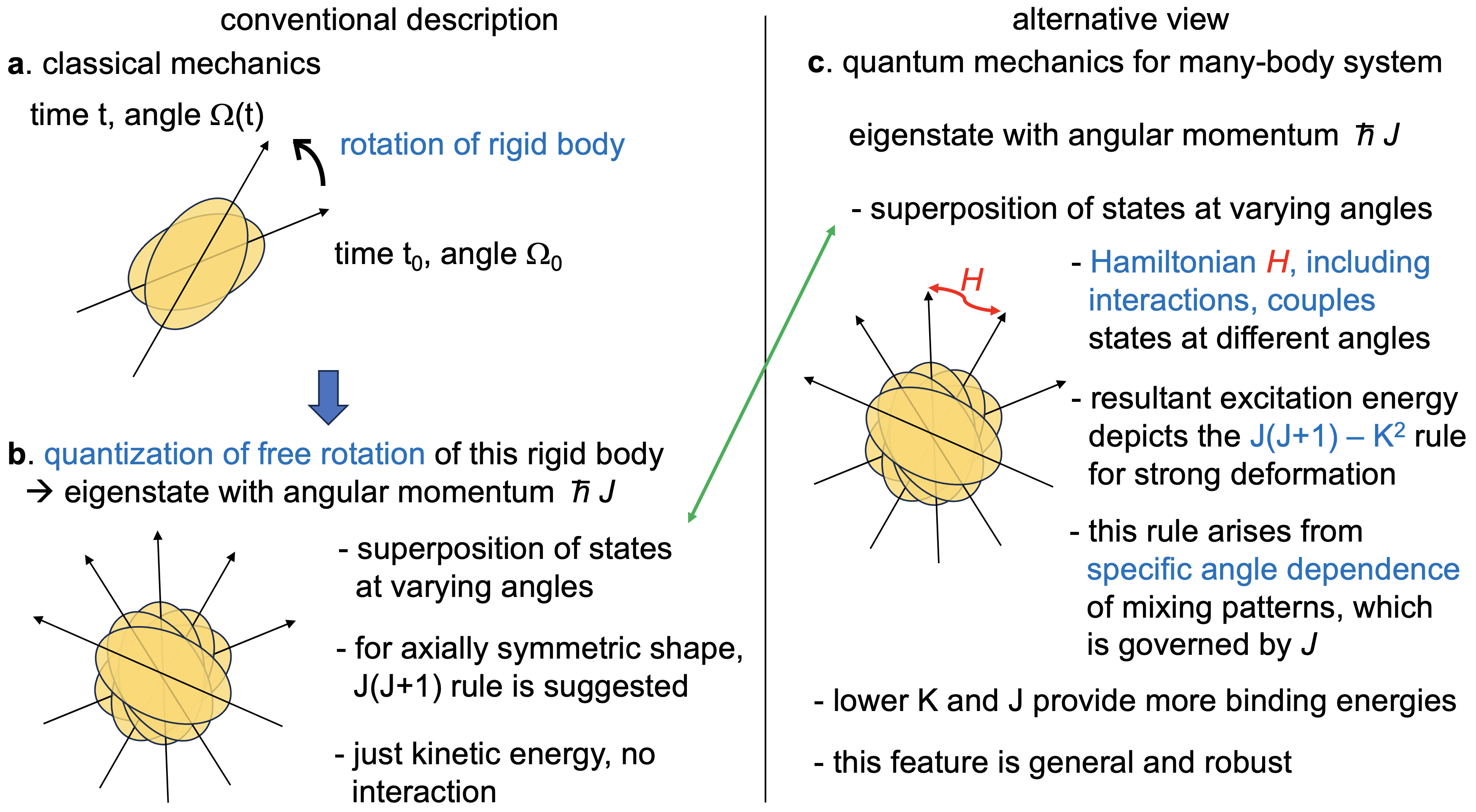}
    \caption{ Pictures of the rotation of objects like atomic nuclei.
    {\bf a.} Classical mechanical view.  {\bf b.} Quantization of freely rotating rigid-body. 
    {\bf c.} View of the quantum mechanical system composed of many ingredients with conserved 
     angular momentum $\hbar J$.  The J(J+1) - K$^2$ rule arises. 
     The green arrow indicates similarity in the wave function, but the energy comes from different 
     origins.  } 
  \label{fig:rotation}  
\end{figure*}  
%%%%%%%%%%%%%%%%%%%%%%%%%%%%%%%

%ZZZZZ

As one of the major points of this paper, we looked into the consequences of the restoration of $J$, total angular momentum.  The excitation energies within rotational bands have been empirically known to follow the $J(J+1)$ rule very well.   However, its explanation in terms of the quantum many-body theory 
\textcolor{black}{
has not been well documented in textbooks and similar literatures:} Traditionally, the argument starts from the classical rotation of a rigid-body object, as shown in panel {\bf a} of Fig.~\ref{fig:rotation} (see, for instance, \cite{rowe_book,deshalit_book,ring_schuck_book}).  The quantization of the free rotation of a rigid-body object with the axial symmetry, like a symmetry top, was adopted as the foundation for the $J(J+1)$ rule (see panel {\bf b} of Fig.~\ref{fig:rotation}), see, for instance, \cite{rowe_book,deshalit_book,ring_schuck_book}.  

\textcolor{black}{The $J(J+1)$ rule has been discussed, to varying extents, from quantum many-body viewpoints in  \cite{peierls_yoccoz_1957,verhaar_1963a,verhaar_1963b,verhaar_1964,kamlah_1968,siemens_book}, as mentioned in subsec.~\ref{subsec:remarks} and Appendix~\ref{Ap_rot_energy}.  
These works, however, have not been widely recognized; For instance, no paper but \cite{peierls_yoccoz_1957} is cited in \cite{bohr_mottelson_book2}.   Panel {\bf c} of Fig.~\ref{fig:rotation} displays the basic underlying idea, which could have been shared with these works at least partly, although it may not have been presented at all.  Panel {\bf c} indicates the following properties.} 
The intrinsic state (represented by ellipsoid) is rotated to various angles shown by radial arrows, and such rotated states are superposed with designated amplitudes so as to be the eigenstate with a given angular momentum $J$.  
The excitation spectrum arises as a consequence of the Hamiltonian acting 
between bra and ket states with different angles.  The integrated effects follow the hierarchy of the power counting of $({\rm cos}\beta -1)^k$, where $\beta$ stands for the usual Euler angle.  The $J(J+1)$ rule arises as the exact solution at the NLO (next-to-leading order), i.e., $k$=1, of this power counting, and can be extended to the $J(J+1)-K^2$ rule for bands with $K \ne 0$. 
The derived formulas enable us to assess the effects of 
Hamiltonians on the band-head energy and on the ``moments of inertia'' of rotational bands, in a transparent and precise way.  Here, the Hamiltonian comprises single-particle energies,  nucleon-nucleon forces, three-nucleon forces, {\it etc}.   The formula is also applicable to various problems of the spin physics of composite systems.

\textcolor{black}{The picture adopted by this work} as to how and why the rotational band appears can be stated as follows.
Short-range attractive nuclear forces favor deformed shapes (as explained in sec.~\ref{sec:ellipsoidal deformation}), providing intrinsic states.   Such deformed intrinsic states break rotational symmetry, and the symmetry restoration occurs by giving good angular momentum to eigenstates (nothing but representation of the corresponding symmetry groups).  The same mechanism works for parity, if relevant.  \textcolor{black}{Before} this stage, $J=0^+$ ground state is not trivial, but it is shown that the projection of intrinsic state onto $J=0^+$ gives maximum additional binding energy.  The projection onto higher $J$'s results in a decrease of this additional binding energy, and the difference from $J=0^+$ ground state almost perfectly follows the $J(J+1)$ rule.  The excitation energy can be decomposed according to the components of nuclear forces: monopole proton-proton and neutron-neutron pairing interactions are major contributors to the excitation energy as well as the monopole component and the quadrupole component of proton-neutron central force.  
\textcolor{black}{Interestingly, the proton-neutron tensor and hexadecupole central forces, which are important for medium triaxiality, show minor (opposite in some cases) effects.  The kinetic energy part of single-particle energies appears to yield very minor (negative in some cases) contributions to excitation energies, in contrast to na\"ive expectation of faster rotation (higher kinetic energy) as $J$ increases.}  
This is a sketch of the understanding of nuclear ``rotation'' from the present study.

The present study on the $K$ and $J$ projections is closely related to the restoration of broken symmetry.  As well known, the Nambu-Goldstone Mode (NGM) has been introduced for the symmetry restoration of internal degrees of freedom \cite{nambu_1960,goldstone_1961,goldstone_1962}.  However, the extension of the NGM to geometrical symmetries like the rotational one is known to be difficult to date.   
\textcolor{black}{The picture illuminated by this work} may contribute to it, for instance, by showing the derivation of the dispersion relation.

%The mixing of states of different $K$ values may occur in principle.  
\textcolor{black}{It is shown that the $K$ quantum number is rather well conserved in strongly deformed heavy nuclei, giving a theoretical background to band structures generated by triaxial rotations classified by such K quantum numbers.   Because of this property, the $J(J+1)-K^2$ rule makes sense including finite triaxialities, and the description of band structures in heavy nuclei in terms of their K quantum numbers is well justified.}   

\textcolor{black}{Based on such a systematic description of the structure of low-lying bands, the prevailing triaxiality in heavy deformed nuclei is addressed.  
The preponderance of prolate shapes in heavy nuclei, stressed by Aage Bohr, has been a textbook item (e.g. \cite{bohr_mottelson_book2,rowe_book,deshalit_book,ring_schuck_book}).  Since the pioneer work by Davydov \cite{davydov1,davydov2}, some questions were raised for heavy nuclei, for instance,  \cite{sun_2002,Sharpey-Schafer_2019}, and some theoretical calculations depicted modest triaxialities of certain heavy nuclei  \cite{li_2010,yang_2021,scamps_2021} (see subsec.~\ref{sec:nuclei_around}).  The preponderance of prolate shapes, however, has remained the general belief or paradigm in the dominant majority of the community to date, as proved by basic/general textbooks stated above.
The whole picture arising from this work is clearly different from it.}  
Consequently, the traditional description of the low-lying 2$^+_2$ state by $\gamma$ vibration will be replaced by the rotation of triaxial ellipsoids.
\textcolor{black}{The present} pictures and concepts are supported by state-of-the-art extra-large-scale CI calculations (i.e., Quasiparticle Vacua Shell Model version \cite{shimizu_2021} of the Monte Carlo Shell Model) with realistic effective interaction.   The agreement with measured spectroscopic data is quite good.  Such salient agreement suggests possible prospects of the present approach applied to other cases, like nuclei with an odd number of nucleons, so-called octupole states, {\it etc}.  \textcolor{black}{In particular, the basic concept of the Nilsson model \cite{nilsson_1955} can be incorporated by coupling the odd particle with the $K^P$=0$^+$-projected core, even if the neighboring even-even nuclei are triaxial.  Such studies are in progress, and the outcome is of great interest.}

It is still a major experimental challenge to explicitly or critically extract the triaxial shape of the ground state, {\it albeit with intrinsic difficulty}.    
Magnetic excitation is a possibility: more excitation modes may appear besides the scissors mode \cite{pietralla_1998}.  Such experiments may be done in facilities like HI$\gamma$S and RCNP, while the outcome may be indirect.  Another intriguing possibility lies in relativistic heavy-ion collision as a new tool for scanning the nuclear shape \cite{giacalone_phdthesis_2020,Ryssens_Giacalone_2023,bally_2022,bally_2022b,jia_2022,dimri_2023,giacalone_2023}.  The experiments in this line have been performed in LHC/CERN or RHIC, for instance, \cite{ adamczyk_2015,alice_2018,sirunyan_2019,aad_2020,abdallah_2022,alice_2022,aad_2023}, and more are expected to come.    
Hyper nuclei containing a $\Lambda$ particile may provide with another opportunity to see the triaxiality, in the form of splitting of the level energies, which can be seen in J-Lab or J-PARC.
Another relevant feature is the chirality.  As a result of substantial triaxiality in the strongly deformed region, nuclear chiral doublet bands \cite{frauendorf_2001} may appear 
more widely than expected, providing new experimental opportunities.

Above all these approaches, significant contributions from multiple Coulomb excitations initiated by Cline {\it et al.} \cite{cline_1986} and extended further by Fahlander {\it et al.} \cite{fahlander_1992} and Werner {\it et al.} \cite{werner_2005} should not be overlooked.  Those experimental data already pointed out, decades ago, the same trend as the present theoretical work, but the meanings may not have been appreciated.  More experiments with improved efficiency and precision will be of interest and use.  For the basic modest triaxiality, the E2 excitation from the ground state to the 2$^+_{g\gamma}$ state seems to be of great interest, for instance, in the case of $^{154}$Sm.

The binding energy gain by the deformation is likely more crucial in heavy and superheavy nuclei, and the possible appearance of triaxiality was shown even without including the present mechanism, for instance, \cite{nature_review_SHE,sobiczewski_2007,kowal_SHE_2010,algerian_2017}.  
\textcolor{black}{The triaxiality was mentioned in a beyond-mean-field calculation with the Gogny interaction for $A \sim$290 nuclei, such as $^{284}$Cn and $^{292}$Lv \cite{egido_2021}.}  
The triaxiality has also been addressed over decades as it may lower the first fission barrier \cite{leander}.  As triaxiality is a robust reality, due to the symmetry restoration and certain characteristics of nuclear forces, rather than an accidental incident, it may also produce more visible effects on such heavy nuclei than previously thought.

Regarding nuclei lighter than $A\sim$ 100, Fig.~\ref{fig:Ex2+} displays a few nuclei with the strong shape deformation fulfilling the criterion given there  ($E_x(4^+_1$)/$E_x(2^+_1) \ge$ 3.00).    
This ratio is 2.99 for $^{12}$C nucleus.   The 0$^+_1$ and 2$^+_1$ states of $^{12}$C were investigated in the {\it ab initio} no-core MCSM calculation \cite{otsuka_2022_nature}, depicting that the $K$=0 intrinsic state extracted from the 0$^+_1$ state almost perfectly reproduces, by the $J$ projection of this $K$=0 state, the wave function of the 2$^+_1$ state obtained by the  MCSM calculation, despite substantial triaxiality in both states \cite{otsuka_2022_nature}.   This feature seems to suggest the applicability of the present idea in such a first-principles approach.  Note that the clustering component is contained in these states, but it does not prevent this feature.

Strongly deformed light- and medium-mass nuclei in Fig.~\ref{fig:Ex2+} depict notable triaxiality in their T-plots, likely implying that the triaxiality can be a global phenomenon.  The poperties of the nuclei $^{24}$Mg, $^{34,36,38}$Mg and $^{42}$S are calculated with the USD \cite{usd}, EEdf1 \cite{tsunoda_2020,otsuka_2022} and SDPF-MU \cite{sdpf-mu} interactions, respectively.  The triaxiality of $^{24}$Mg was noticed decades ago, for instance, in shell-model calculation by the USD interaction \cite{usd} \textcolor{black}{as well as Hartree-Fock calculation with the Kuo interaction \cite{watt_1971}.}   Similarly, a Skyrme-based HFB-GCM calculation displayed triaxiality after $J/K$ projection \cite{bender_2008}, \textcolor{black}{and so did a recent GCM calculation with projections and configuration mixings with the Gogny interaction \cite{rodriguez_2010}.}  However, there can be different aspects in the triaxiality of light nuclei compared to heavy strongly deformed nuclei discussed presently, and further investigations, particularly on possible variations of the \textcolor{black}{small and medium} triaxialities, are of interest.   Very interestingly, exotic nuclei $^{34,36,38}$Mg suggest that the triaxiality seems to be enhanced near driplines, shedding light on the relation between the dripline and the ellipsoidal shapes  \cite{tsunoda_2020,otsuka_2022}.

As already mentioned, a number of medium-mass nuclei also exhibit features of the triaxial shapes, e.g., in $^{66}$Zn \cite{66Zn}, $^{74,76}$Zn \cite{7476Zn}, $^{76}$Ge \cite{76Ge,76Geb} and $^{78}$Se \cite{78Se}, \textcolor{black}{$^{72}$Kr \cite{wimmer_2020}, $^{76}$Kr \cite{yao_2014}, $^{70-98}$Kr \cite{rodriguez_2014}, $^{80}$Zr \cite{rodriguez_2011},  but these nuclei depict weaker ellipsoidal deformation (or no sufficient experimental data),} as not included in Fig.~\ref{fig:Ex2+}.   In addition to existing theoretical studies described in these papers, the structures of these nuclei will be discussed in the near future also in the light of the present study.   
In particular, it is of interest to see in what manner the triaxiality mechanisms evolves or remains in such weakly deformed nuclei with large $\gamma$ values like 30$^{\circ}$.   It is naturally expected that 
\textcolor{black}{shell-model details, such as localized details of single-particle structure or two-body interactions, may play more significant or direct roles than in strongly deformed nuclei. }

%degrees of freedom other than ellipsoidal deformation contribute more in such cases, and some other correlations due to nuclear forces may show up.  The CI calculation with realistic interactions might be advantageous, as another important question arises as to how strongly $K$-mixings occur due to substantially smaller $\beta_2$ (or $E_x$(2$^+_1$) $\gtrapprox$ 0.5 MeV) and larger $\gamma$.

Figure~\ref{fig:Ex2+} suggests that nine nuclei, $^{100}$Sr, $^{102,104,106}$Zr, $^{106}$Mo, $^{122}$Ce, $^{128,130}$Nd, $^{132}$Sm, are strongly deformed in the criterion there.  The triaxiality has already been investigated by MCSM calculations including T-plots for the first five nuclei ($Z<$ 50), as will be reported elsewhere \cite{yanase_2024}, where the triaxiality appears more strongly in quite a few Zr isotopes than in the earlier study \cite{togashi_2016}.   It is of interest that all these five nuclei are exotic neutron-rich isotopes with short lifetimes.  The structure of other weakly deformed nuclei will be studied as well, also from the viewpoint of a shell-model approach to so-called $\gamma$-soft nuclei with $Z\sim$54 \cite{zamfir_1991}.

The applications of the formulas of rotational excitation energies to objects other than atomic nuclei are plausible and indeed of great interest.  We already see that higher-order terms in chemistry \cite{atkins} can be justified.   
It is also of major interest to explore, in other physical systems, applications of the present views may open a new understanding.  The \textcolor{black}{small trixiality} due to the symmetry restoration might emerge in triatomic molecules as an explanation of the bending configuration.  Possible effects similar to the ones by the nuclear tensor force may be found:  the interaction between electric dipole moments, carried possibly by molecules, depicts such similarity and may show up in microclusters of water molecules, metal molecules, {\it etc}.  

The present underlying picture of the shapes and the rotations can be applied not only to fermionic systems but also to bosonic systems.  The Interacting Boson Model 
(IBM) of Arima and Iachello describes the structure of deformed nuclei \cite{ibm_book}.  It is of interest as to how the findings obtained in this work, especially those in Secs.~\ref{sec:K} and \ref{sec:J}, can work with bosons.  For instance, the IBM shows no triaxial minimum in unprojected PES, and it is now an intriguing question as to whether the $K$ projection may change this feature or not.  

We finally address that the preponderance of triaxiality in heavy nuclei was suggested by an Ukrainian physicist, A. S. Davydov (Crimea 1912 - Kyiv 1993).   
This suggestion, especially for heavy strongly deformed nuclei discussed in this paper, has been appreciated, not to an adequate extent in the present view.  Putting the unsuccessful energy predictions by the triaxial-rotor-model aside, his idea of the triaxial deformation as the general trend of nuclear shape turned out to be a superb and correct idea, and can be appreciated more appropriately.    Pioneer experimental works with Multiple Coulomb Excitation by D. Cline are to be better appreciated as well.

%%%%%%%%%%%%%%%%%%%%%%%%%%%%%%%%%%%%%%%%%%%%%

\section*{A\lowercase{cknowledgements}}

The authors are grateful to Drs. P. Van Duppen, N. Pietralla, P. von Neumann-Cosel, A. Tamii, G. Giacalone, P. Ring, D. Vretenar, K. Nishio, Y. Aritomo, T. Azuma, K. Yabana and S. Yamamoto for valuable suggestions and/or discussions.  
TO thanks the Alexander von Humboldt Foundation for the Research Award, as the associated stay in Darmstadt resulted in some parts of this work.  TO is grateful to Prof. N. Nagaosa for generous understanding and precious comments.  The MCSM calculations were performed on the supercomputer Fugaku at RIKEN AICS  (hp190160, hp200130, hp210165, hp220174, hp230207, hp240213).  
This work was supported in part by MEXT as "Priority Issue on Post-K computer" (Elucidation of the Fundamental Laws and Evolution of the Universe) (hp160211, hp170230, hp180179, hp190160) and "Program for Promoting Researches on the Supercomputer Fugaku" (Simulation for basic science: from fundamental laws of particles to creation of nuclei, JPMXP1020200105, Simulation for basic science: approaching the new quantum era, JPMXP1020230411), and by JICFuS.  
This work was supported by JSPS KAKENHI Grant Numbers JP19H05145, JP21H00117, JP21K03564, JP20K03981, JP17K05433  and JP18H05462.  

\section*{Author Contributions}
%\section*{A\lowercase{uthor Contributions}}
T.O. promoted the whole study and wrote the manuscript; Y.T. performed the CI calculations, drew T-plots and contributed to code development and in-depth discussions; N.S. made the main part of the computer codes; Y.U. gave various crucial remarks; T.A. and H.U. made valuable discussions.   All authors discussed the results and commented on the manuscript.

%%%%%%%%%%%%%%%%%%%%%%%%%%%%%%%%%%%%%%%%%%%%%
%%                                                 A P P E N D I  X
%%%%%%%%%%%%%%%%%%%%%%%%%%%%%%%%%%%%%%%%%%%%%

\appendix

\section{Configuration Interaction (CI) calculation or Shell-model calculation}
\label{Ap_shell}

\noindent
We sketch the shell model for atomic nuclei in this Appendix.
The shell-model calculation is one of the standard methods for the nuclear many-body problem \cite{deshalit_book,ring_schuck_book,talmi_book,caurier_2005}.  It belongs to the category of the Configuration Interaction (CI) calculation, which is more familiar to a broad audience and is also used in this article for the meaning of the shell model.   The ingredients of the shell model are (i) single-particle orbitals and their energies, (ii) the numbers of protons and neutrons in these orbitals, (iii) nucleon-nucleon ($NN$) interaction.  Thus, the properties of the nuclear states are determined by them, without other a priori assumptions.  The protons and neutrons can move in these orbitals, scattering each other through the $NN$ interaction adopted.   These protons and neutrons do not include the nucleons in the inert core, as it is a closed shell and is treated as a vacuum.    
The single-particle energies (SPE) and the $NN$ interaction are taken from some models and/or theories.
The matrix elements of the SPE and the $NN$ interaction are expressed, as one- and two-body operators, respectively,  with respect to all possible combinations of single-particle states, which are magnetic substates of each single-particle orbital.  

The Hamiltonian consists of one-body term and two-body term, as usual.
The one-body term is expressed by the SPEs, and the two-body term is expressed in terms of the matrix elements of the $NN$ interaction with respect to antisymmetrized two-nucleon states.  The Hamiltonian is thus constructed, and the actually used Hamiltonian is mentioned in the main text.

The many-body states are described by superpositions of Slater determinants in many {\it conventional} shell-model calculations.
The many-body Schr\"odinger equation is solved for the given Hamiltonian as,  
\begin{equation}
H \, \Psi \,=\, E \, \Psi,             
\label{eq:Schr}
\end{equation}
where $\Psi$ is an eigenstate for an eigenvalue $E$, and the Hamiltonian {\it matrix} for this many-body system is represented by matrix elements of the Hamiltonian {\it operator} for all combinations (bra and ket vectors) of all possible Slater determinants.
By diagonalizing this Hamiltonian matrix, energy eigenvalues and wave functions of eigenstates are obtained.  We can calculate various physical quantities from these wave functions.  
The number of Slater determinants is called the shell model dimension.  
This is the general framework of the shell model, while many-body states can be equivalently expressed otherwise.

The shell model dimension becomes huge in many interesting cases, restricting the actual feasibility of the calculation.  This is the major obstacle of the conventional shell-model calculation, and the current limit of the shell-model dimension is around 10$^{11}$ as of 2019 \cite{shimizu_2019}.  In order to overcome this difficulty, the Monte Carlo Shell Model was introduced, as described in the next Appendix.

The practical setup of the present work is mentioned \cite{otsuka_2019}.  Proton single-particle orbitals are, 1$g_{9/2,7/2}$, 2$d_{5/2,3/2}$, 3$s_{1/2}$, 1$h_{11/2}$, 2$f_{7/2}$, and 3$p_{3/2}$.   The neutron single-particle orbitals are, 1$h_{11/2,9/2}$, 2$f_{7/2,5/2}$, 3$p_{3/2,1/2}$, 1$i_{13/2}$, 2$g_{9/2}$, 3$d_{5/2}$ and 4$s_{1/2}$.  
These orbitals define the model space, which is built on top of the $Z$=40 and $N$=70 magic numbers of the harmonic oscillator potential.
In the case of $^{166}$Er, 28 protons and 28 neutrons are put into the model space formed by these orbitals.  
The model space and the number of nucleons are much larger than those in the usual shell-model calculation so that the shape deformation can be described.  The shell-model dimension becomes as large as 4.8$\times$10$^{33}$ for the case of $^{166}$Er.  This is far beyond the limit of the conventional shell-model calculation but can be overcome by using the MCSM described in the next Appendix.

\section{Monte Carlo Shell Model and T-plot}
\label{Ap_MCSM}

\noindent
In this Appendix, we briefly describe the Monte Carlo Shell Model (MCSM).
The MCSM was initially proposed in Ref. \cite{mcsm_1995}.  A prototype of its present version was shown in Ref. \cite{mcsm_1998}.  The method and applications of the MCSM were reviewed, for instance, in Refs. \cite{mcsm_2001,mcsm_2012,mcsm_2017}.  
The MCSM is the methodology fully exploited in this work.
%, by incorporating very recent developments.  
Although this work uses the advanced version of MCSM, we here outline its earlier version, for simplicity.   It is formulated with Slater determinants as the basis vectors, similar to the conventional shell-model calculation.  
However, the Slater determinants are not the same as those used in the conventional shell-model calculation.   In the conventional one, each Slater determinant is a direct product of some single-particle states, each of which is a magnetic substate of the single-particle orbital.   

\textcolor{black}{
An MCSM eigenstate is written as (see \cite{otsuka_2022_emerging} for more detailed but pedagogical concise explanation),
\begin{equation}
 |\Psi\rangle=\sum_{i,K} f_{i,K}{\mathcal P}_{J,M,K} |\phi^{(i)}\rangle
%%%%%%\Psi \, = \, \sum_k \, f_k \, \hat{{\mathcal P}}_{J^{P}} \, \phi_k \,\,, 
\label{eq:mcsm_psi}
\end{equation}
where $f_{i,K}$ denotes the amplitude, ${\mathcal P}_{J,M,K}$ means the projection operator on to the  $J, M, K$ (parity projection is also included, but not explicitly shown for brevity), and $\phi^{(i)}$ stands for the $i$-th MCSM basis vector. The integer index $K$ runs from $-J$ to $J$, because such $K$ values generate independent basis vectors for a given set of $J$ and $M$.  
In the MCSM with Slater determinants, 
$\phi^{(i)}$ = $\Pi_k \, a^{(i)\dagger}_k \,|0 \rangle$.  Here, $|0 \rangle$ is the inert core (closed shell), 
$a^{(i)\dagger}_k$ refers to a superposition, 
\begin{equation}
\label{eq:mcsm_bv} 
a^{(i)\dagger}_k \,=\, \sum_n \, D^{(i)}_{n,k} \, c^{\dagger}_n \,\, ,
\end{equation}
with $c^{\dagger}_n$ being the creation operator of a usual single-particle state, and $D^{(i)}_{n,k}$ denoting a matrix element.  
}

By choosing an optimal matrix $D^{(i)}$, we can select $\phi^{(i)}$ so that such $\phi^{(i)}$ better contributes to the lowering of the corresponding energy eigenvalue.  
The value of the amplitude $f_{i,K}$ is determined through the diagonalization of $H$ with obtained basis vectors.  
Thus, the determination of $D^{(k)}$ is the core of the MCSM calculation. The index $k$ runs up to 50-100, but can be more.  These are much smaller than the dimension of the many-body Hilbert space, which is 4.8$\times$10$^{33}$ for the case of $^{166}$Er, as already mentioned.

Thus, the basis vectors of the MCSM calculation are composed of ``stochastically - variationally deformed'' single-particle states.  The adopted basis vectors are mutually independent, otherwise there is no energy gain.    By having a set of these MCSM basis vectors thus fixed, we diagonalize the Hamiltonian and obtain energy eigenvalues and eigenstates.  
A large number of MCSM calculations have been performed as exemplified in Refs. \cite{tsunoda_2014,togashi_2016,otsuka_2016,leoni_2017,marsh_2018,togashi_2018,ichikawa_2019,otsuka_2019,taniuchi_2019,tsunoda_2020,marginean_2020,abe_2021,otsuka_2022_nature}.

\textcolor{black}{ 
For the analysis of the K mixing (see subsec.~\ref{subsec:Kmix}), $K$-projected state $|\Psi_K\rangle$ is defined as
\begin{equation}
    |\Psi_K\rangle={\mathcal N}_K\sum_{i} f_{i,K} {\mathcal P}_{J,M,K}| \phi^{(i)} \rangle,
    \label{eq:Psi_K}
\end{equation}
where ${\mathcal N}_{K}$ is a normalization factor for $\langle \Psi_{K} | \Psi_{K} \,\rangle$=1.
After fixing the values of $f_{i,K}$, $\phi^{(i)}$ is rotated so that the longest ellipsoidal axis of each $\phi^{(i)}$ is aligned in the same direction, called the $z$ axis.  The amplitudes $f_{i,K}$ are consistently transformed.  This operation does not change MCSM results of observables.  
The longest axis can be rotated by 180$^{\circ}$.    
In order to incorporate this possibility, $| \, \tilde{\Psi}_{K} \, \rangle$ is defined with $K \ge 0$, similarly to eq.~(\ref{eq:Psi_K}), as
\begin{equation}
    | \, \tilde{\Psi}_{K} \, \rangle={\tilde{\mathcal N}}_{K}
    \sum_{i,K'=\pm K} f_{i,K'}{\mathcal P}_{J,M,K'}|\phi^{(i)}\rangle,
\end{equation}
while obviousely $K'$=0 if $K$=0.
Table II shows the overlap probability $|\langle\Psi \, | \, \tilde{\Psi}_{K} \rangle|^2$, for certain $K$ values, by using the QVSM instead of the MCSM with Slater determinants.   Note that the overlap probability does not depend on $M$.  
}

%%%%%%%%%%  original  tex text  %%%%%%%%%%%%%%%%%%%%%
%%MCSM wave function $|\Psi\rangle$ is written as
%%\begin{equation}
%%    |\Psi\rangle=\sum_{i,K}f_{i,K}{\mathcal P}_{J,M,K}|\phi_i\rangle
%%\end{equation}
%%$K$-projected state $|\Psi_K\rangle$ is defined as
%%\begin{equation}
%%    |\Psi_K\rangle={\mathcal N}_K\sum_{i}f_{i,K}{\mathcal P}_{J,M,K}|\phi_i\rangle
%%\end{equation}
%%where ${\mathcal N}_K$ is a normalization factor
%%satisfying $\langle\Psi_K|\Psi_K\rangle=1$.
%%Similarly, $|\Psi_{|K|=K'}\rangle$ is defined as
%%\begin{equation}
%%    |\Psi_{|K|=K'}\rangle={\mathcal N}_{|K|=K'}
%%    \sum_{i,K=\pm K'}f_{i,K}{\mathcal P}_{J,M,K}|\phi_i\rangle
%%\end{equation}
%%Table II shows overlap probability $|\langle\Psi|\Psi_{|K|=K'}\rangle|^2$.
%%%%%%%%%%%%%%%%%%%%%%%%%%%%%%%%%%%%%%%%%

Besides the breakthrough in the computational limit, the MCSM also has the advantage of providing a very useful way to visualize the intrinsic shape of each MCSM eigenstate through what is called the T-plot \cite{tsunoda_2014,otsuka_2016}.  
Because the MCSM basis vector is a deformed Slater determinant, 
\textcolor{black}{the matrix elements of quadrupole operator in the polar coordinates,
\begin{equation} 
    Q_k=\sqrt{16\pi/5} \, r^2Y_{2k},
\end{equation}%\begin{equation} 
can be given in the Cartesian coordinates as,
%    Q=
%    \left(
%    \begin{array}{ccc}
%        {\rm Re}(Q_2)-\frac{1}{\sqrt{6}}Q_0&{\rm Im}(Q_2)&-{\rm Re}(Q_1)\\
%        {\rm Im}(Q_2)&-{\rm Re}(Q_2)-\frac{1}{\sqrt{6}}Q_0&-{\rm Im}(Q_1)\\
%        -{\rm Re}(Q_1)&-{\rm Im}(Q_1)&\frac{2}{\sqrt{6}}Q_0
%    \end{array}
%    \right)
%\end{equation}
\begin{eqnarray} 
    Q&=&\frac{3}{2}
    \left(
    \begin{array}{ccc}
        x^2-\frac{1}{3}r^2&xy&xz\\
        xy&y^2-\frac{1}{3}r^2&yz\\
        xz&yz&z^2-\frac{1}{3}r^2
    \end{array}
    \right)  %\nonumber\\
\end{eqnarray}    
where each matrix element stands for the expectation value of the displayed operator with respect to the MCSM basis vector being considered.  
%This matrix can be transformed back to one in the polar coordinates as
%\begin{eqnarray}    
%    Q&=&\frac{\sqrt{6}}{4}
%   \left(
%    \begin{array}{ccc}
%        {\rm Re}(Q_2)-\frac{Q_0}{\sqrt{6}}&{\rm Im}(Q_2)&-{\rm Re}(Q_1)\\
%        {\rm Im}(Q_2)&-{\rm Re}(Q_2)-\frac{Q_0}{\sqrt{6}}&-{\rm Im}(Q_1)\\
%        -{\rm Re}(Q_1)&-{\rm Im}(Q_1)&\frac{2Q_0}{\sqrt{6}}
%    \end{array}
%    \right)
%\end{eqnarray}
%\begin{eqnarray} 
%    Q&=& \frac{3}{2} \, 
%        \left(
%    \begin{array}{ccc}
%        x^2-\frac{1}{3}r^2&xy&xz\\
%        xy&y^2-\frac{1}{3}r^2&yz\\
%        xz&yz&z^2-\frac{1}{3}r^2
%    \end{array}
%    \right) \nonumber\\
%    &=&
%    \left(
%    \begin{array}{ccc}
%        {\rm Re}(Q_2)-\frac{1}{\sqrt{6}}Q_0&{\rm Im}(Q_2)&-{\rm Re}(Q_1)\\
%        {\rm Im}(Q_2)&-{\rm Re}(Q_2)-\frac{1}{\sqrt{6}}Q_0&-{\rm Im}(Q_1)\\
%        -{\rm Re}(Q_1)&-{\rm Im}(Q_1)&\frac{2}{\sqrt{6}}Q_0
%    \end{array}
%    \right) \nonumber\\
%%%%%%%%%%%    &\propto&
%\end{eqnarray}
By diagonalizing this matrix, we obtain three eigenvalues in the order
$Q_{zz}\ge Q_{xx}\ge Q_{yy} $ by choosing coordinate axes.
%\begin{equation} 
%    Q=
%    \left(
%    \begin{array}{ccc}
%        Q_{xx}&0&0\\
%        0&Q_{yy}&0\\
%        0&0&Q_{zz}
%    \end{array}
%    \right)
%\end{equation}
%By choosing coordinate axes so that the following relations are fulfilled, 
%\begin{equation} 
%    Q_{zz}\ge Q_{xx}\ge Q_{yy} ,
%\end{equation}
We obtain 
\begin{equation} 
%    Q_0=\frac{\sqrt{6}}{2}Q_{zz},\quad Q_2=\frac{1}{2}(Q_{xx}-Q_{yy}).
    Q_0=2Q_{zz},\quad Q_2=\frac{2}{\sqrt{6}}(Q_{xx}-Q_{yy})
\end{equation}
}

\textcolor{black}{
The quantities $Q_0$ and $Q_2$ are expressed by two parameters $\beta_2$ and $\gamma$, as described in the main text (see also \cite{utsuno_2015} and \cite{otsuka_2022}).  Such $\beta_2$ and $\gamma$ are used to pin down each MCSM basis vector by a certain symbol on the so-called $\beta-\gamma$ plane.  This process and the resulting set of the symbols are called T-plot.}   The importance of each MCSM basis vector to a given eigenstate (its overlap probability in the MCSM eigenstate) is represented by the size (area) of its circular symbol in the T-plot.  The T-plot can be made on the $\beta_2$-$\gamma$ plane, but is usually on the PES, and intuitively exhibits the underlying physical pictures for the states of interest, as demonstrated in a variety of studies, {\it e.g.} in Refs. \cite{leoni_2017,marsh_2018,ichikawa_2019,togashi_2016,togashi_2018,otsuka_2019,marginean_2020,otsuka_2022_nature}.

We stress that the T-plot played a major role in the present study, displaying not only the triaxial shapes of the eigenstates but also their rigidity over different eigenstates.   Some details of T-plot are also found in \cite{mcsm_2017}.

As mentioned at the beginning of this Appendix, the present calculation was performed by the most advanced methodology of the MCSM.  This is called Quasiparticle Vacua Shell Model (QVSM) \cite{shimizu_2021}.  In the original version of the MCSM, the pairing correlations are mainly incorporated by superposing different MCSM basis vectors, which are deformed Slater determinants as stated above.  The QVSM basis vectors are somewhat like a Hartree-Fock-Bogoliubov ground state (which is a generalization of the BCS ground state), and this feature enables each QVSM basis vector to contain both effects of the deformed mean field and effects of the pairing correlations.  This advantage is particularly exploitable for heavy nuclei where the pairing correlations over different single-particle orbitals become more important than for lighter nuclei.  In the original MCSM, the pairing correlations are largely carried by superpositions of different MCSM basis vectors.  As the QVSM calculation is computationally heavier, this merit makes sense for some nuclei heavier than $A\sim$100. 
\textcolor{black}{The T-plot analysis can be extended to the QVSM straightforwardly.}

The spurious center-of-mass motion is removed to a sufficient extent by the Lawson method \cite{lawson_method}.

%%%%%%%%%%%%%%%%%%%%%%%%%%%%%%%%%%%%%%%%%%%%%%%%

%\def\thesection{\Alph{section}}
\section{K projection}
\label{Ap_Kproj}

We here present some detailed discussion on the $K$ projection.  
The $K$-projected state was defined in eq.~(\ref{eq:K-proj}) as,
\begin{equation}
\Phi \bigl[\phi, K\bigr] \,=\, \frac{1}{{\mathcal N_K}} \frac{1}{2\pi}  \int_0^{2\pi} d\gamma \, e^{i\gamma (\hat{J}_z - K)} \, \phi ,
\label{eq:K-proj+}
\end{equation}
where $\gamma$ is the angle in the $xy$ plane, and $\mathcal{N}_K$ is a normalization constant with an arbitrary phase factor,
\begin{equation}
|{\mathcal N_K}|^2  \,=\, \frac{1}{2\pi}  \int_0^{2\pi} d\gamma \, \langle \, \phi \,|\, 
e^{i\gamma (\hat{J}_z - K)} \,  | \, \phi \rangle .
\label{eq:N0}
\end{equation} 
We introduce the norm kernel,
\begin{equation}
n(\gamma)  \,=\,  \langle \, \phi \,|\, e^{i\gamma \hat{J}_z} \,  | \, \phi \rangle.
\label{eq:Nkernel}
\end{equation} 

Although $n(\gamma)$ looks like a complex number, it is a real number with the time reversal symmetry by the following reason.
The intrinsic state is expanded by components of definite $K$ values, $K$= 0, $\pm$1, $\pm$2, ...  In calculating the norm kernel, 
the magnitudes of amplitudes in this expansion are considered to remain unchanged for the exchange of the signs of $K$ values, i.e., $K$ and -$K$. This property originates in the time reversal symmetry which is conserved in eigenstates.  
%Although the time-reversal symmetry is conserved in eigenstates, it can be violated in individual basis vectors.   Nevertheless, this symmetry is well preserved in major basis vectors, otherwise its restoration would be difficult.  It can be proved by simple mathematical manipulations that this property makes the norm kernel real.
Likewise, matrix elements of the Hamiltonian to be discussed shortly are real.

The cases with $\gamma$ and $2\pi-\gamma$ are equivalent because of the same relative angle, and only a phase factor may come in.
For the real norm kernel, $n(\gamma)$=$n(2\pi-\gamma)$ holds because of no phase ambiguity for a real number.  Eq.~(\ref{eq:N0}) is then rewritten as
\begin{equation}
|{\mathcal N_K}|^2 = \frac{1}{2\pi}  \int_0^{2\pi} d\gamma \, e^{- i\gamma K} \, n(\gamma) %\nonumber \\ 
 = \frac{1}{2\pi}  \int_0^{2\pi} d\gamma \, {\rm cos}(K\,\gamma) \, n(\gamma),
\label{eq:NK2}
\end{equation} 
%\begin{eqnarray}
%&|{\mathcal N_K}|^2  &\,=\, \frac{1}{2\pi}  \int_0^{2\pi} d\gamma \, e^{- i\gamma K} \, n(\gamma), \nonumber \\ 
%& & \,=\, \frac{1}{2\pi}  \int_0^{2\pi} d\gamma \, {\rm cos}(K\,\gamma) \, n(\gamma),
%\label{eq:NK2}
%\end{eqnarray} 
with $|{\mathcal N_K}|$ being the same for both signs of $K$.  
The state $\phi$ is assumed, in this study, to have a positive parity, and the projection actually implies the $K^+$ projection.  

The weighting factor in the integral of eq.~(\ref{eq:NK2}) is ${\rm cos}(K\,\gamma)$.  The $K$=0 projection is simple, where contributions from all $\gamma$ angles are equally superposed as
\begin{equation}
\Phi \bigl[\phi, K=0\bigr] \,=\, \frac{1}{{\mathcal N_{0}}} \frac{1}{2\pi}  \int_0^{2\pi} d\gamma \, e^{i\gamma \hat{J}_z} \, \phi  . 
\label{eq:K=0}
\end{equation}
For $K$=2, however, this factor takes values such as 1, 0, -1, 0, 1, respectively, for $\gamma = 0, \pi/4, \pi/2, 3\pi/4, \pi$.  This is in contrast to the unity factor for $K$=0 (see eq.~(\ref{eq:K=0})).

Considering that $H$ is invariant under the rotation, 
the expectation value of the Hamiltonian $H$ is given for $K$=0 by 
\begin{eqnarray}
& h_{K=0} \,&=\, \langle \Phi \bigl[\phi, K=0\bigr] \, | H \, | \Phi \bigl[\phi, K=0\bigr] \rangle \nonumber \\
 & &=\, \frac{1}{|{\mathcal N}_0|^2} \frac{1}{{2\pi}} \, \int_0^{2\pi} d\gamma \, \langle \phi \, | \, H \, | \, e^{i\gamma \hat{J}_z} \, \phi \,\rangle . 
\label{eq:H K=0 pre}
\end{eqnarray}
Introducing the energy kernel similarly to the norm kernel, 
%\begin{equation}
$h(\gamma)  \,=\,  \langle \, \phi \,|\, H e^{i\gamma \hat{J}_z} \,  | \, \phi \rangle$,
%\label{eq:Ekernel}
%\end{equation} 
we obtain
\begin{equation}
h_{K=0} \,=\, \frac{1}{|{\mathcal N}_0|^2} \frac{1}{{2\pi}} \, \int_0^{2\pi} d\gamma \, h(\gamma). 
\label{eq:H K=0}
\end{equation}
Like the norm kernel, $h(\gamma)$ is a real quantity if the time-reversal symmetry holds.
 
For $K \ne$ 0, we obtain 
\begin{eqnarray}
& h_{K} \,&=\langle  \Phi \bigl[\phi, K\bigr] \, | H \, | \Phi \bigl[\phi, K\bigr] \rangle 
\nonumber \\
 & &=\frac{1}{|{\mathcal N}_K|^2} \, \frac{1}{2\pi} \, \int_0^{2\pi} d\gamma \,\langle \phi \, | H \, | \,  e^{i\gamma (\hat{J}_z - K)} \,  \phi \rangle  \nonumber \\ 
%% & &=\frac{1}{|{\mathcal N}_K|^2} \, \frac{1}{{2\pi}} \, \int_0^{2\pi} d\gamma \, e^{i K \gamma}  \,  h(\gamma),  \nonumber \\ 
 & &=\frac{1}{|{\mathcal N}_K|^2} \, \frac{1}{{2\pi}} \, \int_0^{2\pi} d\gamma \, {\rm cos}(K\,\gamma) \, h(\gamma).
\label{eq:H K pre}
\end{eqnarray}
where a similar argument to eq.~(\ref{eq:NK2}) is used.  

We look into contributions in eq.~(\ref{eq:H K=0}).  By introducing a complete set of states orthogonal to $\phi$ denoted by $\zeta$'s, we decompose $h_{K=0}$ into the diagonal and off-diagonal contributions as
\begin{eqnarray}
&h_{K=0} &= \frac{1}{|{\mathcal N_0}|^2} \frac{1}{2 \pi} \int_0^{2\pi} d\gamma \nonumber \\ 
& &  \,\,\,\,\,\,\, \Bigl\{ \langle \phi \, | \, H \, | \, \phi \, \rangle \, \langle \phi  | e^{i\gamma \hat{J}_z} \, \phi \,\rangle 
+ \sum_{\zeta}  \langle \phi \, | \, H \, | \, \zeta \, \rangle \, \langle \zeta | e^{i\gamma \hat{J}_z} \, \phi \,\rangle \Bigr\}  \nonumber \\ 
%&& = \, \langle \phi \, | \, H \, | \, \phi \, \rangle \nonumber \\
%& &\,\,\,+ \frac{1}{|{\mathcal N_0}|^2} \frac{1}{2 \pi} \, \int_0^{2\pi} d\gamma \, \sum_{\zeta} \langle \phi \, | \, H \, | \, \zeta \, \rangle \, \langle \zeta | e^{i\gamma \hat{J}_z} \, |\,  \phi \,\rangle \Bigr\} \nonumber \\
& & =  \langle \phi  |  H  |  \phi \rangle + \frac{1}{2 \pi |{\mathcal N_0}|^2} \int_0^{2\pi} d\gamma \Bigl\{h(\gamma) - h(\gamma=0) n(\gamma) \Bigr\}. 
\label{eq:H K=0decomp}
\end{eqnarray}
The first term on the right-hand-side (RHS) is the expectation value with respect to $\phi$ and provides a major contribution.  It is often called unprojected energy.  The state $\phi$ is, however, not an eigenstate of $H$ in general, and the second term then does not vanish.   The only exception arises for $\gamma$=0; no $\zeta$ is generated by the operation $e^{i\gamma \hat{J}_z} \, \phi$.

%%%%%%%%%%%%  FIGURE 25   (earlier 26)  %%%%%%%%%%%%%
% Fig: kernels

\begin{figure}[bt]
  \centering
  \includegraphics[width=8.5cm]{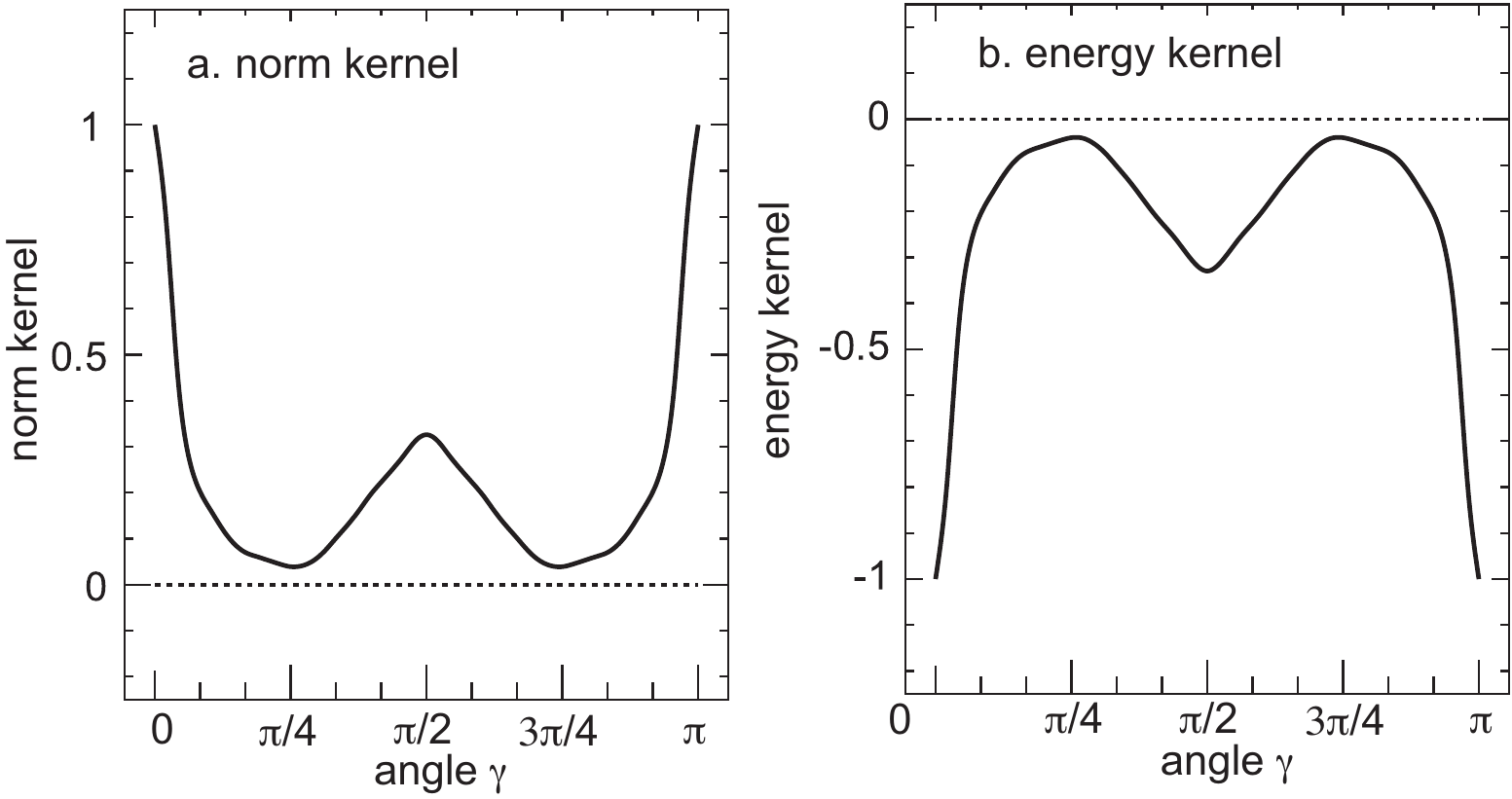}
    \caption{{\bf a} norm kernel and {\bf b} energy kernel as functions of the angle $\gamma$
    in radian.
    Each kernel value is scaled by the magnitude of the original value at $\gamma$=0.
    The case of $^{166}$Er is taken.  
} 
  \label{fig:n_h}  
\end{figure}  

%%%%%%%%%%%%%%%%%%%%%%%%%%%%%%%%

Figure \ref{fig:n_h} shows actual values of the norm kernel, $n(\gamma)$, and the energy kernel, $h(\gamma)$, obtained from the state $\xi_0$ in the discussion for $^{166}$Er in Sec.~\ref{sec:gamma band} (the most important basis vector for the ground state).  As expected, $n(\gamma)$ decreases from $n(\gamma=0)$ as $\gamma$ increases.  Likewise, the magnitude of $h(\gamma)$ decreases, but the $\gamma$ dependence is not equal between the two.

\begin{figure}[tb]
  \centering
  \includegraphics[width=8.5cm]{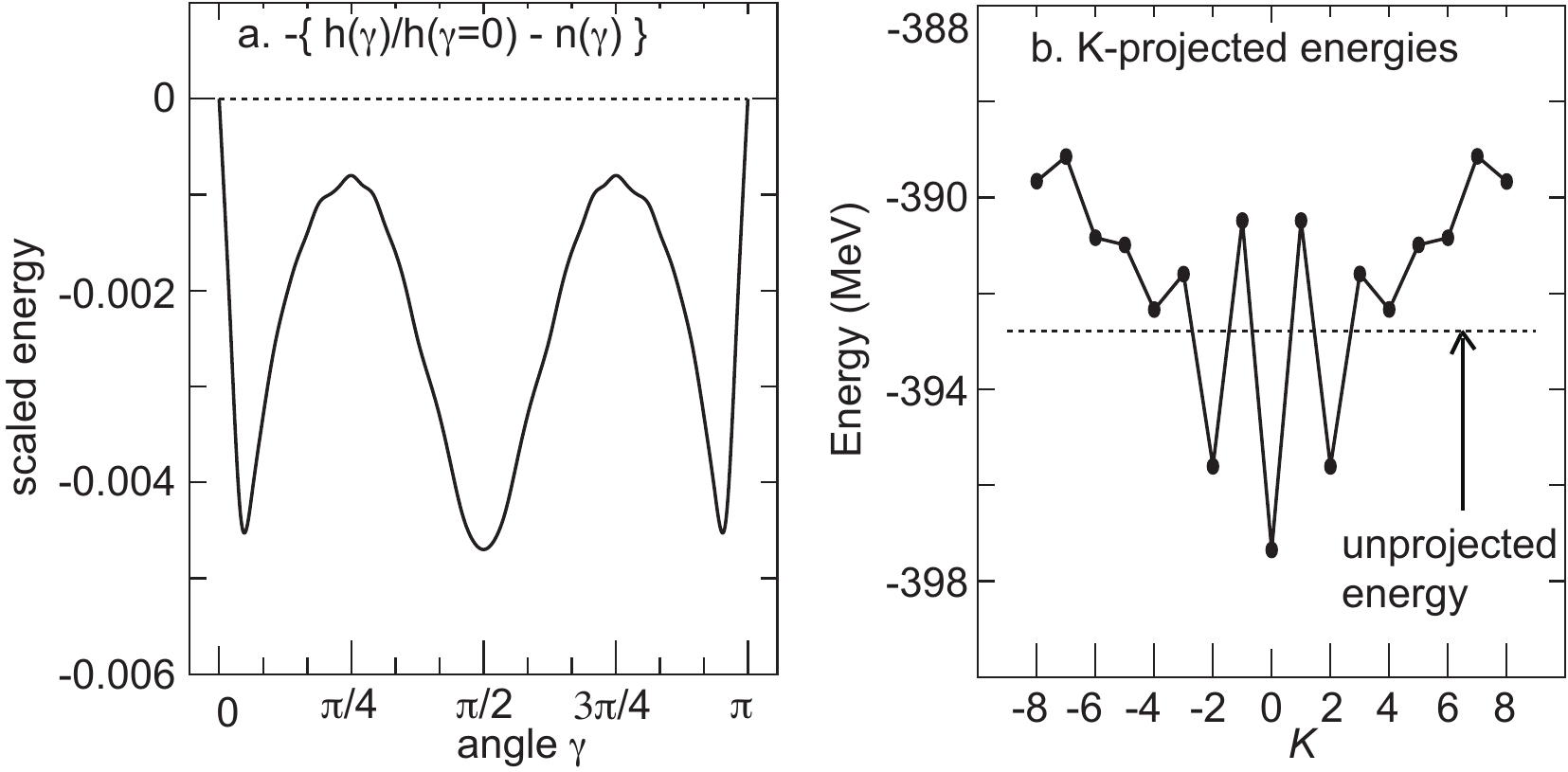}
    \caption{a. $-\{h(\gamma)/h(0)-n(\gamma)\}$ with parity projection and b. energies of $K^+$-projected states.
    The unprojected energy (with parity projection) is shown by a dotted horizontal line.  
    } 
  \label{fig:h-hn}  
\end{figure}

Figure \ref{fig:h-hn} (a) shows the quantity $h(\gamma) - h(\gamma=0) \, n(\gamma)$, which appears in the integrand of the RHS of eq.~(\ref{eq:H K=0decomp}).  This quantity remains negative (more binding) for the whole region of $\gamma$.  This gives more binding energy to the $K$=0 state, as an average of this quantity becomes the change of the energy of the $K$=0 state.  
Figure \ref{fig:h-hn} (b) shows the energies of $K$-projected states.
It indicates that the $K$=0 state gains the binding energy by about 4 MeV from the unprojected energy.    

The contribution represented by the quantity, $h(\gamma)-h(\gamma=0) \, n(\gamma)$, thus provides additional substantial binding energy to triaxial states, while it vanishes for the states of no triaxiality (as far as the quadrupole degree of freedom is concerned).  
The underlying origin of this binding energy gain is different $\gamma$ dependences of the two kernels, $n(\gamma)$ and $h(\gamma)$.  
The magnitude of the latter decreases more gradually as a function of $\gamma$ than the former, probably because various correlations from the $NN$ interaction, including finite-range attraction, blur displacement effects between bra and ket vectors, and hence weaken the effects of decreasing overlaps represented by $n(\gamma)$.  We note that this is an effect for a given triaxiality or a given value of the deformation parameter $\gamma$, but also that the amount of this effect varies as the triaxiality changes.   

%%%%%%%%%%%%  FIGURE 26   (earlier 27)  %%%%%%%%%%%%%

%%%%%%%%%%%%%%%%%%%%%%%%%%%%%%%%

It is of interest what contribution the same term produces for $K \ne$0. 
The quantity $h_K$ is rewritten in a similar way to eq.~(\ref{eq:H K=0decomp}) as
\begin{eqnarray}
&h_{K} &= \langle \phi \, | \, H \, | \, \phi \, \rangle \, + \frac{1}{|{\mathcal N_K}|^2} \frac{1}{2 \pi} \nonumber \\
& & \,\,\,\int_0^{2\pi} d\gamma \,  {\rm cos}(K\gamma) \Bigl\{h(\gamma) -h(\gamma=0) n(\gamma)  \Bigr\} .  
\label{eq:H K}
\end{eqnarray}
The first term on the RHS is again the unprojected energy, and the $K$ dependence is carried by the second term.

We first discuss the $K$=2 case.  The factor ${\rm cos}(2\gamma)$ in eq.~(\ref{eq:H K}) varies from 1 for $\gamma=0$, to
0 for $\gamma=\pi/2$ and $-1$ for $\gamma=\pi$.  This is in stark contrast to a constant factor eq.~(\ref{eq:H K=0decomp}).  The quantity, $-\{h(\gamma) /h(\gamma=0)\,-\, n(\gamma)\}$, is shown in Fig.~\ref{fig:h-hn} (a).  This quantity keeps the sign of $h(\gamma)$, and exhibits deviations of the relative change, $h(\gamma) /h(\gamma=0)$, from $n(\gamma)$, as $n(\gamma=0)$=1.   It remains negative (attractive) for the entire region of $\gamma$, but is multiplied by ${\rm cos}(2\gamma)$, which flips between opposite signs.  Thus, a certain cancellation occurs, and the net effect becomes less attractive.  This pushes up the excitation energy of the $K$=2 (projected) state relative to the $K$=0 state, being consistent with Fig.~\ref{fig:Kproj}.  Fig.~\ref{fig:h-hn} (b) also displays the energy of the $K$=2 (projected) state, which is about 2 MeV above the $K$=0 (projected) state.  On the other side, the $K$=2 energy is well below the unprojected energy, implying the triaxiality gives a substantial amount of the binding energy also to the $K$=2 state.

For states of other $K$ values, the energies can be similarly calculated, with more frequently changing multiplication factors.  Figure~\ref{fig:h-hn} (b) shows the energies of such $K$-projected states.  For $K$=4, the ${\rm cos}(4\gamma)$ factor causes cancellations so severe as to result in almost no net additional binding energy.  For $K$=1, there are cancellations between contributions from $\gamma \sim0$ and $\gamma \sim\pi$.  For higher $K$ values, the factors make the contribution even more positive (repulsive), placing the energies of such states above the unprojected ones.

%%%%%%%%%%%%%%%%%%%%%%%%%%%%%%%%%%%%%%%%%%%%%
%%%  Rotational  Energies
%%%%%%%%%%%%%%%%%%%%%%%%%%%%%%%%%%%%%%%%%%%%%

\section{Remarks on rotational energies} 
\label{Ap_rot_energy}

\textcolor{black}{This is Appendix presenting details following the introduction in subsec.~\ref{subsec:remarks}. 
As a pioneering work on the multi-nucleon picture of rotational bands, we recall} 
the generator coordinate method (GCM) applied by Peierls and Yoccoz \cite{peierls_yoccoz_1957}.
The following trial wave function in the laboratory frame is considered,
\begin{equation}
\Psi \,=\,  \int d\theta \, d\phi \, \chi(\theta, \phi) \, \Phi(R_{\theta\phi} {\bf x}) ,
\label{eq:peierls}
\end{equation} 
where $\theta$ and $\phi$ are angles, $\chi$ stands for the generating function to be chosen, $\Phi$ is the wave function in the body-fixed frame with many-body coordinates shown collectively by ${\bf x}$, and $R_{\theta\phi}$ means the rotation operator for angles $\theta$ and $\phi$. 
\textcolor{black}{This is clearly a GCM idea:}
the generating function was taken to be proportional to spherical harmonics $Y_{lm} (\theta, \phi)$ so that stationary solutions are obtained for the variational equation.  Although the quantity $l$ is discussed as if it is the angular momentum of the nucleus, the angular momentum is carried by the coordinates ${\bf x}$.  So, this expression does not seem to correspond to the present work.
On the other hand, as $Y_{lm}$ happens to be equal to the function $d^l_{m,0}$, the projection onto $l$ incidentally occurs in this expression, with no such statement in \cite{peierls_yoccoz_1957}.  This method was limited to $K$=0, and the general formula for the J(J+1)-K$^2$ rule is outside the \textcolor{black}{view} of this approach.  \textcolor{black}{After all,} there seems to be no idea of the application of $J$ projection 
\textcolor{black}{of an intrinsic state} in \cite{peierls_yoccoz_1957}.   
We also demonstrated that the J(J+1) rule arises exactly at the NLO in the power counting with the $({\rm cos}\beta - 1)^k$ terms, in contrast to the arguments on the polynomial expansion of $\beta$ in \cite{peierls_yoccoz_1957}.  Nevertheless, it is somewhat of interest that the result exhibits an apparent resemblance to a part of the present results.

As another pioneering approach, the so-called Kamlah expansion was introduced by Kamlah in \cite{kamlah_1968} by assuming that the energy kernel $h_y$ can be represented well by the norm kernel $n_y$ and its derivatives with 
some factors adjusted at $\beta$=0.  The important feature found by the present work is the difference between $n_y$ and $h_y$ apart from their scales (and units).  In fact, in the present work, no relations between them are assumed. In this sense, the present work may not have any direct connection to the Kamlah expansion.  The J(J+1)-K$^2$ rule does not come up in the approach \cite{kamlah_1968}.
 
%The above two works were known as referred in some papers/books (\cite{ring_schuck_book,} for both, \cite{bohr_mottelson_book2} for \cite{peierls_yoccoz_1957}, \cite{mang_1975} for \cite{kamlah_1968}).    
Another pioneering work is the one by Verhaar published in 1963-4 \cite{verhaar_1963a,verhaar_1963b,verhaar_1964}.  This work may have not been widely recognized for unknown reasons.  
%Although it was referred by \cite{ring_schuck_book} but not directly for the rotational spectrum of even-even nuclei.  Other papers do not refer to the papers by Verhaar.  
This work \textcolor{black}{is closest to the present work, as it} showed results which correspond to the approximation of the present results by replacing  (cos$\beta$ -1) terms in sec.~\ref{sec:J} by -1/2 $\beta^2$.  
\textcolor{black}{It was obtained by a Taylor expansion of Wigner's $d$ function.  As we discussed in sec.~\ref{sec:J}, the $d$ function is the Legendre polynomial comprising terms, (cos$\beta$ -1)$^n$ (n=0, 1, ...), and the (cos$\beta$ -1) term is the next lowest-power part of it.  The (-1/2 $\beta^2$) term is an approximation neglecting higher powers of $\beta$ in (cos$\beta$ -1).}   Namely, the two steps, the truncation at the lowest two terms in the Legendre polynomial and the approximation of cos$\beta$ by $\beta^2$, were carried out at once in the method of Verhaar \cite{verhaar_1963a,verhaar_1963b,verhaar_1964}.  We mention that the second step does not reduce numerical task at all, and that its meaning or advantage is unclear.  The same formulas as those shown in Verhaar work \cite{verhaar_1963a,verhaar_1963b,verhaar_1964} were presented in the textbook of Siemens and Jensen \cite{siemens_book}, without citing Verhaar papers.  It was stated in \cite{siemens_book} that the formulas being considered are not quantitatively good, but we were unable to clarify the actual meaning of this message. 
%We note that the above works were overviewed in \cite{wong_1975}.   }  

\textcolor{black}{We noticed that a good amount of efforts were devoted for finding possible relations between the cranking idea and rotational spectrum (see for instance \cite{ring_schuck_book}).   This objective likely remains open.}

%%%%%%%%%%%%%%%%%%%%%%%%%%%%%%%%%%%%%%%%%%%%%%%%%%%%%%
%%%      
%%%%%%%%%%%%%%%%%%%%%%%%%%%%%%%%%%%%%%%%%%%%%%%%%%%%%%

\section{Large triaxiality in weakly deformed nuclei} % and the self-organization}
\label{Ap_large}

\textcolor{black}{
The triaxiality has been studied in many works, especially for the cases of large triaxiality.  As discussed in subsec.~\ref{subsec:Kmix} and subsec.~\ref{subsec:large}, large triaxiality likely occur in weakly deformed nuclei.} 

\textcolor{black}{
For weakly deformed nuclei, the projection on angular momentum and parity can be important because the deformed mean field dominates the structure for such nuclei to lesser extent than for strongly deformed nuclei.     
A series of attempts were made by Enami {\it et al.} \cite{enami_2000,enami_2001,enami_2002a,enami_2002b}, basically claiming that ``spin-projected potential energy surface'' is a better choice for exploring nuclear shape, with various models from single-j-shell to realistic shells.   The pairing plus QQ interaction was used.   The spin-projection here means the $J$=$K$=0 projection without separated discussions for K and J projections.}

\textcolor{black}{
As mentioned in sec.~\ref{subsec:166Er_Tplot}, triaxial shapes have been studied extensively for medium-mass nuclei with weaker ellipsoidal deformations, such as $^{74,76}$Zn \cite{7476Zn}, $^{76}$Ge \cite{76Ge,76Geb} or $^{78}$Se \cite{78Se}, $^{76}$Kr \cite{yao_2014}.  This trend continues to heavier regions, for instance, to Kr isotopes \cite{rodriguez_2014,wimmer_2020} and $^{80}$Zr \cite{rodriguez_2011}, where the Gogny interaction was treated with projections.}

%%%%%%%%%%%%%%%%%%%%%%%%%%%%%%%%%%%%%%%%%%%%%

%%%%%%%%%%%%%%%%%%%%%%%%%%%%%%%%%%%%%%%%%%%%%

\makeatletter
\renewcommand\@biblabel[1]{#1.}
\makeatother

\def\bibsection{\section*{\bf references}}
%%%%%%%%%%


\begin{thebibliography}{99}
\bibitem{note_circle}
Here, by deformation parameter $\gamma$=0$^{\circ}$, we symbolically mean a perfect invariance for the rotation about the $z$ axis.  If other deformation modes occur, more conditions are needed.  Such details are omitted for simplicity.
\end{thebibliography}
\end{document}